\documentstyle[11pt,aaspp4]{article}
\begin{document}
\renewcommand{\textfraction}{0}
\newcommand{\beq}{\begin{equation}}
\newcommand{\eeq}{\end{equation}}
\newcommand{\beqa}{\begin{eqnarray}}
\newcommand{\eeqa}{\end{eqnarray}}
\newcommand{\grad}{\mbox{${\bf \nabla}$}}
\newcommand{\Interp}[1]{\mbox{$\left\langle #1 \right\rangle$}}
\newcommand{\Sub}[2]{\mbox{$#1_{\mbox{\scriptsize #2}}$}}
\newcommand{\Sup}[2]{\mbox{$#1^{\mbox{\scriptsize #2}}$}}
\newcommand{\SupSub}[3]{\mbox{$#1_{#3}^{\mbox{\scriptsize #2}}$}}
\newcommand{\Wspline}[1]{\mbox{$W^{#1 \mbox{\scriptsize -D}}_
                              {\mbox{\scriptsize spline}}$}}
\newcommand{\Aspline}[1]{\mbox{$A^{#1 \mbox{\scriptsize -D}}_
                              {\mbox{\scriptsize spline}}$}}
\newcommand{\WgaussP}[1]{\mbox{$W^{#1 \mbox{\scriptsize -D}}_
                              {\mbox{\scriptsize Gauss2}}$}}
\newcommand{\AgaussP}[1]{\mbox{$A^{#1 \mbox{\scriptsize -D}}_
                              {\mbox{\scriptsize Gauss2}}$}}
\newcommand{\Wi}{\mbox{$W({\bf r}_i - {\bf r}_j,h)$}}
\newcommand{\Wijfull}{\mbox{$\frac{1}{2} [W({\bf r}_i - {\bf r}_j,h_i) +
                        W({\bf r}_i - {\bf r}_j,h_j)]$}}
\newcommand{\gWi}{\mbox{$\grad_i W({\bf r}_i - {\bf r}_j,h)$}}
\newcommand{\gWijfull}{\mbox{$\frac{1}{2}
                         [\grad_i W({\bf r}_i - {\bf r}_j,h_i) + 
                          \grad_i W({\bf r}_i - {\bf r}_j,h_j)]$}}
\newcommand{\Wij}{\mbox{W$_{ij}$}}
\newcommand{\gWij}{\mbox{$\grad \mbox{W}_{ij}$}}
\newcommand{\gxWij}{\mbox{$\grad_x \mbox{W}_{ij}$}}
\newcommand{\WPij}{\mbox{W$^\Pi_{ij}$}}
\newcommand{\gWPij}{\mbox{$\grad \mbox{W}^\Pi_{ij}$}}
\newcommand{\gxWPij}{\mbox{$\grad_x \mbox{W}^\Pi_{ij}$}}
\newcommand{\rij}{\mbox{${\bf r}_{ij}$}}
\newcommand{\vij}{\mbox{${\bf v}_{ij}$}}
\newcommand{\xij}{\mbox{${\bf x}_{ij}$}}
\newcommand{\wij}{\mbox{${\bf w}_{ij}$}}
\newcommand{\eps}{\mbox{$\varepsilon$}}
\newcommand{\hx}{\mbox{$h _{1}$}}
\newcommand{\hy}{\mbox{$h _{2}$}}
\newcommand{\hz}{\mbox{$h _{3}$}}
\newcommand{\hix}{\mbox{$h _{1}^{-1}$}}
\newcommand{\hiy}{\mbox{$h _{2}^{-1}$}}
\newcommand{\hiz}{\mbox{$h _{3}^{-1}$}}
\newcommand{\vxx}{\mbox{$\frac{\partial v_x}{\partial x}$}}
\newcommand{\vxy}{\mbox{$\frac{\partial v_x}{\partial y}$}}
\newcommand{\vxz}{\mbox{$\frac{\partial v_x}{\partial z}$}}
\newcommand{\vyx}{\mbox{$\frac{\partial v_y}{\partial x}$}}
\newcommand{\vyy}{\mbox{$\frac{\partial v_y}{\partial y}$}}
\newcommand{\vyz}{\mbox{$\frac{\partial v_y}{\partial z}$}}
\newcommand{\vzx}{\mbox{$\frac{\partial v_z}{\partial x}$}}
\newcommand{\vzy}{\mbox{$\frac{\partial v_z}{\partial y}$}}
\newcommand{\vzz}{\mbox{$\frac{\partial v_z}{\partial z}$}}
\newcommand{\Jt}{\mbox{${\bf J}$}}
\newcommand{\Ht}{\mbox{${\bf H}$}}
\newcommand{\Gt}{\mbox{${\bf G}$}}
\newcommand{\SGt}{\mbox{$\langle {\bf G} \rangle$}}
\newcommand{\Tr}{\mbox{${\bf T_r}$}}
\newcommand{\Gtp}{\mbox{${\bf G}'$}}
\newcommand{\Rt}{\mbox{${\bf R}$}}
\newcommand{\Bt}{\mbox{${\bf B}$}}
\newcommand{\dlxx}{\mbox{$\delta l_{11}$}}
\newcommand{\dlxy}{\mbox{$\delta l_{12}$}}
\newcommand{\dlxz}{\mbox{$\delta l_{13}$}}
\newcommand{\dlyx}{\mbox{$\delta l_{21}$}}
\newcommand{\dlyy}{\mbox{$\delta l_{22}$}}
\newcommand{\dlyz}{\mbox{$\delta l_{23}$}}
\newcommand{\dlzx}{\mbox{$\delta l_{31}$}}
\newcommand{\dlzy}{\mbox{$\delta l_{32}$}}
\newcommand{\dlzz}{\mbox{$\delta l_{33}$}}
\newcommand{\sigt}{\mbox{\boldmath$\sigma$}}
\newcommand{\veta}{\mbox{\boldmath$\eta$}}
\newcommand{\dgamma}{\mbox{$\dot{\gamma}$}}
\newcommand{\dtheta}{\mbox{$\dot{\theta}$}}
\newcommand{\dphi}{\mbox{$\dot{\phi}$}}
\newcommand{\lp}{\left(}
\newcommand{\rp}{\right)}
\newcommand{\TrG}{\mbox{Tr(\Gt)}}
\newcommand{\etal}{{\frenchspacing et al.}}
\newcommand{\ie}{{\frenchspacing i.e.}}
\newcommand{\eg}{{\frenchspacing e.g.}}
\title{Adaptive Smoothed Particle Hydrodynamics: Methodology II}
\author{J. Michael Owen \altaffilmark{1}}
\affil{Department of Astronomy, Ohio State University, Columbus, OH
43210, USA \\ 
Email: mikeowen@llnl.gov}
\altaffiltext{1}{Current Address:  LLNL, L-16, Livermore, CA 94551}
\author{Jens V. Villumsen}
\affil{Max Planck Institut f\"{u}r Astrophysik, Karl Schwarzschild Strasse 1,
85740 Garching bei Munchen, Germany \\
Email: victoria@infinet.com}
\author{Paul R. Shapiro and Hugo Martel}
\affil{Department of Astronomy, University of Texas at Austin, Austin, TX
78712, USA \\
Email: shapiro@astro.as.utexas.edu, hugo@sagredo.as.utexas.edu}
%\date{\today}

\begin{abstract}
Further development and additional details and tests of Adaptive Smoothed
Particle Hydrodynamics (ASPH), the new version of Smoothed Particle
Hydrodynamics (SPH) described in Shapiro \etal\ (1996; Paper I) are
presented.  The ASPH method replaces the isotropic smoothing algorithm of
standard SPH, in which interpolation is performed with spherical kernels of
radius given by a scalar smoothing length, with anisotropic smoothing
involving ellipsoidal kernels and tensor smoothing lengths.  In standard
SPH the smoothing length for each particle represents the spatial
resolution scale in the vicinity of that particle, and is typically allowed
to vary in space and time so as to reflect the local value of the mean
interparticle spacing.  This isotropic approach is not optimal, however, in
the presence of strongly anisotropic volume changes such as occur naturally
in a wide range of astrophysical flows, including gravitational collapse,
cosmological structure formation, cloud-cloud collisions, and radiative
shocks.  In such cases, the local mean interparticle spacing varies not
only in time and space, but in {\em direction} as well.  This problem is
remedied in ASPH, where each axis of the ellipsoidal smoothing kernel for a
given particle is adjusted so as to reflect the different mean
interparticle spacings along different directions in the vicinity of that
particle.  By deforming and rotating these ellipsoidal kernels so as to
follow the anisotropy of volume changes local to each particle, ASPH adapts
its spatial resolution scale in time, space, and direction.  This
significantly improves the spatial resolving power of the method over that
of standard SPH at fixed particle number per simulation.

This paper presents an alternative formulation of the ASPH algorithm for
evolving anisotropic smoothing kernels, in which the geometric approach of
Paper I, based upon the Lagrangian deformation of ellipsoidal fluid
elements surrounding each particle, is replaced by an approach involving a
local transformation of coordinates to those in which the underlying
anisotropic volume changes appear to be isotropic.  Using this formulation
the ASPH method is presented in 2D and 3D, including a number of details
not previously included in Paper I, some of which represent either advances
or different choices with respect to the ASPH method detailed in Paper I.
Among the advances included here are an asynchronous time-integration
scheme with different time steps for different particles and the
generalization of the ASPH method to 3D.  In the category of different
choices, the shock-tracking algorithm described in Paper I for locally
adapting the artificial viscosity to restrict viscous heating just to
particles encountering shocks, is not included here.  Instead, we adopt a
different interpolation kernel for use with the artificial viscosity, which
has the effect of spatially localizing effects of the artificial viscosity.
This version of the ASPH method in 2D and 3D is then applied to a series of
1D, 2D, and 3D test problems, and the results are compared to those of
standard SPH applied to the same problems.  These include the problem of
cosmological pancake collapse, the Riemann shock tube, cylindrical and
spherical Sedov blast waves, the collision of two strong shocks, and
problems involving shearing disks intended to test the angular momentum
conservation properties of the method.  These results further support the
idea that ASPH has significantly better resolving power than standard SPH
for a wide range of problems, including that of cosmological structure
formation.
\end{abstract}


\section{Introduction}
As reviewed by Monaghan (1992), the Smoothed Particle Hydrodynamics (SPH)
method has proved extremely useful and versatile in simulating a wide range
of astrophysical flow problems, from the formation of stars, planets, and
moons, to the supernova explosions of massive stars, to the collision of
stars, to the formation of galaxies and large-scale structure in cosmology,
to name a few. In recent years, the SPH method, in combination with one of
a variety of gravitational N-body solvers, has played a leading role in
simulations of cosmological gas dynamics (e.g. Evrard 1988; Hernquist \&
Katz 1989, hereafter HK89; Steinmetz \& M\"uller 1993; Navarro \& White
1993; Couchman, Thomas, \& Pearce 1995; Tissera, Lambas, \& Abadi 1997).
For such flow problems, which involve gravitational collapse, supersonic
velocities, and the build-up of extremely nonlinear density and pressure
contrasts, the SPH method has several advantages over traditional Eulerian
finite-difference methods. As a Lagrangian method, SPH is naturally able to
adjust its length resolution scale so as to follow the mass as it flows to
create a highly inhomogeneous density. In addition, SPH replaces the
spatial grid used in traditional finite-difference methods by an
interpolation scheme based upon the smoothing of numerical data over
irregularly spaced, neighboring mass points -- the ``particles'' in SPH --
thereby eliminating the problem which traditional grid-based, Lagrangian
schemes have when the grid distorts too much and twists upon itself. A
large body of work now exists to show that the SPH method, which is based
upon the standard conservation equations of compressible gas dynamics, does
indeed yield correct solutions of those equations, as long as the simulated
gas is represented by a large enough number of particles, just as Eulerian
finite-difference schemes yield correct solutions if the number of grid
cells is large enough.

The great complexity and inhomogeneity of astrophysical flows like those 
studied in cosmology, however, requires an enormous number of particles (or,
equivalently, grid cells in the case of Eulerian schemes) in order to resolve
the multi-scale phenomena of structure formation arising from a field of
Gaussian-random-noise, initial density perturbations, as typically assumed.
Unfortunately, current computer hardware is inadequate to handle the
required minimum number of SPH particles, in general, so some sacrifice of
accuracy is inevitable. As a result, any effort to improve the accuracy and
resolving power of the SPH method at fixed number of particles is extremely
worthwhile. One such attempt was presented by us in Shapiro et al. (1996,
hereafter Paper I), in which we replaced the isotropic smoothing algorithm
of SPH, which utilizes spherical interpolation kernels characterized by a
scalar smoothing length which varies in space and time according to the local
variations of the gas density, by an anisotropic smoothing algorithm which uses
ellipsoidal kernels characterized by a different smoothing length along each
axis of the ellipsoid and varies these three axes so as to follow the value
of the local mean separation of particles surrounding each particle, as it
changes in time, space, and {\it direction}. A scheme was devised whereby the
ellipsoidal kernels of this anisotropic smoothing algorithm were evolved
continuously during a simulation so as to track the local deformations and
rotations of the fluid. By analogy with the concept of adaptive mesh 
refinement introduced in Eulerian finite-difference methods in order to replace
the traditional spatial grid with one which varies the cell size in time and
space automatically so as to respond to the local need for higher spatial
resolving power, we termed our modified SPH algorithm with evolving
ellipsoidal kernels, ``Adaptive Smoothed Particle Hydrodynamics (ASPH).''

The particular motivation for this algorithmic advance was the realization
that the isotropic SPH smoothing algorithm, in which the smoothing length is
traditionally adjusted in proportion to $\rho^{-1/3}$ (where $\rho$ is
the gas density), is adequate for isotropic volume changes but is seriously
mismatched to the generic anisotropic volume changes which occur in problems
involving gravitational collapse, the collision and merging of subclumps, and
strong shock waves, all common to cosmic structure formation. This problem was
first discussed by Shapiro (1989) and Shapiro, Kang, \& Villumsen (1991),
in which the first use of anisotropic smoothing in SPH for cosmological gas
dynamics was made, in order to simulate a 1D, planar, cosmological pancake
collapse with a 3D, SPH code. There, the anisotropic kernels were all fixed
in orientation and, for each, the axis perpendicular to the pancake plane was
adjusted in proportion to $\rho^{-1}$ in order to reflect the 1D, planar 
geometry of collapse. The result was a significant improvement in the 
resolution of the pancake shocks and postshock flow compared to the 
traditional isotropic smoothing approach. A second problem which standard
SPH has with flows like this was also noted. As had been discussed by Shapiro
and Struck-Marcell (1985) before, in their 1D, Lagrangian treatment of pancake
hydrodynamics, standard artificial viscosity causes gas undergoing supersonic
collapse to be strongly heated, as in the pancake problem, in which gas 
outside the two strong shocks near the central plane, which parallel the 
central plane on either side, falls supersonically toward that plane. In
principle, however, this heating should be restricted to gas which is
actually encountering the shock. It was found to be necessary, in fact, to
suppress the unphysical viscous heating of collapsing gas prior to shock
formation, caused by the presence of artificial viscosity, in order to
resolve the shock and postshock flow correctly. This second problem led 
Shapiro et al. (1996) to a second modification of SPH which Paper I describes
as part of the ASPH method, that of automating the suppression of viscous
heating for gas not undergoing a shock transition. In Paper I, this was
accomplished by using the anisotropic kernel calculations to predict particle
caustic formation as a ``shock-tracking'' algorithm. Only particles which were
thereby identified as those in imminent danger of
encountering a shock were allowed to experience
artificial viscous heating. As demonstrated in detail in Paper I, the new 
algorithms of ASPH resulted in a substantial improvement of resolving power 
over the standard SPH method, at fixed particle number. The reader is referred
to Paper I for the detailed demonstration of the relative shortcomings of
SPH in comparison with ASPH, including several tests of ASPH, as well as a
description of the algorithmic details.

The idea of using anisotropic sampling kernels with SPH dates back to
Bicknell \& Gingold (1983), who first utilized ellipsoidal SPH kernels in
order to study the tidal flattening of stars involved in close-encounters
with black holes.  They exploited the special geometry of that problem,
simplifying the evolution of the flattened kernel shapes.  Shapiro, Martel,
\& Villumsen first began seriously investigating a generalized approach to
the problem of using ellipsoidal sampling in SPH (Shapiro \etal\ 1993;
Shapiro, Martel, \& Villumsen 1994; Martel \etal\ 1994), motivated by the 
shortcomings noted above in SPH simulations of idealized problems typical of
cosmological structure formation (\ie, the Zel'dovich pancake).  This
investigation ultimately led to the ASPH formulation presented in Paper I,
in which the scalar smoothing length $h$ of standard SPH is generalized to
a smoothing tensor \Ht, whose 9 components define three vectors ${\bf h}_k$
which correspond to the three axes of the smoothing ellipsoid for each
particle (in 2D, the \Ht\ tensor has four components, defining the axes of
a smoothing ellipse).  In principle, these ellipsoid axes ${\bf h}_k$ point
along the principal axes of the local particle density distribution, with
lengths proportional to the mean separation of particles along these
directions.  In practice, it is necessary to provide a mathematical
prescription for evolving the \Ht\ tensor for each particle simultaneously
with the time integration of the SPH dynamical equations.  In Paper I the
mathematical prescription given for calculating and evolving the H tensor
is a time-integration based upon the components of the deformation tensor
($\partial v_\alpha/\partial x_\beta$).  A geometric approach is taken to
evolve the three axes ${\bf h}_k$ of the smoothing ellipsoid at time $t$
into new axes at $t + \Delta t$, by treating the ellipsoid as a Lagrangian
fluid element whose axes are given by ${\bf h}_k$ and using the tensor
$\partial v_\alpha/\partial x_\beta$ explicitly to evolve the ellipsoidal
fluid element.  To first order in $\Delta t$ and $\Delta x$, if the surface
bounding a fluid element at time $t$ is an ellipsoid, then the new surface
which bounds the same fluid element after a time $\Delta t$ has elapsed
must also be an ellipsoid.  This approach is used explicitly to derive the
evolution equations for the components of the \Ht\ tensor for the case of
2D flows.  The generalization of the mathematical derivation to 3D, while
straightforward, was left for a future publication.

In this paper we present an alternative mathematical formalism for evolving
the ASPH smoothing scale.  We adopt the viewpoint that the anisotropic
volume changes represented by the smoothing ellipsoid can be expressed as a
local, linear transformation of coordinates to those in which the
underlying anisotropic volume changes appear to be isotropic.  This
coordinate transformation, expressed in terms of a tensor \Gt\ with units
of inverse length, represents the (direction dependent) amount by which the
actual interparticle distances must be rescaled in order to make the mean
interparticle separations independent of direction in the transformed
coordinates.  The components of \Gt\ are uniquely and simply related to
those of the \Ht\ tensor defined in Paper I, since the two prescriptions
are mathematically equivalent to first-order in $\Delta t$ and $\Delta x$.
In fact, the directions of the eigenvectors of \Gt\ and \Ht\ are identical,
while the associated eigenvalues of \Gt\ are the inverse of those for \Ht.
Expressed in the frame in which \Gt\ and \Ht\ are both diagonal, that is,
the diagonal elements of G are just $h_k^{-1}$, where $h_k$ are the
diagonal elements of \Ht.  In order to evolve \Gt\ in time simultaneously
with the solution of the hydrodynamical conservation equations, we evaluate
\Gt\ by equating the inverse square of \Gt\ with a quantity similar to the
inertia tensor of the smoothing ellipsoid.  This yields a set of
first-order differential equations for the time evolution of \Gt, once
again involving the components of the deformation tensor, $\partial
v_\alpha/\partial x_\beta$.  The mathematical derivation for the evolution
of \Gt\ is valid for arbitrary dimension, and we explicitly evaluate the
form of both \Gt\ and its time derivative $D\Gt/Dt$ in 2D and 3D.  This
represents the first time that the mathematical prescription for evolving
ellipsoidal smoothing kernels has been presented explicitly in 3D.

There are a number of minor implementation differences between this work
and that of Paper I.  For instance, we do not use the same scheme for
suppressing the artificial viscosity.  The approach to the artificial
viscosity presented here does not require violating energy conservation,
and does not include any free parameters which are tweaked to suit a
particular problem.  On the downside, the algorithm presented here is not
as successful at suppressing artificial viscous preheating in the
Zel'dovich pancake problem as that of Paper I, though it does improve over
the standard implementation of artificial viscosity used with SPH and is
general in application, allowing us to test a wide variety of problems.  We
have also developed a number of efficient numerical algorithms in order to
make ASPH applicable to large scale problems, including an adaptive
asynchronous time integrator, such that each ASPH node is allowed to
possess its own timestep and current time, described in detail in Appendix
\ref{Spheral.app}.  We investigate a number of test problems, including: the
Zel'dovich pancake problem (to compare the formalism presented here with
that of Paper I); the Sedov Blastwave solution; standard hydrodynamical
tests such as the Riemann shocktube and the problem of two interacting
strong shocks as described in Woodward (1982) and Woodward \& Colella
(1984); and a pair rotational problems designed to test the conservation
of angular momentum under ASPH.

This paper is organized as follows.  In \S \ref{SPH.sec} and \S
\ref{ASPH.sec} we describe the techniques of SPH and ASPH, deriving ASPH
based upon SPH.  In \S \ref{tests.sec} we present various tests and
comparisons of these techniques.  In \S \ref{Disc.sec} we summarize the
results of this paper and discuss directions for future work.  In Appendix
\ref{Dyneqs.app} we present the detailed numerical (A)SPH dynamical
equations (both for proper and comoving coordinates).  Appendix
\ref{Genderiv.app} presents the full mathematical formalism and derivation
for defining and evolving the ASPH smoothing tensor.  Finally, Appendix
\ref{Spheral.app} outlines some of the major algorithms developed in order
to efficiently implement ASPH numerically.

\section{Standard SPH}
\label{SPH.sec}
ASPH is a generalization of SPH, and therefore throughout this paper we
refer to ``Standard'' SPH as both a starting point and comparison for ASPH.
Unfortunately such a standard version does not exist, as there are a
variety of subtly different ways to implement SPH.  Therefore in Appendix
\ref{Dyneqs.app} we present the formalism for what we call Standard SPH.
This discussion, in combination with what is discussed in the body of this
paper, is complete, but brief.  The reader who wishes a more in-depth
introduction to SPH is referred to the reviews of Monaghan (1992) or Benz
(1990), either of which provide an excellent introduction to this subject.
For the experienced practitioner of SPH, we simply state that our
implementation of SPH consists of evolving the momentum and specific
thermal energy for each node based upon the Lagrangian conservation
equations, accounting for pressure and gravitational forces as well as
radiative cooling.  The mass density is updated using the summation
approach.  We implement a standard Monaghan-Gingold artificial viscosity
(Monaghan \& Gingold 1983) defined on a pairwise basis, with the shearing
correction suggested by Balsara (1995).  We have derived the (A)SPH
dynamical equations based upon these choices for proper time-domain
coordinates, as well as comoving coordinates expressed as a function of a
power of the cosmological expansion factor (for use in cosmological
simulations).  Finally, we allow a variable smoothing scale which evolves
in accordance with the continuity equation as outlined by Benz (1990).  We
discuss this SPH method of evolving the smoothing scale in \S
\ref{DhsDt.sec}.

\section{SPH vs.\ ASPH}
\label{ASPH.sec}
We will now introduce the basic notation and concepts necessary to
understand ASPH and how it relates to SPH.  As SPH and ASPH are so similar,
much of our discussion is applicable to both.  In such ambiguous situations
we refer to the technique as (A)SPH.  The major distinction between SPH and
ASPH is the manner in which smoothing scales are defined and evolved, and
therefore we will concentrate on this subject.  This section is organized
as follows: in \S \ref{interp.sec} we discuss how smoothing scales are
defined under both SPH and ASPH, as well as how they are used in order to
make interpolated estimates of local quantities; in \S \ref{DhsDt.sec} we
discuss how a variable smoothing scale is evolved under SPH; finally in \S
\ref{DGDt.Sec} we describe how the anisotropic smoothing of ASPH is
justified and evolved.  Note that we discuss ASPH descriptively here,
simply presenting the resulting 2D evolution equations without their
derivation.  We defer a detailed discussion of the mathematics upon which
ASPH is formulated until Appendix \ref{Genderiv.app}, where the complete
2D and 3D derivations may also be found.

\subsection{Interpolation under SPH \& ASPH}
\label{interp.sec}
(A)SPH functions by numerically solving the Lagrangian conservation
equations at a series of discrete points or nodes, which are interrelated
through an interpolation scheme.  In this way (A)SPH resembles the various
flavors of Lagrangian finite-difference methods, the major distinction
being the method of interpolation.  Under a traditional Lagrangian
finite-difference approach the interpolation nodes are required to be
arranged on an underlying geometry or grid, while (A)SPH's interpolation
scheme makes no such restrictions.

Under the SPH formalism, the interpolated value of a quantity $F$ at some
spatial position ${\bf r}$ is defined through the integral
\beq
  \label{Wint.eq}
  \Interp{F({\bf r})} = \int F({\bf r}') W({\bf r} - {\bf r}', h) \, 
                        d{\bf r}'.
\eeq
The function $W({\bf r},h)$ is called the interpolation kernel, where $h$
represents the SPH smoothing scale.  The interpolation kernel has the
properties
\beq
  \label{Norm.eq}
  \int W({\bf r} - {\bf r}',h) \, d{\bf r}' = 1,
\eeq
\beq
  \label{Wlim.eq}
  \lim_{h \to 0} W({\bf r} - {\bf r}', h) = \delta ({\bf r} - {\bf r}').
\eeq
Note from equation (\ref{Wlim.eq}) that although the integration of
equation (\ref{Wint.eq}) formally extends over all space, in fact the most
important contributions to the integral occur within a few $h$ of
the position ${\bf r}$.  This reflects the localized nature of hydrodynamic
interactions.

In SPH applications, $h$ is defined as a scalar quantity associated with
each discrete SPH node.  Thus, SPH interpolation about any given node using
such a scalar smoothing scale is isotropic in nature.  In other words, each
SPH node samples a spherical volume about itself, regardless of the
physical conditions in which it is embedded.  As we discuss in \S
\ref{DGDt.Sec}, in general this is not optimal.  In order to implement an
anisotropic smoothing scheme, we require a more generalized methodology for
defining smoothing scales.  We begin by noting that the interpolation
kernel can be written as a function of $\veta \equiv {\bf r}/h$, such
that we can restate $W({\bf r},h) = W({\bf r}/h) = W(\veta)$.  In
general we refer to $\veta$ as the normalized position vector.

We can now introduce a generalized method of mapping from real to
normalized position space (${\bf r} \to \veta$), which we define
through a linear transformation \Gt.  In comparison with SPH, this relation
is
\begin{center}
  SPH:  $\veta = {\bf r}/h \quad \to \quad$
  ASPH: $\veta = \Gt {\bf r}$.
\end{center}
Clearly the \Gt\ tensor has the units of an inverse length-scale.  The
simplest generalization of the spherical smoothing implemented by SPH is to
allow smoothing in elliptical (2D) or ellipsoidal (3D) volumes.  Placing
such a restriction upon \Gt\ implies that it must be a real, symmetric
matrix.  Our scheme under ASPH then is to associate such an ellipsoidal
\Gt\ tensor with each computational node, taking the place of the scalar
smoothing scale $h$ of SPH.  Under this formalism, SPH can be thought of as
a special case of ASPH, where the \Gt\ tensor is diagonal and each diagonal
element is equal to $1/h$.

Under this notation it is simple to define an ASPH kernel estimate.  We
need only replace the SPH method of defining $\veta$ with the ASPH
method in the kernel function $W(\veta) = W(\Gt {\bf r})$.  However,
we also require the spatial gradient of the interpolation kernel $\grad W$
in order to implement the (A)SPH dynamical equations.  This can be
expressed through equation (\ref{gW.eq}), so long as we are neglecting
$\grad \Gt$ terms (an important assumption we will return to in \S
\ref{SHASPH.sec})
\beq
  \label{gW.eq}
  \grad W(\Gt {\bf r}) = \frac{\partial W(\Gt {\bf r})}{\partial {\bf r}} 
          = \frac{\partial \veta}{\partial {\bf r}} 
            \frac{\partial W}{\partial \veta} 
          = \Gt \frac{\veta}{\eta} \frac{\partial W}{\partial \eta}.
\eeq

Before we can go on to express how to use \Gt\ in making kernel
interpolations, we must deal with how to symmetrize such estimates.  This
issue arises for SPH with a variable smoothing scale as well.  The problem
can be understood by examining equation (\ref{Wint.eq}), which expresses
the interpolated value of some general quantity $F$ in terms of a volume
integral involving $h$.  This equation makes no mention of how to deal with
a smoothing scale which varies spatially $h({\bf r})$.  One could choose to
either use $h({\bf r})$ (known as a ``gather'' formalism) or $h({\bf r}')$
(``scatter'').  However, it is advantageous to spatially symmetrize the
kernel estimation process such that $W({\bf r},{\bf r}', h({\bf r}),
h({\bf r}')) = W({\bf r}', {\bf r}, h({\bf r}'), h({\bf r}))$ in order to
symmetrize the (A)SPH dynamical equations, thereby ensuring rigorous
conservation of quantities such as linear momentum.  In order to accomplish
this we adapt the symmetrization scheme of Hernquist \& Katz (HK89), which
defines a symmetrized kernel estimate as the average of both the gather and
scatter approaches.  In terms of two discrete positions ${\bf r}_i$ and
${\bf r}_j$, the symmetrized kernel function \Wij\ is
\beq
  \label{Wij.eq}
  \Wij \equiv \frac{1}{2} \left[ W(\veta_i) + W(\veta_j) \right],
\eeq
where
\beq
  \veta_i \equiv \Gt_i \rij, \quad \veta_j \equiv \Gt_j \rij,
\eeq
\beq
  \rij \equiv {\bf r}_i - {\bf r}_j.
\eeq
A symmetrized gradient of the kernel can be similarly defined as
\beq
  \label{gWij.eq}
  \gWij \equiv \frac{1}{2} \left[ \grad W(\veta_i) + \grad
               W(\veta_j) \right].
\eeq
These forms also arise naturally from a derivation of (A)SPH based upon the
variational principle.

The interpolation relation of equation (\ref{Wint.eq}) can now be
represented numerically by assigning known values for the general quantity
$F({\bf r})$ on a series of discrete nodes at positions ${\bf r}_j$ (each
with an associated mass $m_j$, mass density $\rho_j$, and number density
$n_j = \rho_j/m_j$), such that we have a discrete set $F_j$.  The
interpolated value of $F$ at position ${\bf r}_i$ is
\beq
  \label{Wint_d.eq}
  F_i \equiv \Interp{F({\bf r}_i)} = \sum_j \frac{F_j}{n_j} \Wij 
                        = \sum_j F_j \frac{m_j}{\rho_j} \Wij.
\eeq
Note that equation (\ref{Wint_d.eq}) basically represents a monte-carlo
interpretation of equation (\ref{Wint.eq}).  With this machinery in place,
it is now possible to derive ASPH dynamical equations using the same
approach as SPH.  So long as we use the notational conventions outlined
above (expressing quantities in terms of the normalized position vector
$\veta$ rather than explicitly using $h$) the SPH and ASPH dynamical
equations are identical.  We present our complete set of (A)SPH dynamical
equations derived in this fashion in Appendix \ref{Dyneqs.app}.

The form of the \Gt\ tensor depends upon the dimensionality in which it
implemented.  The \Gt\ tensor can be defined in terms of the underlying
geometry as follows.  In 2D, consider a unit normalized position
isocontour associated with a given \Gt.  In general such an isocontour
represents an arbitrary ellipse in real position space, which is uniquely
defined by a semi-major axis \hx, semi-minor axis \hy, and position angle
$\psi$ associated with the semi-major axis.  Since we are considering a
unit isocontour, \hx\ and \hy\ are the smoothing scales along the primary
axes of the ellipse.  In terms of these geometrical quantities, the
elements of the \Gt\ tensor are (see Appendix \ref{Genderiv.app})
\beq
  \label{2dG.eq}
  \Gt = \lp
  \begin{array}{cc}
     G_{11} & G_{21} \\ G_{21} & G_{22}
  \end{array} \rp = \lp
  \begin{array}{cc}
     \hix \cos ^{2} \psi + \hiy \sin ^{2} \psi & 
     ( \hix - \hiy) \cos \psi \sin \psi       \\
     ( \hix - \hiy) \cos \psi \sin \psi       & 
   \hix \sin ^{2} \psi + \hiy \cos ^{2} \psi
  \end{array} \rp.
\eeq
The 3D \Gt\ tensor can be similarly derived, as discussed in appendix
\ref{3dasph.app}, resulting in equation (\ref{3dG.eq}).

We have discussed the interpolation kernel $W$ completely generally to this
point.  There are many possible choices for such a function, so long as the
criteria of equations (\ref{Norm.eq}) and (\ref{Wlim.eq}) are met and they
can be expressed as functions of the normalized position $\veta$.  The
most popular forms currently in use with SPH are the Bi-Cubic Spline and
the Gaussian kernels.  In our investigation we use the Bi-Cubic Spline,
which in $\nu$ dimension is
\beq
  \label{spline.eq}
  \Wspline{\nu}(\veta) = \Aspline{\nu} \left\{ 
    \begin{array}{l@{\quad}l}
      1 - 3/2 \, \eta^2 + 3/4 \, \eta^3, & 0 \le \eta \le 1; \\
      1/4 \, (2 - \eta)^3, & 1 < \eta \le 2; \\
      0, & \eta > 2,
    \end{array} \right.
\eeq
\beq
  \grad \Wspline{\nu}(\veta) = \Aspline{\nu} \Gt
    \frac{\veta}{\eta} \left\{
    \begin{array}{l@{\quad}l}
      -3 \eta + 9/4 \, \eta^2, & 0 \le \eta \le 1; \\
      -3/4 \, (2 - \eta)^2, & 1 < \eta \le 2; \\
      0, & \eta > 2, 
    \end{array} \right.
\eeq
\beq
  \Aspline{1} = \frac{2}{3} |\Gt|, \quad
  \Aspline{2} = \frac{10}{7 \pi} |\Gt|, \quad
  \Aspline{3} = \frac{1}{\pi} |\Gt|,
\eeq
where \Aspline{\nu} represents the normalization constant required to
meet the condition of equation (\ref{Norm.eq}).

\subsection{Evolving the Smoothing Scale under SPH}
\label{DhsDt.sec}
Before going on to discuss how \Gt\ is evolved under ASPH, we must begin
with the rationale for evolving the variable smoothing scale under SPH.
Most modern implementations of SPH allow an individual, time variable
smoothing scale to be associated with each SPH node, such that
$h({\bf r},t)$.  The justification for implementing and evolving such a
variable smoothing scale is based upon the philosophy that each SPH node
should sample roughly the same number of ``significant neighbors'' (that
is, the number of neighboring SPH nodes within some critical threshold
distance, usually expressed as a multiple of the smoothing scale $\eta \le
\Sup{\eta}{cut}$).  This implies that each SPH node will always sample
roughly the same amount of mass, which is consistent with SPH's Lagrangian
nature.  This results in higher resolutions in dense regions, while still
maintaining meaningful (if poorly resolved) measurements in low density
regions.  This condition can be stated mathematically in $\nu$ dimensions
as $h_i \propto \rho_i^{-1/\nu}$ for a given node $i$.

A well-defined standard method for evolving the SPH smoothing scale in
order to meet this criterion is based upon the continuity equation (Benz
1990 and references therein), and can be expressed as
\beq
  \label{Sphh.eq}
  \frac{Dh_i}{Dt} = -\frac{1}{\nu} \frac{h_i}{\rho_i} \frac{D\rho_i}{Dt}
	= \frac{h_i}{\nu} (\grad \cdot {\bf v})_i
	= -\frac{1}{\nu} \frac{h_i}{\rho_i} \sum_j m_j \vij \cdot \gWij.
\eeq

\subsection{Evolving \Gt\ under ASPH}
\label{DGDt.Sec}
In order to understand the motivation for developing ASPH, it is necessary
to understand the potential inefficiencies of the SPH approach as defined
by equation (\ref{Sphh.eq}).  Consider a system undergoing planar collapse
in some arbitrary direction.  In 3D, the standard SPH smoothing scale will
adapt to this collapse process as $h \propto \rho^{-1/3}$.  However, a
purely planar collapse is really a 1D problem, and ideally smoothing scales
perpendicular to the plane of collapse should evolve as $h_\bot \propto
\rho^{-1}$, while smoothing scales parallel to the collapse should remain
unchanged.  In such a situation, the approach in 3D SPH will lead to less
than optimal resolution along the direction of collapse (since $h_\bot$ is
shrinking too slowly), while nodes in the collapsing region will lose
contact with neighbors parallel to the collapse (since $h_\|$ is shrinking,
but the internode spacing parallel to the plane of collapse remains
unchanged).  In general, problems of computational interest are not
isotropic in nature, and in particular gravitational clustering scenarios
generically result in strongly anisotropic density evolution.  ASPH seeks
to address this problem by allowing an anisotropic definition of the
smoothing process, as well as a self-consistent method for evolving this
anisotropic smoothing function.  The ASPH algorithm can be viewed as trying
to adapt to the physical or intrinsic dimensionality of a problem, rather
than the imposed geometrical dimensionality.

This line of reasoning suggests a straightforward approach to evolving the
ASPH smoothing transformation \Gt.  We can generalize the SPH philosophy of
attempting to maintain the same number of neighbors per node into an
attempt to maintain the same number of neighbors {\em in all directions}
for each node.  This can be rephrased to state that we will try to keep
the distribution of neighboring nodes isotropic in normalized
($\veta$) space.  In more physical terms each ASPH node attempts to
always sample the same {\em Lagrangian} volume as the system evolves, which
is appropriate for ASPH's Lagrangian nature.  This can be viewed as an
attempt to track the local deformation of a fluid element with an idealized
shape corresponding to the underlying geometry of the \Gt\ tensor.  At best
such a picture represents an analogy, as a true fluid element is not
constrained to remain ellipsoidal, and the volumes defined by the
collection of $\Gt_i$'s interpenetrate.  Nevertheless, the fluid element
analogy is useful in order to gain an intuitive sense as to the behaviour
of the \Gt\ tensors.

The deformation tensor \sigt\ ($\sigma_{\alpha \beta} \equiv \partial
v_\alpha/\partial r_\beta$ where $(\alpha,\beta)$ refer to spatial
directions) indicates how the velocity field varies spatially to
first-order, such that ${\bf v}({\bf r} + d{\bf r}) \approx
{\bf v}({\bf r}) + \sigt ~d{\bf r}$.  This quantity predicts how a local
volume of the fluid will deform with time.  If we visualize the ASPH
interpolation volume as an embedded volume within the fluid, \sigt\
predicts how this volume should be deformed by the local velocity field,
such that the enclosed Lagrangian volume remains constant.  As this is a
first-order transformation, an initially ellipsoidal volume will be in
general be mapped to a new ellipsoidal volume, guaranteeing that our
ellipsoidal transformation for \Gt\ will remain appropriate.  Under the
(A)SPH formalism the deformation tensor can be estimated by
\beq
  \label{sigker.eq}
  \Interp{\sigma_{\alpha \beta}}_i = -\sum_j m_j (v_{ij})_\alpha
  \frac{\partial \Wij}{\partial r_\beta}.
\eeq

We will defer a rigorous mathematical derivation of \Gt\ and its evolution
to appendix \ref{Genderiv.app}, and simply present the results for the 2D
case here.  The evolution of the 2D \Gt\ tensor (eq. [\ref{2dG.eq}]) is
given by
\beqa
  \label{2dGevolu.eq}
  \frac{D\Gt}{Dt} &=& \lp 
  \begin{array}{cc}
    DG_{11}/Dt & DG_{21}/Dt \\
    DG_{21}/Dt & DG_{22}/Dt
  \end{array} \rp \\ 
  &=& \lp
  \begin{array}{cc}
    G_{21} (\dot{\theta} - \sigma_{21}) - G_{11} \sigma_{11} &
    G_{22} \dot{\theta} - G_{11} \sigma_{12} - G_{21} \sigma_{22} \\
    -G_{11} \dot{\theta} - G_{21} \sigma_{11} - G_{22} \sigma_{21} &
    -G_{21} (\dot{\theta} + \sigma_{12}) - G_{22} \sigma_{22}
  \end{array} \rp, \nonumber \\
  \dot{\theta} &=& \frac{G_{11} \sigma_{12} - G_{22} \sigma_{21} -
                       G_{21} (\sigma_{11} - \sigma_{22})}
                      {G_{11} + G_{22}}.
\eeqa

In order to demonstrate the connection of this evolution equation for
\Gt\ with the SPH relation for $h$, consider the evolution of the
determinant $|\Gt|$.
\beqa
  \label{Hproptorho.eq} 
  \frac{D|\Gt|}{Dt} &=& G_{11}\frac{DG_{22}}{Dt} +
                        G_{22}\frac{DG_{11}}{Dt} - 
                        2 G_{21}\frac{DG_{21}}{Dt} \\ 
                    &=& -(G_{11} G_{22} - G_{21}^2)(\sigma_{11} + 
                         \sigma_{22}) \nonumber \\ 
                    &=& -|\Gt| \grad \cdot {\bf v} \nonumber \\ 
                    &=& \frac{|\Gt|}{\rho} \frac{D\rho}{Dt}. \nonumber
\eeqa
For the special case of SPH, the determinant reduces to $\Sup{|\Gt|}{SPH} =
h^{-\nu}$ in $\nu$ dimensions.  In 2D, the SPH evolution equation for
$h$ (eq. [\ref{Sphh.eq}]) yields
\beq
  \frac{D h^{-2}}{Dt} 
     = -2 h^{-3} \frac{Dh}{Dt} 
     = \frac{h^{-2}}{\rho} \frac{D\rho}{Dt}.
\eeq
Equation (\ref{Hproptorho.eq}) shows that $|\Gt|$ is directly proportional
to the density $(|\Gt| \propto \rho)$.  This also proves that the smoothing
volume represented by $|\Gt|^{-1}$ under ASPH evolves identically with its
SPH counterpart, demonstrating that for the case of isotropic evolution
ASPH formally reduces to SPH.  Though derived here in 2D, this result
applies in general.

\subsection{Stabilizing ASPH: Smoothing the \Gt\ Tensor Field}
\label{SHASPH.sec}
The evolution equation for the \Gt\ tensor is derived based upon
first-order arguments about the local velocity field -- specifically, that
the local velocity can be approximated by ${\bf v}({\bf r} + d{\bf r})
\approx {\bf v}({\bf r}) + \sigt ~d{\bf r}$.  Clearly, in complex
simulations one can encounter situations where this approximation is not
valid on ``local'' scales (scales of order a few $h$).  A cosmological
example of such a situation is the formation of a poorly resolved cluster
at the intersection of several filaments, each of which contains gas
streaming into the cluster.  Ideally one would wish to resolve all mass
scales adequately such that the ``local'' velocity field will always be
well represented by \sigt\ on scales of $h$, but in reality computational
limitations make this infeasible.  Any simulation will always have a lower
mass cutoff in the resolved mass distribution.  In particular, a Cold Dark
Matter like initial density fluctuation power-spectrum will generically
have substantial power on small scales, and it is reasonable to assume that
in such situations one will have to deal with nonlinear velocity fields
down to scales of $h$.

If we examine the assumptions underlying our derivation of the evolution
equations for \Gt, we find that we implicitly require the \Gt\ tensor field
to be well-behaved on scales of a few smoothing lengths.  This point can be
demonstrated in two ways.  First, consider the expression for the gradient
of the kernel $\grad W$ (eq. [\ref{gW.eq}]).  As with SPH, we neglect
any $\grad \Gt$ terms, which formally should be included.  Including these
terms is somewhat problematic as they must be estimated numerically, and we
have found that attempting to include them does little more than introduce
noise.  Neglecting such terms implies that we require $\grad \Gt$ to be
negligible on scales of a few characteristic smoothing lengths.  Another
way of viewing this problem emerges if we use our idealized fluid element
analogy.  The smoothing volumes represented by neighboring \Gt\ tensors
must necessarily overlap -- this implies that the \Gt\ tensors of closely
neighboring nodes should represent the same local idealized fluid element.
Strong disorder in this \Gt\ tensor field on local scales is therefore
inconsistent.

Arguments such as these lead us to conclude that we require the \Gt\ tensor
field to be almost uniform on scales of a few smoothing lengths, and
smoothly varying on larger scales.  If this condition is not met, how would
we expect this inconsistency to affect ASPH?  Under our standard
formulation of SPH, it will formally conserve mass (with the summation
definition for the density), linear and angular momentum (because all pair
interactions are symmetric and radial), while energy is conserved to
second-order.  ASPH utilizes the same dynamical equations as SPH, including
the use of the symmetrized kernel function \Wij\ (eq. [\ref{Wij.eq}]).
Therefore, ASPH will rigorously conserve mass and linear momentum, while
energy will be conserved to second-order (see Appendix \ref{Dyneqs.app} for
a discussion of the dynamical equations and these conservation properties).
However, because the gradient of the kernel $\grad W$ is not necessarily
radial under ASPH, forces between interacting pairs of nodes will not in
general be radial and therefore angular momentum will not be rigorously
conserved.  This makes the conservation of angular momentum under ASPH
analogous to the conservation of energy -- angular momentum will only be
conserved to the order that the system is being solved (in general to
second-order).  ASPH is therefore vulnerable to errors in the angular
momentum as well as the energy, and we might expect that as an ASPH
simulation breaks down non-conservation of these quantities could be a
symptom.  As we discuss in \S \ref{2dtests.sec}, we find precisely this
sort of behaviour.

We therefore require a method of ensuring that the \Gt\ tensor field is
well-behaved.  The most obvious step to take is to smooth the \Gt\ tensor
field using the ASPH formalism.  We must be cautious in the implementation
of such a scheme, however.  Recall that the \Gt\ tensor has units of an
inverse smoothing scale.  Taking a straight ASPH estimate of \Gt\ is
equivalent to taking the harmonic mean of the smoothing scales, which can
be unstable toward small smoothing scales.  Additionally, we would like to
preserve the property of the determinant $|\Gt|$ such that it evolves
smoothly in accordance with the local density, as demonstrated in equation
(\ref{Hproptorho.eq}).  Discontinuous changes in \Gt\ are equivalent to
discontinuously changing an individual nodes contribution to the local
density.  After exploring several possible implementations for a smoothing
scheme, we have settled on the following approach.  Periodically each
$\Gt_i$ tensor is replaced by an averaged $\Gt_i'$ calculated as
\beq
  \label{Hsmooth.eq}
  \Interp{\Gt^{-1}}_i = \frac{\sum_j \Gt_j^{-1} \Wij}{\sum_j \Wij},
  \quad
  \Gt_i' = |\Gt_i|\,\left| \Interp{\Gt^{-1}}_i \right| \Interp{\Gt^{-1}}_i^{-1}.
\eeq
Note this scheme represents three modifications of an ordinary ASPH
average.  First, we average the quantity $\Gt^{-1}$ in order to avoid the
problems of a harmonic mean on the smoothing scale.  Second, we force the
determinant to be preserved $|\Gt_i'| = |\Gt_i|$.  Finally, we force the
normalization of the average to be unity by dividing by $\sum_j \Wij$.
Formally this sum should be unity (eq. [\ref{Norm.eq}]), but in
practice because we only sum over a finite number of nodes, this quantity
typically deviates from that ideal.  We have found that normalizing the
average in this way increases the stability of the technique.  The
frequency with which we must enforce this smoothing process must be
determined experimentally.  We have found that smoothing roughly once or
twice every characteristic timescale is generally adequate (where by
characteristic timescale we mean the timescale setting the current timestep
-- see appendix \ref{timestep.app}).

It is worth noting that there is an alternate method of viewing this
problem and its solution.  The evolution of the \Gt\ tensor is based upon
attempting to follow the deformation of the local velocity field based on
using the deformation tensor $\sigma_{\alpha \beta} = \partial
v_\alpha/\partial x_\beta$, which is only correct to a first-order.
If the local velocity field is not this obligingly simple,
then the arguments we base our \Gt\ evolution equations on break down and
there is no guarantee trouble will not ensue.  This leads to the idea that
smoothing on the deformation tensor \Interp{\sigt} could also be an
equivalent method of dealing with this problem.  Such a solution seems
intuitively pleasing, as it would no longer involve directly fiddling with
the \Gt\ tensors themselves.  However, this is basically equivalent to our
adopted method of smoothing on the \Gt\ tensors, which has the additional
advantage of directly guaranteeing the good behaviour of the \Gt\ tensor
field.  We should also point out that formally SPH also requires that
$\grad h$ be negligible on scales of a few $h$, and therefore standard SPH
schemes utilizing a spatially variable smoothing scale should also ensure
this behaviour in some way.  Steinmetz \& M\"{u}ller (1993) find that the
stability of SPH with a variable smoothing scale is indeed improved by
spatially smoothing the $h$ field.

\subsection{Artificial Viscosity under ASPH}
\label{AV.sec}
Cosmological simulations often show evidence of nonphysical preheating in
shock forming regions, particularly during collapse situations.  This can
result in poor resolution of shockfronts and related phenomena.  This
behaviour is due to the use of an artificial viscosity such as that denoted
by $\Pi$ in the (A)SPH dynamical equations [\ref{Sphmom.eq}] \&
[\ref{Sphtherm.eq}]).  Such an artificial viscosity term is required in the
(A)SPH formalism both for stability and because without it (A)SPH is
insufficiently dissipative to prevent the interpenetration of converging
streams of gas, resulting in a poor representation of shock conditions.
However, the artificial viscosity is by definition an artificial term, and
ideally its use should be restricted solely to ongoing shocks, where it is
required.  This problem is particularly accentuated in cosmological
studies, since the kinetic energy is typically much larger than the
thermal, and the artificial viscosity functions by converting kinetic to
thermal energy.  Inspection of the standard Monaghan-Gingold (1983) form of
the artificial viscosity (eqs. [\ref{Sphvisc.eq}] \& [\ref{Sphmu.eq}])
shows that traditionally $\Pi$ is restricted to only be active for
convergent flows within the material being modeled.  While a convergent
flow is a minimal requirement for the presence of shocks, clearly not all
convergent flows necessarily result in the formation of shocks (\eg\
homologous collapse).  This overuse of $\Pi$ is what leads to the excessive
heating of the material around shockfronts, a problem which can be
especially troublesome in gravitational collapse scenarios, where the
spurious preheating of the gas can become acute enough to interfere with
the collapse process itself.  As this is precisely the sort of scenario we
are concerned with modeling well, we would like to improve upon this
algorithm.

This line of reasoning led to the development of the algorithm outlined in
Paper I for the suppression of the artificial viscosity.  Through
experimentation we determined that it is possible to delay the turn-on time
for the artificial viscosity in the energy equation (which is the source of
the spurious heating), but it is necessary to keep the artificial viscosity
active in the momentum equation in order to prevent interpenetration of the
(A)SPH nodes.  That scheme yields excellent results insofar as suppressing
the artificial viscous preheating of material, particularly in the
Zel'dovich pancake scenario, for which it was developed.  In this paper we
adopt a different prescription for suppressing the artificial viscosity.
Rather than attempting to delay the turn-on time for $\Pi$ and treating it
distinctly between the momentum and energy equations, we instead adopt a
different interpolation kernel to be used with $\Pi$, which we denote as
\WPij.  We choose a form for \WPij\ such that it is more spatially compact
and has a sharper gradient than \Wij.  We use the standard artificial
viscosity turn-on criteria (given in eq. [\ref{Sphvisc.eq}]), and make no
distinction between $\Pi$ in the momentum and energy equations.  In effect,
this approach can be thought of as restricting the influence of the
artificial viscosity spatially, rather than temporally as the scheme of
Paper I.

There are many possible choices that could be used for \WPij.  In order to
successfully suppress the preheating problem while still stopping
interpenetration, we want a kernel which is more spatially compact and
possesses a stronger gradient as compared with the Spline kernel (eq.
[\ref{spline.eq}]).  We use a simple variant of the Gaussian kernel, as
given by
\beq
  \label{WgaussP.eq}
  \WgaussP{\nu}(\veta) =  \AgaussP{\nu} \exp \lp -K \eta^4 \rp,
\eeq
where the appropriate normalization constants are
\beq
  \AgaussP{1} = \frac{2 K^{1/4} |\Gt|}{\Gamma (1/4)}, \quad
  \AgaussP{2} = \frac{2 K^{1/2} |\Gt|}{\pi^{3/2}}, \quad
  \AgaussP{3} = \frac{K^{3/4} |\Gt|}{\pi \Gamma (3/4)}.
\eeq
The gradient of this kernel is
\beq
  \grad \WgaussP{\nu}(\veta) = \Gt \frac{\veta}{\eta} \frac{\partial
  \WgaussP{\nu}}{\partial \eta} = -4K \WgaussP{\nu} \eta^2 \Gt \veta.
\eeq
Through experimentation we have found setting $K = 1.5^4$ is a safe choice.
Using this kernel as \WPij\ successfully prevents interpenetration in
all of our test cases, while still reducing the preheating of the ASPH gas
in collapse simulations (see the Zel'dovich pancake tests in \S
\ref{Zeldovich.sec}).  Comparison of these results with similar 2D pancake
runs in Paper I reveal that this form of $\Pi$ suppression is not as
effective as that presented in Paper I, at least for this class of
problems.  However, this scheme does have the advantages that it is
applicable in all situations (there are no problem dependent parameters to
fiddle), and because $\Pi$ is treated identically in the momentum and
energy equations energy conservation is preserved.

Finally, we should point out that while we find using this modified
artificial viscosity works well under the ASPH tests presented in this
paper, we also find in similar SPH experiments that it has little effect
when used in conjunction with SPH.  The precise reason for this remains
unclear, but it is possible that the planar shocks in these problems are
simply not well enough resolved in the SPH models we perform.  One could
therefore infer if the SPH resolution were improved by using a larger
number of particles, we might ultimately find some benefit to this scheme
under SPH.

\section{Tests}
\label{tests.sec}
In this section we discuss a set of test cases performed under both ASPH
and SPH.  These tests have been selected to compare the relative advantages
of the two techniques, as well as to test questions about the validity of
ASPH (in particular the angular momentum issue).  We concentrate on 2D
simulations, for the pragmatic reason that we must run problems in at least
2D for there to be a distinction between ASPH and SPH (in 1D the two are
formally identical), while 3D simulations are much more computationally
expensive.  We present a few low-resolution 3D versions of some of our
tests, to begin exploring the validity of the technique in 3D.  We defer
extensive testing of the 3D formalism for now.  All simulation results
presented as SPH utilize our ``Standard'' SPH formalism, outlined in
Appendix \ref{Dyneqs.app}.  All ASPH simulations have been performed using
both smoothing of the \Gt\ tensor field (\S \ref{SHASPH.sec}) and special
treatment of the artificial viscosity (\S \ref{AV.sec}), unless otherwise
noted.  In appendix \ref{Spheral.app} we present a brief discussion of the
specific algorithms used to perform these simulations.

We present two broad classes of test cases: those which are physically
1D (such as the Zel'dovich pancake problem, the Sedov Blastwave, and a
few flavors of the Riemann Shocktube), and those which are physically
2D (rotating tests such as the Pseudo-Keplerian Disk and a collapsing
disk with angular momentum).

\subsection{1D Test Cases}
\label{1dtests.sec}
This class of test problems, although performed in higher dimensional
frameworks, possess symmetries which allow them to considered physically
1D.  These tests are chosen for two reasons.  First, they possess
analytical solutions, allowing objective comparison and judgment of the
results.  Secondly, since they are physically 1D they represent cases
where ASPH can be expected to have an advantage over SPH, in that ASPH can
recognize and adapt to the physical dimensionality of the problem, whereas
SPH cannot.  We present four examples of this class of problem: the
Zel'dovich pancake (\S \ref{Zeldovich.sec}), the Sedov blastwave (\S
\ref{Sedov.sec}), a simple 4:1 Riemann shocktube (\S \ref{Shocktube.sec}),
and a double-shocktube with two interacting strong shockfronts (\S
\ref{DoubleBlast.sec}).  The first two cases are chosen to typify the sorts
of problems we are interested in solving as well as demonstrating the
distinction between ASPH and SPH, while the last two represent standard
hydrodynamic test cases.

\subsubsection{Zel'dovich Pancake Test}
\label{Zeldovich.sec}
This class of problems consists of setting up 1D plane-wave perturbations
in an arbitrary dimension cosmological scenario (Zel'dovich 1970), and has
been well-studied previously in scenarios which incorporate gas dynamics as
well as collisionless dark matter (Shapiro \& Struck-Marcell 1985).  The
Zel'dovich problem represents a standard test case for cosmological codes
(Efstathiou \etal\ 1985; Villumsen 1989).  As this class of problems has
been discussed extensively in Paper I as well as Shapiro \& Struck-Marcell
(1985), we need not go over the analytical properties here.  We reexamine
this problem primarily because the implementation of ASPH used for this
work differs from that of Paper I, and because here we present 3D as well
as 2D results.  Pancake collapse represents a simplified example of the
sort of cosmological structure formation scenarios we wish to investigate
in general.

All of the Zel'dovich Pancake simulations we present are performed under an
Einstein-de Sitter cosmology ($\Omega=1$, $\Lambda=0$), with equal baryonic
and dark matter mass fractions (\Sup{\Omega}{bary} = \Sup{\Omega}{dm} =
0.5).  The baryonic gas is assumed to be a pure hydrogen, adiabatic gas
($\mu = 1$, $\gamma = 5/3$).  Since these simulations are carried out in a
cosmological framework, they are evolved in comoving coordinates using a
power of the expansion factor ($p = a^\alpha$), rather than time, as the
integration variable (see appendix \ref{codyneqs.app}).  The system is
simulated in a periodic unit volume, using a 2D Particle-Mesh (PM) gravity
calculation to solve for the self-gravitation.  We present 2D examples and
a low-resolution 3D example of this problem.  In 2D, the gravity obeys a
$1/r$ force law, such that each pair of nodes interact gravitationally as
though they are a pair of infinite, thin, parallel rods in 3D.  In this way
the 2D simulations can be thought of as a ``slice'' through an infinite 3D
simulation.  Table \ref{Zeldovich.tab} summarizes the major simulation
parameters we use for our Zel'dovich pancake simulations.

Under this framework, we present ASPH and SPH simulations of 2D Zel'dovich
pancakes with $(k_x=0,k_y=1)$ and $(k_x=2,k_y=1)$, and a 3D version with
$(k_x=0,k_y=0,k_z=1)$.  All \Gt\ tensors are initialized as SPH spherical
tensors with the determinant $|\Gt|$ scaled appropriately for the local
density.  This is somewhat inconsistent for ASPH runs, since the linear
evolution of an ASPH \Gt\ tensor should only affect the geometry of the
\Gt\ tensors perpendicularly to the plane of collapse.  However, at $a = 1$
the initial conditions are almost uniform (the perturbations are small),
and starting in this manner allows both the SPH and ASPH runs to use
identical initial conditions.

\paragraph{Single-Wavelength 2D Zel'dovich Pancakes:}
\label{2dpan01.sec}
This case represents the simplest possible class of this problem: a
single-wavelength pancake along one principal axis.  Here we choose to use
$(k_x=0, k_y=1)$, which corresponds to a collapse proceeding in the $y$
direction -- by symmetry, there should be no evolution in the $x$
direction.

\begin{figure}[htbp]
\begin{minipage}[t]{\hsize}
\plottwo{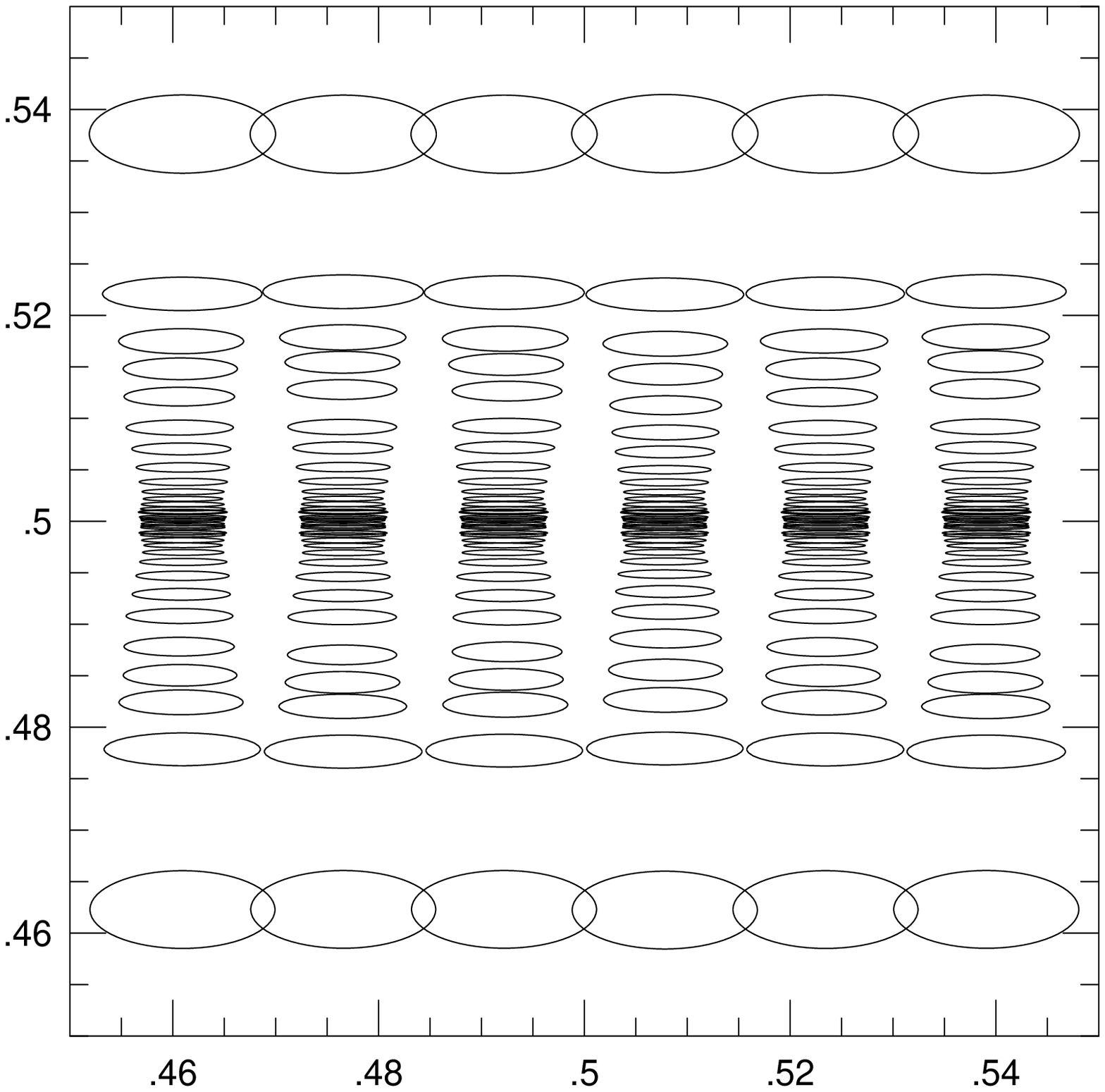}{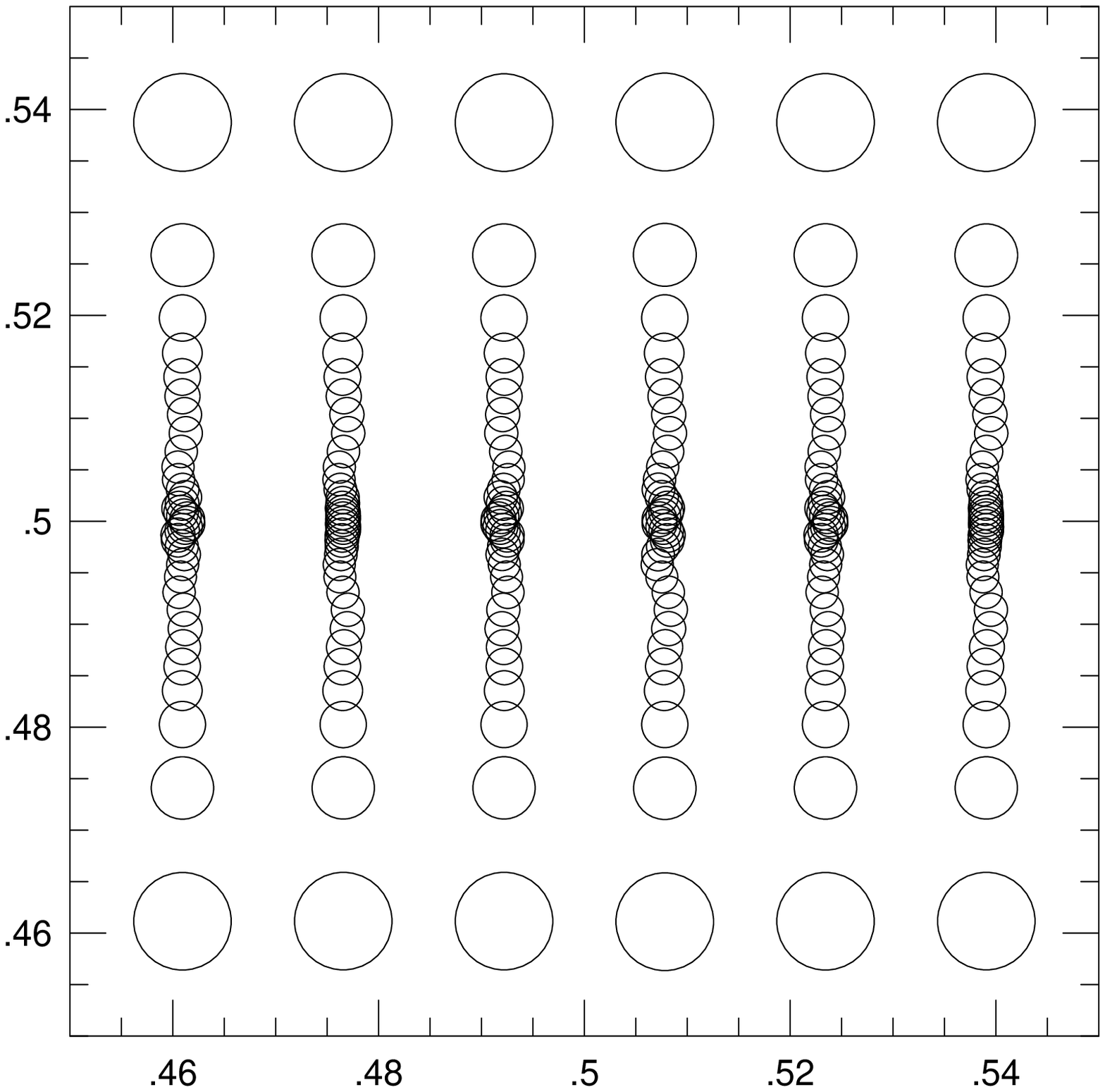}
\end{minipage}
\begin{minipage}{\hsize}
\plottwo{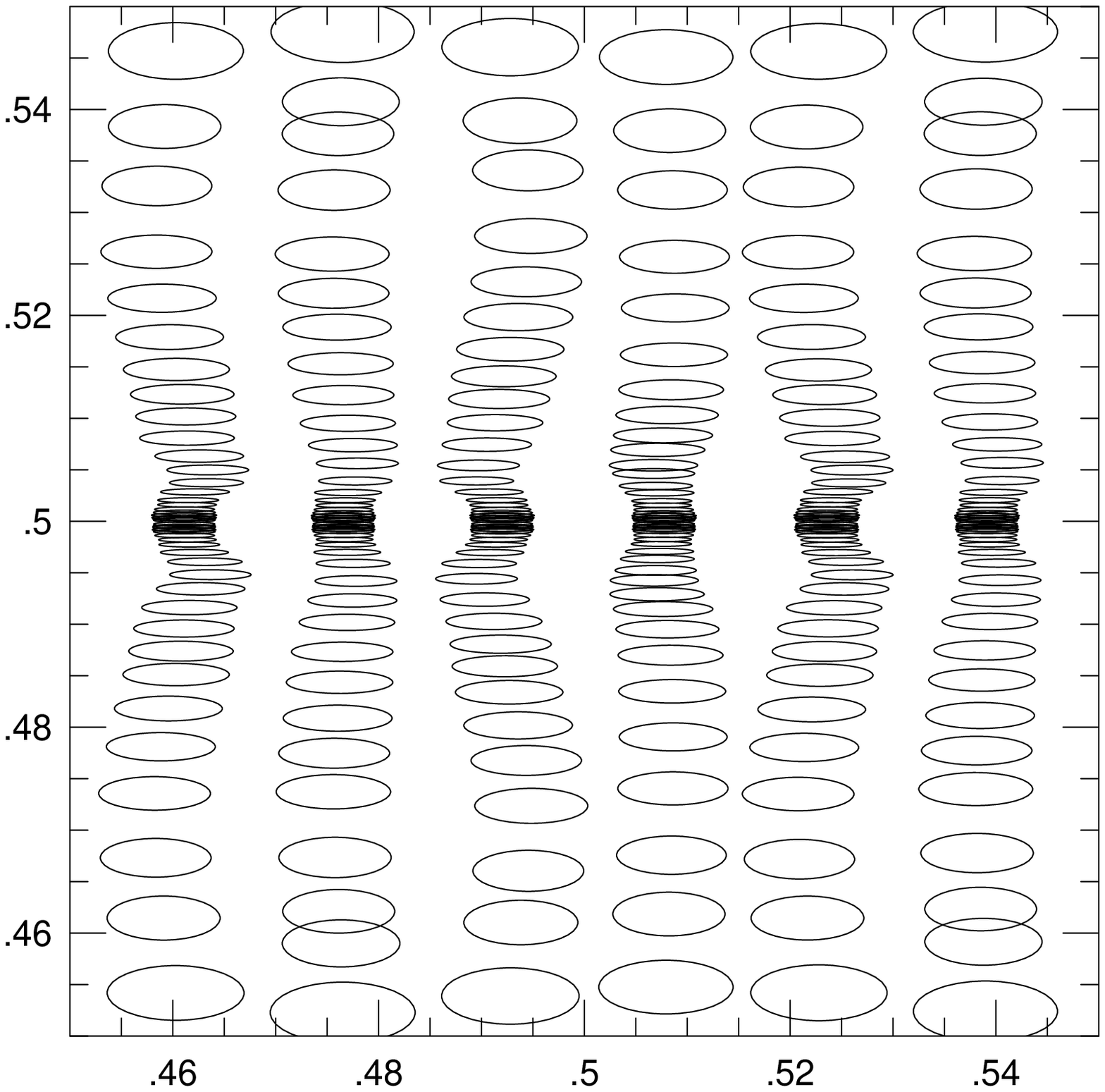}{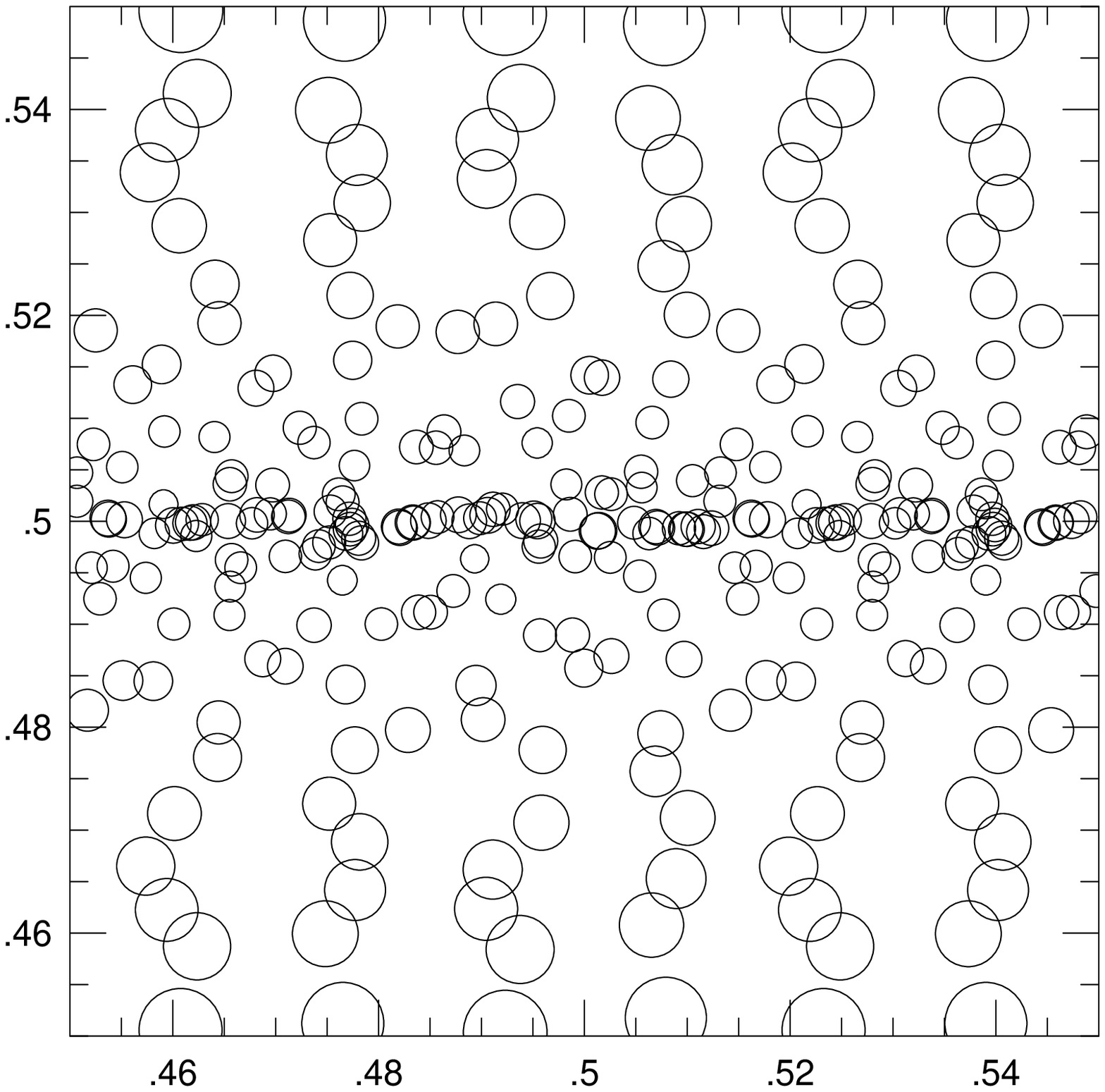}
\end{minipage}
\caption{``Kernel plots'' ($h = 0.2$ smoothing scale isocontours) for the
central region in the $(k_x=0, k_y=1)$ 2D Zel'dovich pancake simulations.
The upper and lower panels on the left show the ASPH simulation at
expansions $a/a_c=1.5$ and $a/a_c=2.5$, respectively, while the right-hand
panels show the SPH simulation at the same times.}
\label{2dpan01_ker.fig}
\end{figure}
Figure \ref{2dpan01_ker.fig} gives examples of what we refer to as ``kernel
plots''.  These are plots of smoothing scale isocontours about each (A)SPH
node in a given region.  In 2D, SPH kernel plots are circles about each
nodes position, whereas in general ASPH kernel plots are ellipses.  We
generically choose to plot the $h=0.2$ isocontour.  Since the spline kernel
extends to a cutoff radius of $\Sup{\eta}{cut}=2$, this implies each node
``sees'' neighbors out to a contour 10 times that shown.  We find that
these sorts of figures are quite useful in order to gain an intuitive feel
for how the \Gt\ tensor field is adapting to the local fluid flow.  In this
case, we have plotted subregions centered about the pancake midplane at
expansion factors $a/a_c=1.5$ and $a/a_c=2.5$.  We can see that ASPH is
better able to deal with this sort of 1D flow, since the ASPH \Gt\ tensors
can adapt the smoothing scales fully in the $y$ direction, whereas SPH must
adapt smoothing scales isotropically.  In other words, the ASPH smoothing
scale in the $y$ direction is better able to evolve as the ideal
$\rho^{-1}$ (since the collapse is a 1D process), whereas the SPH smoothing
scales are constrained to evolve as $\rho^{-1/2}$.  This implies not only
that ASPH is better able to sample along the physically interesting
dimension where the collapse is occurring, but also avoids the undesirable
effect of losing neighbor information parallel to the plane of the pancake.
Since the SPH smoothing scale must shrink isotropically as each node falls
into the pancake, while the internode spacing parallel to the pancake is
not changing, communication between the nodes parallel to the pancake
begins to break down.  This makes SPH more unstable to perturbations in the
$x$ direction, as is evident in the SPH kernel plot at $a/a_c=2.5$ (the
bottom right panel in Figure \ref{2dpan01_ker.fig}).  This disorder in the
SPH node positions is purely a hydrodynamical problem, since the dark
matter does not suffer from such deviations.  In the ASPH plots we can also
see that, while formally there should be no evolution of the ASPH smoothing
scale parallel to the pancake plane, this quantity is changing slightly.
This is due to the fact that we preserve the determinant $|\Gt|$ when we
smooth the \Gt\ tensor field.  Without such smoothing, the ASPH $\Gt$
tensors do not demonstrate any evolution parallel to the pancake plane.

\begin{figure}[htbp]
\plottwo{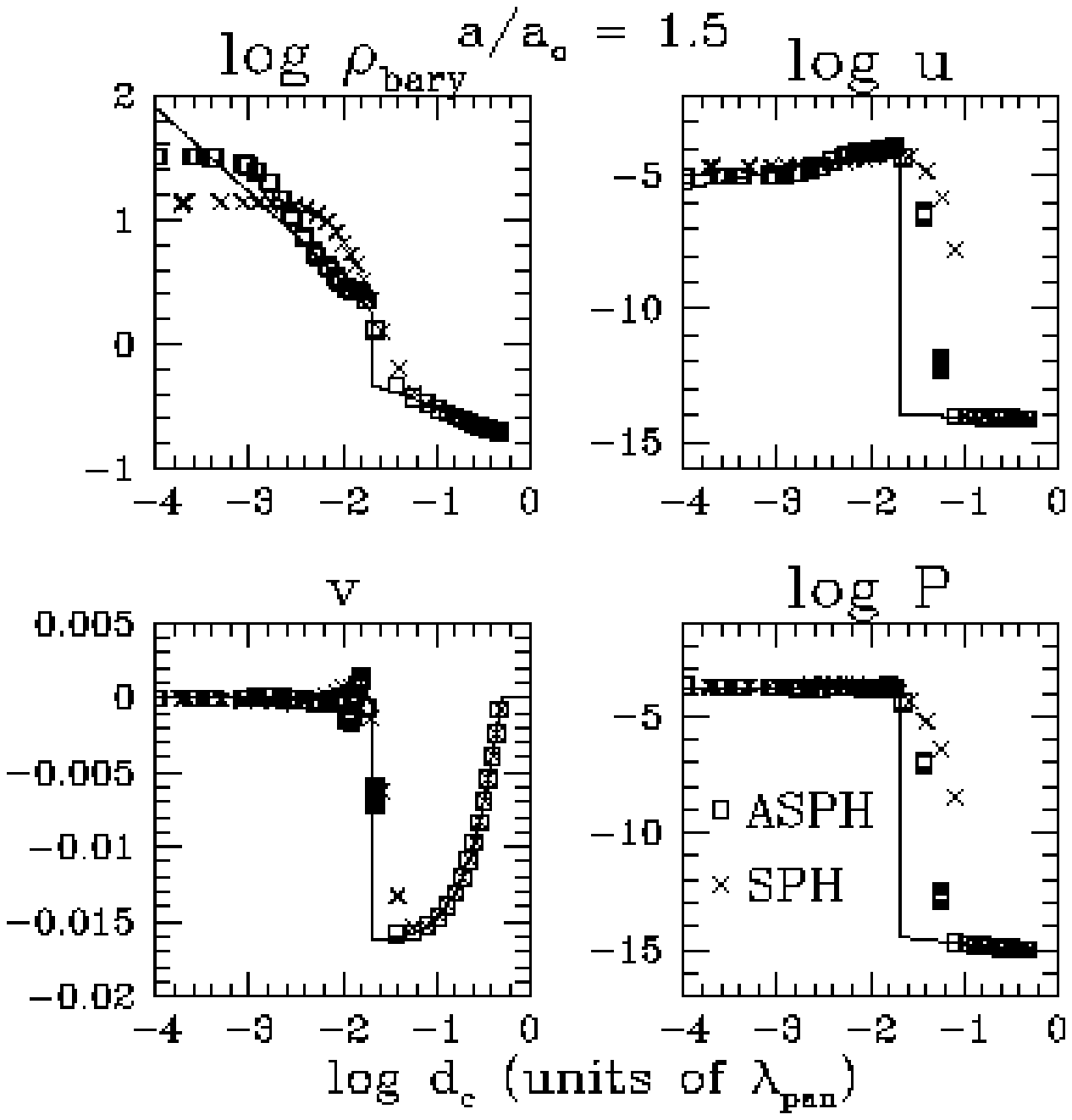}{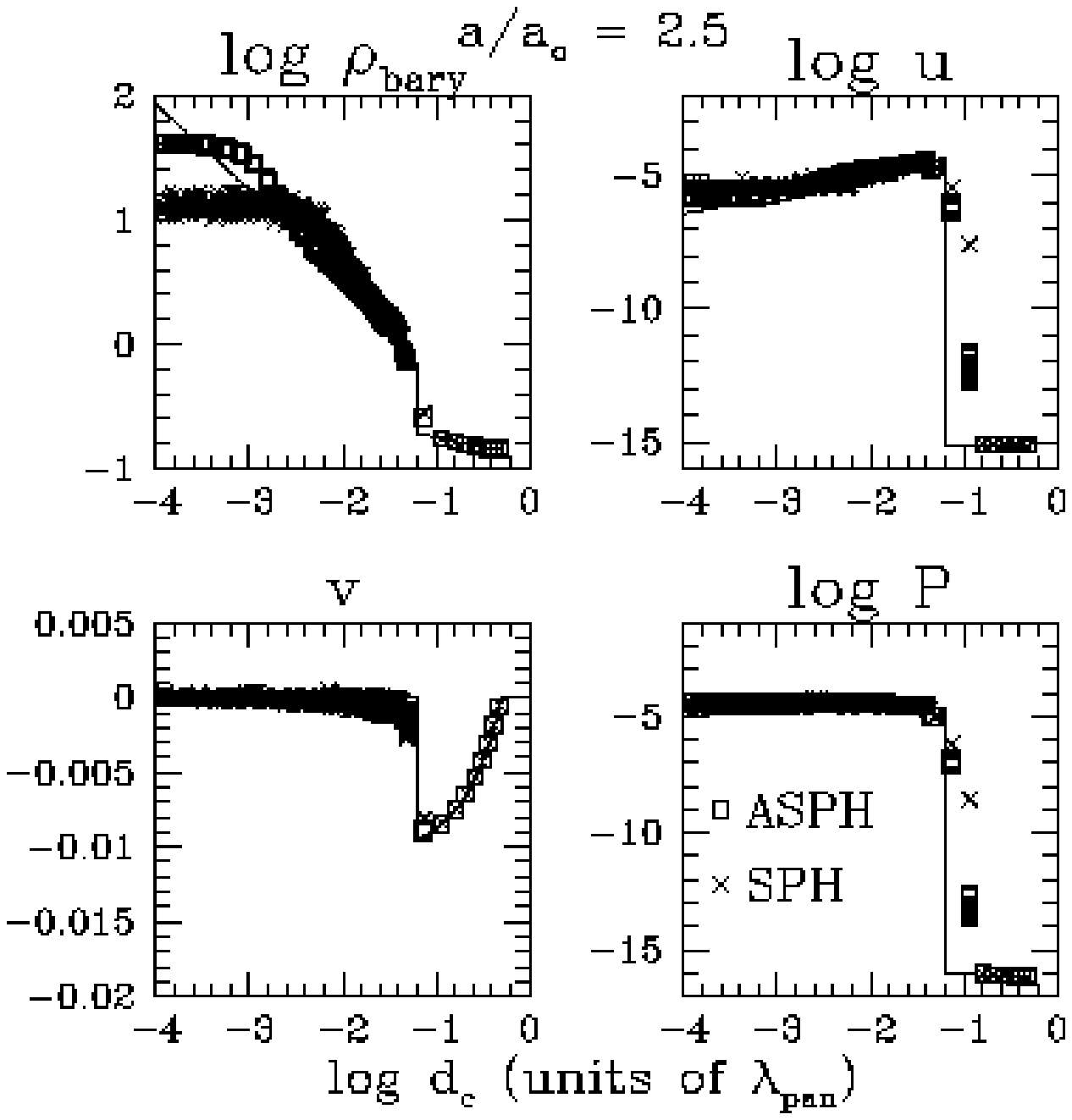}
\caption{Half-wavelength profiles of density $\rho$, specific thermal
energy $u$, velocity $v$, and pressure $P$ as a function of distance from
the pancake caustic $d_c$ for the $(k_x=0, k_y=1)$ 2D Zel'dovich pancake
simulations.  All quantities are converted to proper coordinates and are
expressed in units of the critical density, pancake wavelength, and the
Hubble time at the beginning of the simulation.  The solid lines are the
analytical expectations, the square points ASPH results, and the crossed
points SPH. The left panel is at $a/a_c=1.5$, while the right shows
$a/a_c=2.5$.}
\label{2dpan01_prof.fig}
\end{figure}
Figure \ref{2dpan01_prof.fig} shows half-wavelength profiles for the mass
density $\rho$, specific thermal energy $u$, velocity $v$, and pressure $P$
for the baryonic component of these simulations, plotted as a function of
distance from the pancake caustic $d_c$.  This figure represents the system
for $y \ge 0.5$, and can be thought of as looking parallel to the pancake
in the $x$ direction through the system.  Since there are $64^2$ (A)SPH
nodes in these simulations, there are effectively 64 nodes per wavelength
(since this is a physically 1D problem), or 32 nodes per half-wavelength as
plotted here.  Bear in mind that all nodes in the region $y \ge 0.5$ are
plotted, and therefore each of the ``points'' seen actually represents 64
overlapping points which share the same $x$ coordinate.  It is clear where
the symmetry in these simulations breaks down, as these points begin to
diverge.  This symmetry breaking is particularly evident in the final
results at $a/a_c = 2.5$ in the SPH simulation.  ASPH's finer resolution
along the direction of collapse allows a better representation of the
physical state of the system.  In particular, ASPH resolves peak densities
in the plane of collapse $\sim 3$ times what SPH is capable of with the
same number of nodes.  Additionally, ASPH proves much more stable against
symmetry breaking than SPH, since the ASPH nodes are able to sample
effectively parallel to the pancake even under extreme collapse.  The
careful reader will note, however, that the gravitational softening due to
the PM grid becomes important for scales below $\log d_c/\Sub{\lambda}{pan}
\lesssim -2.1$, and the scales where we reach our peak densities ($\log
d_c/\Sub{\lambda}{pan} \sim -3$ in the ASPH case) are below this threshold.
The scales at which the shock transitions themselves occur are certainly
resolved by the expansions considered here, but just how far in the results
can be trusted is somewhat unclear.  For this problem the gravitational
softening does not necessarily represent the scale at which numerics begins
to dominate the solution, because the trajectories of the mass elements are
essentially ballistic as they approach the pancake caustic, and therefore
the gravitational acceleration at that point is of secondary importance.
This conclusion is supported by the fact these profiles are unchanged when
the models are rerun with four times as many PM gridcells ($256^2$ rather
than $128^2$), and therefore twice the linear gravitational resolution.

\begin{figure}[htbp]
\epsscale{0.5}
\plotone{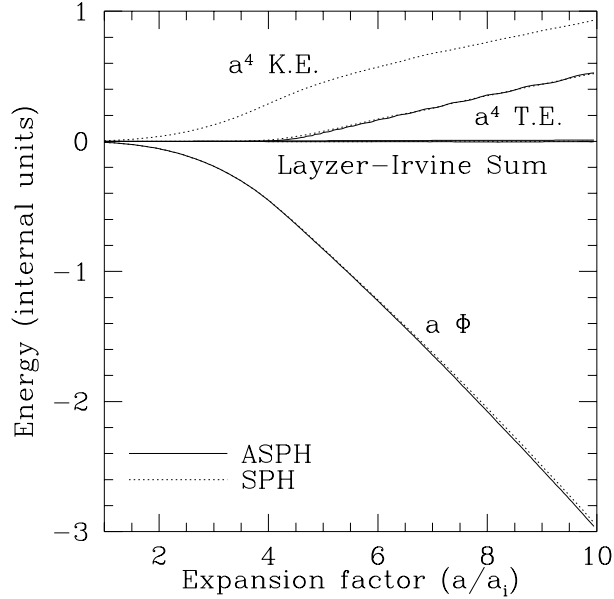}
\caption{Evolution of the global energies (kinetic, thermal, potential,
and Layzer-Irvine sum) for the $(k_x=0, k_y=1)$ 2D Zel'dovich pancake
simulations.}
\label{2dpan01_cons.fig}
\end{figure}
Figure \ref{2dpan01_cons.fig} plots the evolution of the global energies
(kinetic, thermal, potential, and the Layzer-Irvine sum) throughout these
simulations.  The Layzer-Irvine energy sum, $L.I. = a^4(K.E. + T.E.) + a
\Phi - \int \Phi \, da$ (Efstathiou \etal\ 1985), demonstrates that both
ASPH and SPH conserve global energy equivalently, with fluctuations in this
total of less than a percent the total fluctuation in its components.

A natural question which might arise at this point is: how many particles
would an SPH simulation require to treat this problem roughly as well as
ASPH?  We could, for instance, choose to match the maximum resolved
density, which differs by about a factor of 3.  The SPH smoothing scale is
related to the density and total number of particles as $\Sub{h}{SPH}
\propto \rho^{-1/\nu} \propto N^{-1/\nu}$ in $\nu$ dimensions, implying that
the resolved density varies linearly with the number of particles.  We can
therefore estimate that an SPH simulation would require of order 3 times as
many particles in order to resolve the peak density equivalently to ASPH in
this problem.  The ASPH algorithm itself (\eg, using and evolving the \Gt\
tensors) involves very little computational penalty compared with straight
SPH -- the majority of the increased computational effort is due the
smaller timesteps required by ASPH's increased resolution.  Since an SPH
simulation with equivalent resolution would also require such smaller
timesteps, the overall computational penalty for running an SPH simulation
with equivalent resolution to our ASPH example is roughly given by the
increased number of particles, or again roughly a factor of 3 in this case.
Alternately, if we instead choose to try and match the maximum linear
resolution achieved, experiments such as those shown in Paper I indicate
SPH would require of order 10 times as many particles per dimension to
match ASPH, implying 100 times as many particles in 2D or 1000 times
in 3D.

However, it is important to consider these sorts of comparisons with
caution, as the results depend upon both the problem in question and
precisely what it is we are trying to match.  Because both ASPH and SPH are
Lagrangian techniques their intrinsic resolutions are most naturally
expressed in terms of a {\em mass} resolution, fundamentally represented by
the discretization of the mass in the simulation into particles.  The
actual mass resolution achieved is a convolution of the particle mass scale
with the local (A)SPH interpolation kernel, which is where the spatial
adaptability of the technique comes into play.  The theoretical ideal mass
resolution is set by the particle discretization, and the degree to which
this resolution is actually realized at each point in a problem is
determined by how well the local spatial filtering adapts to the distorting
particle distribution, making this a complicated, problem dependent
question.  All the comparisons in this paper relate SPH and ASPH
experiments using the same number of particles and the same number of
neighbors per particle, and therefore formally have the same ultimate
limiting mass resolution.  We can improve the fidelity of the SPH runs by
using more particles until the quantitative SPH results for some chosen
parameter (such as the peak density for a particular structure) match those
of ASPH.  However, by increasing the number of particles we are in fact
using an SPH experiment with a potentially much better mass resolution than
the ASPH example.  The results of such intrinsically unbalanced comparisons
can be misleading.  As an example, consider the collapse of a planar
perturbation such as the Zel'dovich pancake. 
Suppose this pancake is subject to some instability which causes it
to depart from planar symmetry after the pancake shocks form, leading
to the formation of clumps and filaments in the postshock layer.
Suppose the mass of the clumps which are physically produced by this
instability is too small to be resolved by either the SPH or ASPH simulations
when they have the same particle number, because that mass is below the
mean mass per smoothing volume, for example.  In that case, one would
not expect to see the production of the clumps in either method.
If one were to try to offset
the poorer spatial resolving power of the SPH method at the same particle
number as the ASPH method, by increasing the particle number of the SPH method
by some substantial factor, however,
then the nominal mass resolution of the SPH method increases as well.
It might then be possible for the SPH method to produce clumps as
expected from the existence of the instability, while the ASPH simulation
with fewer particles resolves the pancake shocks just as well but
is not capable of producing transverse clumping.
In this example we would find qualitatively different results because we
have fundamentally altered the underlying resolution of the experiment, and
are therefore probing different physical regimes.

These sorts of problem dependent, confusing comparisons can be avoided by
recognizing that the fundamental resolution of these techniques is best
expressed as a mass per smoothing volume, or more precisely as the mass per
smoothing scale within each representative volume.  ASPH is better able to
maintain a fixed mass per smoothing scale in the presence of evolving
physical anisotropies than SPH, and thereby strives to better and more
consistently realize the theoretical mass resolution set by the particle
discretization throughout the computational volume, regardless of how the
particle distribution is distorted.  A more appropriate way to compare the
two techniques is that, for physically anisotropic situations such as
demonstrated here, ASPH allows a more effective representation of the
underlying physics than an SPH model for a given number of particles at
relatively little computational penalty.  The distinction is not unlike
that of comparing SPH with a variable smoothing scale to the case with
constant smoothing: the resolution of SPH with a constant smoothing scale
depends on how many particles happen to fall within a given smoothing
volume at any point during the problem; SPH with variable smoothing
improves upon this by trying to ensure that each particle samples the same
amount of mass at all times; ASPH goes further still in that it tries to
ensure that at all times each particle will sample the same amount of mass
in every direction, so that the mass per smoothing scale is held constant.
ASPH is not a substitute for improving the fundamental mass resolution
through increasing the number of particles, but rather strives to maintain
the best possible realization of the ideal Lagrangian resolution for a
given number of particles.

We should also point out that comparison of the profiles for the Zel'dovich
pancake shown here with analogous figures in Paper I demonstrates that,
while the results are fundamentally the same (\ie, the relative improvement
offered by ASPH over SPH), the shock transition is more tightly constrained
using the formalism of Paper I.  This distinction is due to the difference
in the methods used to suppress the artificial viscosity.  The method
described in Paper I is clearly more effective at capturing the sharp shock
transition in this problem.

\paragraph{Multiple-Wavelength 2D Zel'dovich pancakes:}
\begin{figure}[htbp]
\plottwo{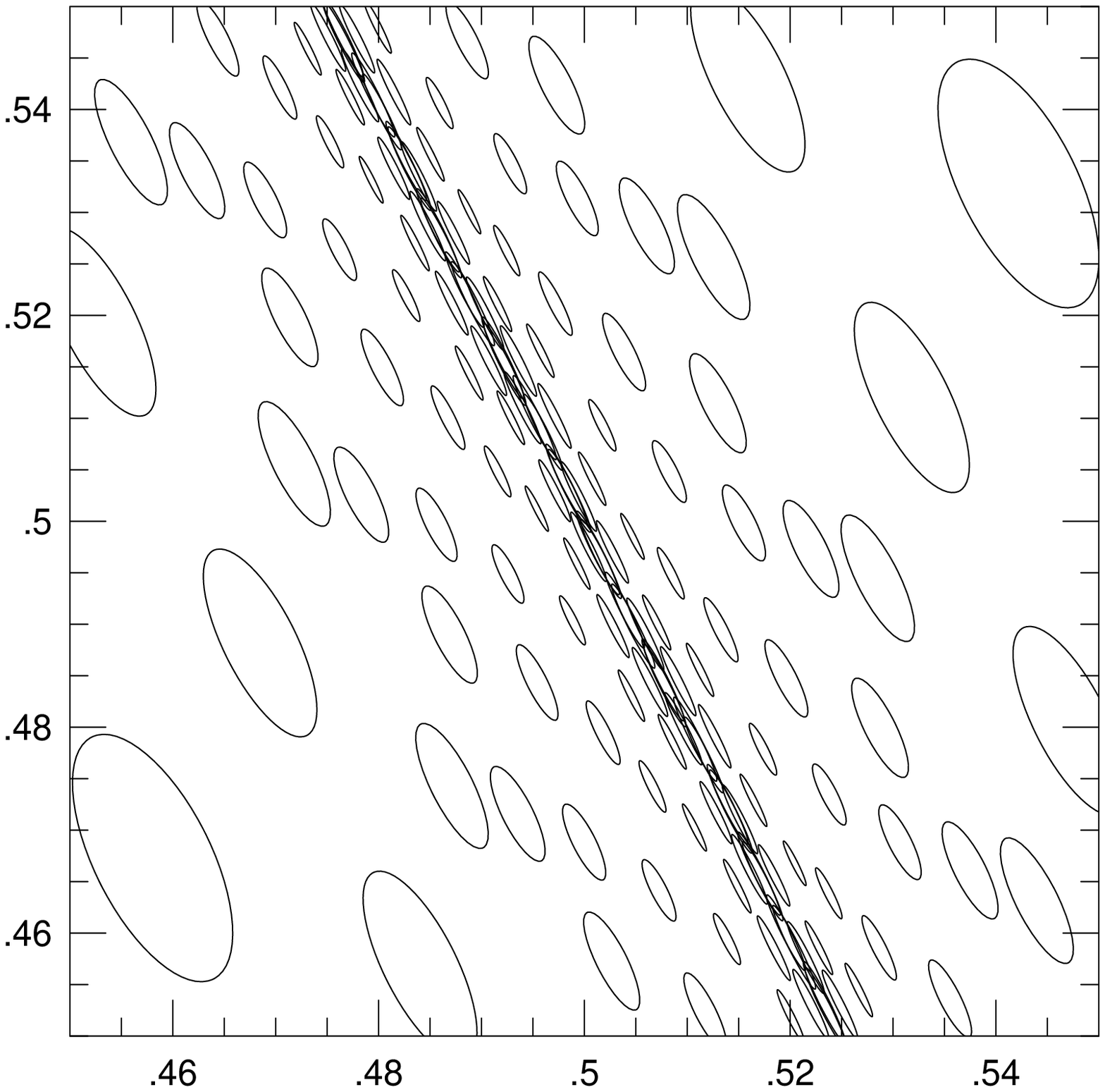}{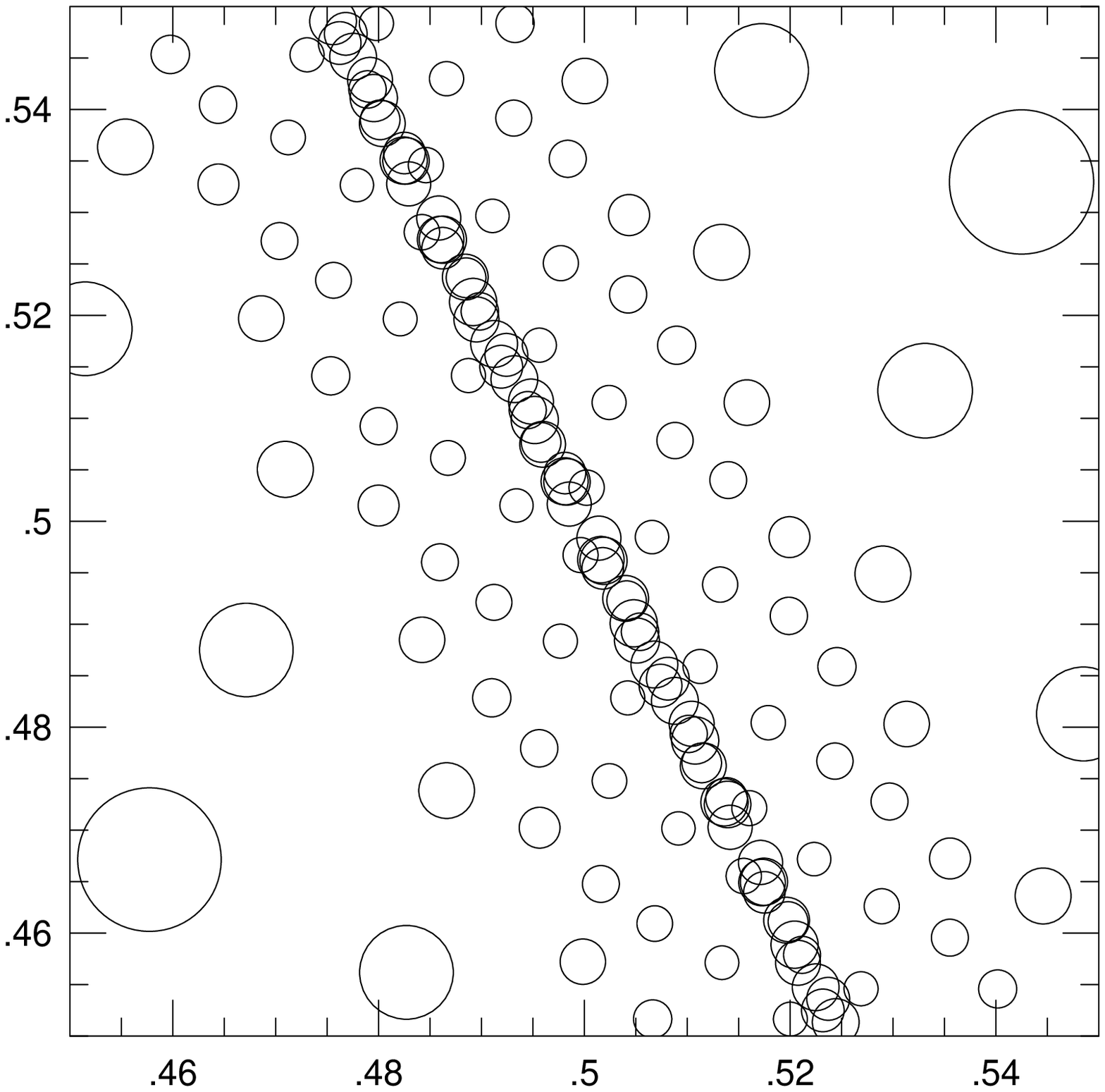}
\caption{Kernel plots for the multiple-wavelength $(k_x=2, k_y=1)$ 2D
Zel'dovich pancake simulations @ $a/a_c = 2.5$.  The left panel shows the
ASPH run, and the right SPH.}
\label{2dpan21_ker.fig}
\end{figure}

\begin{figure}[htbp]
\plottwo{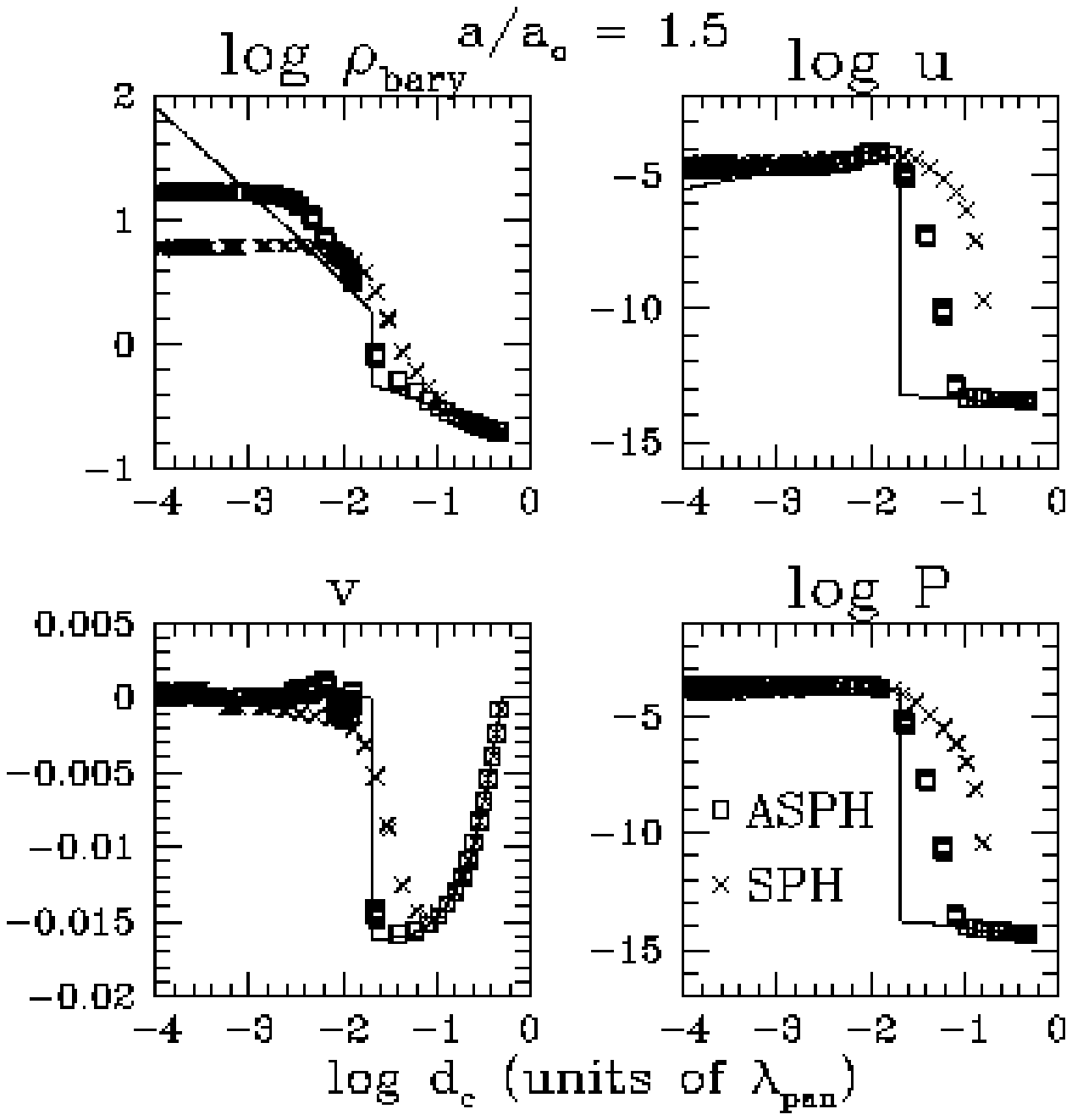}{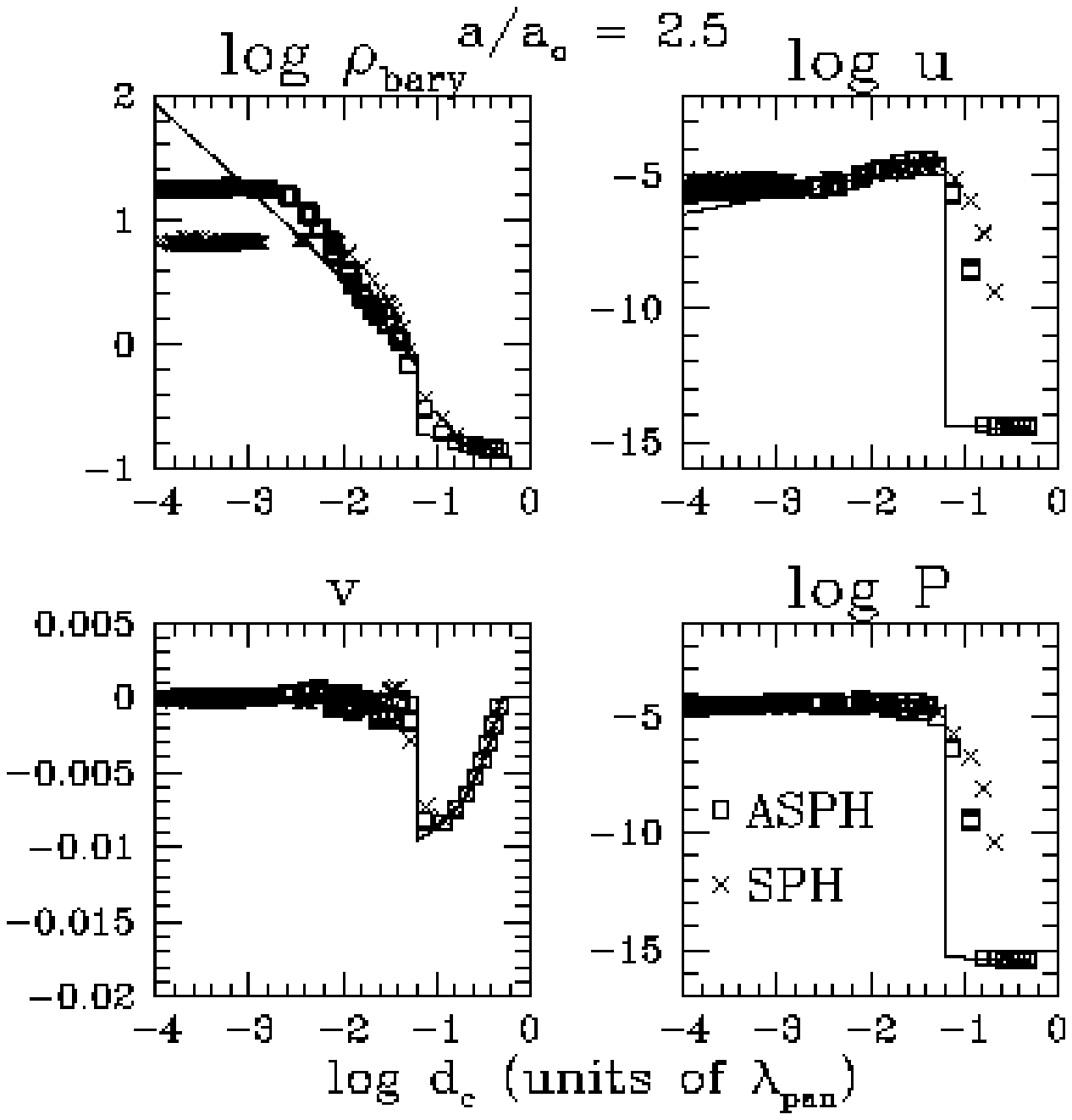}
\caption{Half-wavelength physical profiles of density $\rho$, specific
thermal energy $u$, velocity $v$, and pressure $P$ as a function of
distance from the pancake caustic $d_c$ for the $(k_x=2, k_y=1)$ 2D
Zel'dovich pancake simulations.  Units and plotting conventions are the
same as used in Figure \protect\ref{2dpan01_prof.fig}.  The left panel
shows results for $a/a_c=1.5$, and the right shows $a/a_c=2.5$.}
\label{2dpan21_prof.fig}
\end{figure}

\begin{figure}[htbp]
\epsscale{0.5}
\plotone{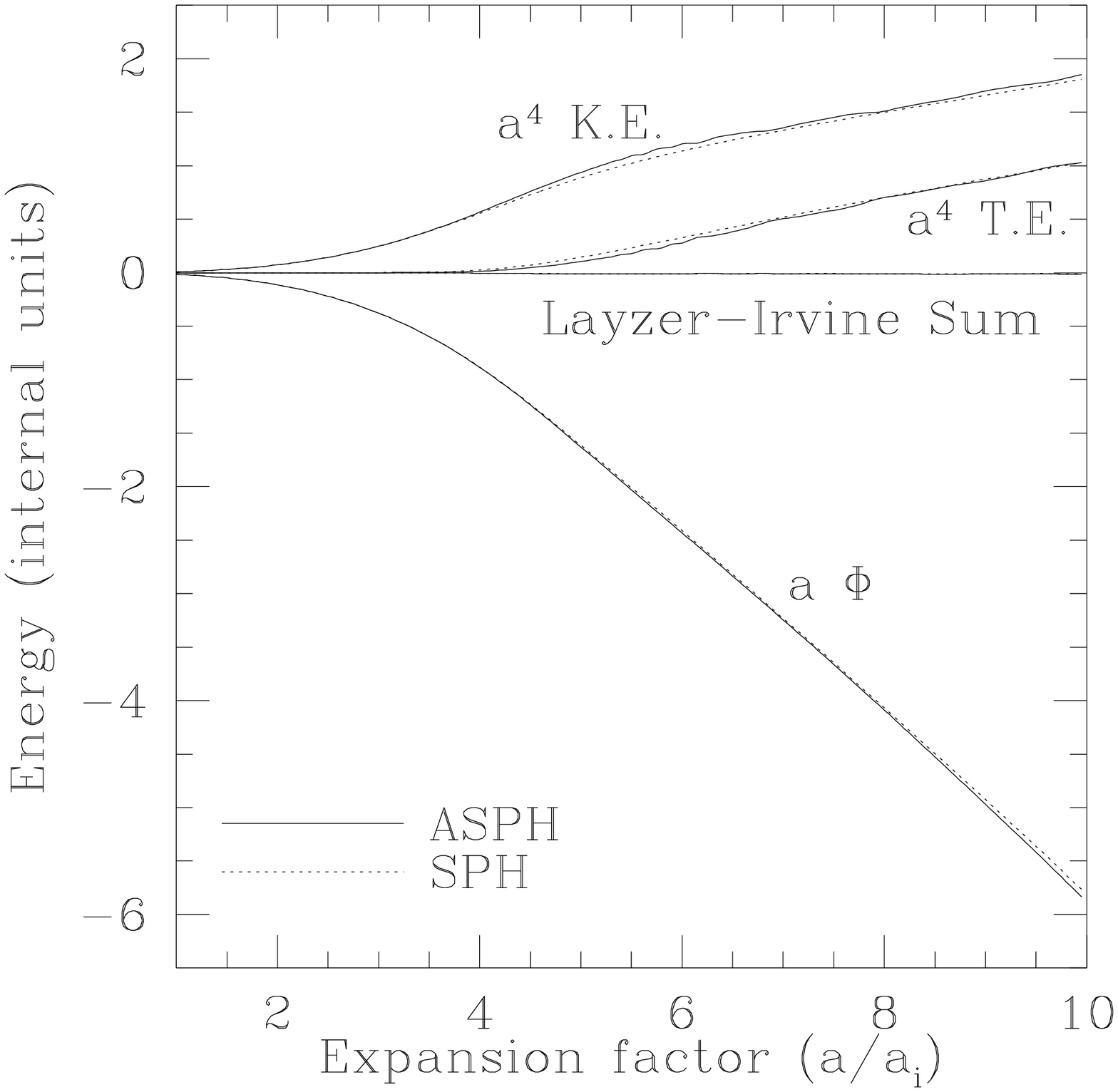}
\caption{Evolution of the global energies (kinetic, thermal, potential,
and Layzer-Irvine sum) for the multiple-wavelength $(k_x=2, k_y=1)$ 2D
Zel'dovich pancake simulations.}
\label{2dpan21_cons.fig}
\end{figure}
We now present a slightly modified pancaking problem: that of
$(k_x=2,k_y=1)$.  In this case there are $5^{1/2}$ wavelengths in our
computational volume, each tilted with respect to the principal axes.  This
problem represents no fundamental change from the previous example, but
rather provides an example where the physical problem does not align with
any special simulation symmetry (\ie, there is no alignment with either a
simulation axis or any special direction with respect to the initial node
seeding).  The only real difference between this and the previous
single-wavelength example is that now the resolution per wavelength is
effectively reduced, since we have increased the number of waves in the box
with the same number of particles.  There are now effectively $64/5^{1/2}
\sim 28$ nodes per wavelength, or $\sim 14$ nodes per half-wavelength.

Figure \ref{2dpan21_ker.fig} presents kernel plots for both the SPH and
ASPH simulations at expansion $a/a_c=2.5$.  It is clear that ASPH has
successfully adapted to the tilted geometry of this problem.  Additionally,
there is no evidence of the transverse perturbations evident in the
single-wavelength case (as is particularly notable for SPH at $a/a_c =
2.5$).  This difference is due to the fact that in the $(k_x=0,k_y=1)$ case
we are trying to stack up nodes exactly one on top of the other.  Any
deviation from this perfect line-up will be amplified, and lead to a
breakdown in the symmetry.  In the $(k_x=2,k_y=1)$ case we have broken that
symmetry, and therefore the instability is greatly reduced.  Figure
\ref{2dpan21_prof.fig} presents half-wavelength profiles for both
simulations at expansion factors $a/a_c=1.5$ and $a/a_c=2.5$.  The results
are fundamentally the same as shown in Figure \ref{2dpan01_prof.fig} --
ASPH again resolves central densities $\sim 3$ times those in the
equivalent SPH experiment.  We can therefore conclude that it would take
SPH 3 times as many particles (and CPU work) to produce equivalent fidelity
for this problem, though with the same caveats as outlined for such
comparisons in \S \ref{2dpan01.sec}.  Finally, Figure
\ref{2dpan21_cons.fig} shows the evolution of the global energies for these
simulations.  Again there is little qualitative change from the
single-wavelength case, with a total energy conservation violation of less
than a percent.

\paragraph{Single-Wavelength 3D Zel'dovich Pancakes:}
\begin{figure}[htbp]
\plottwo{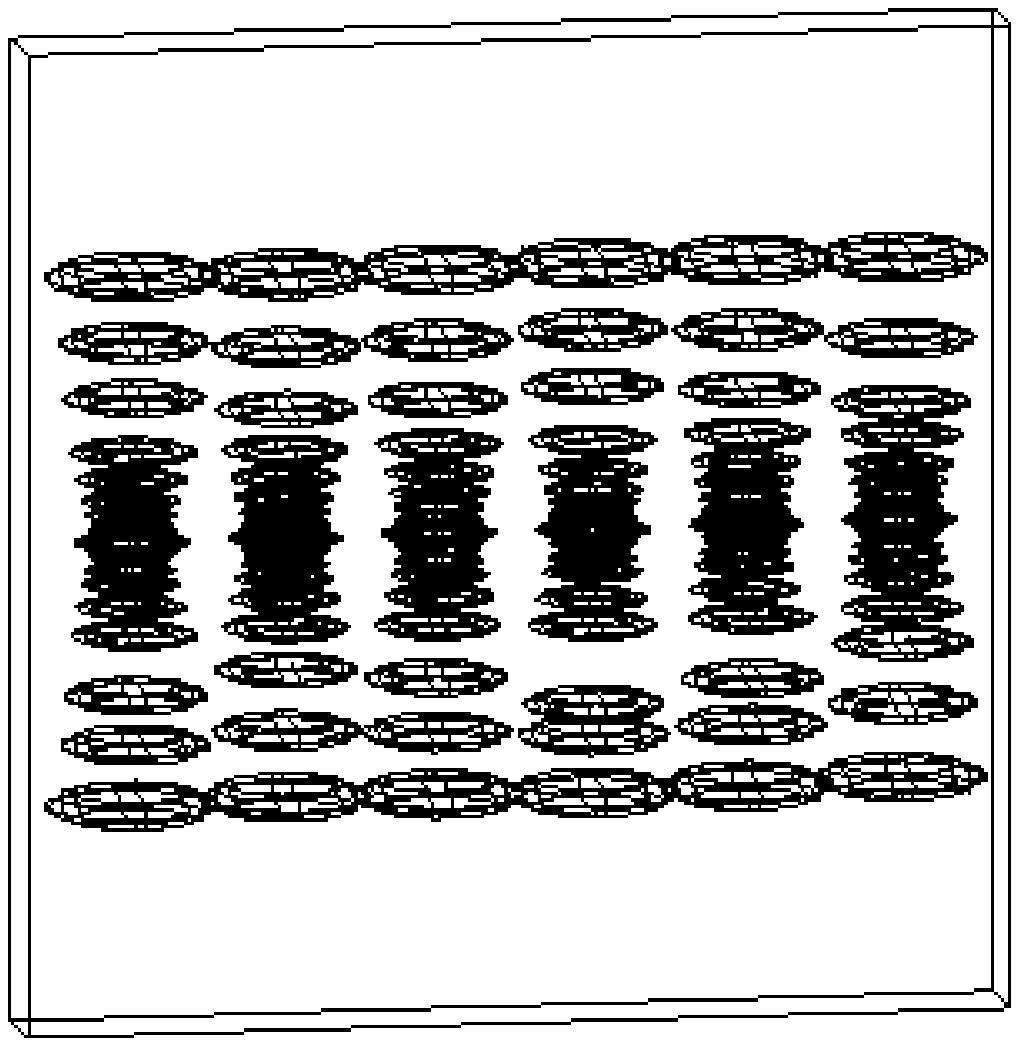}{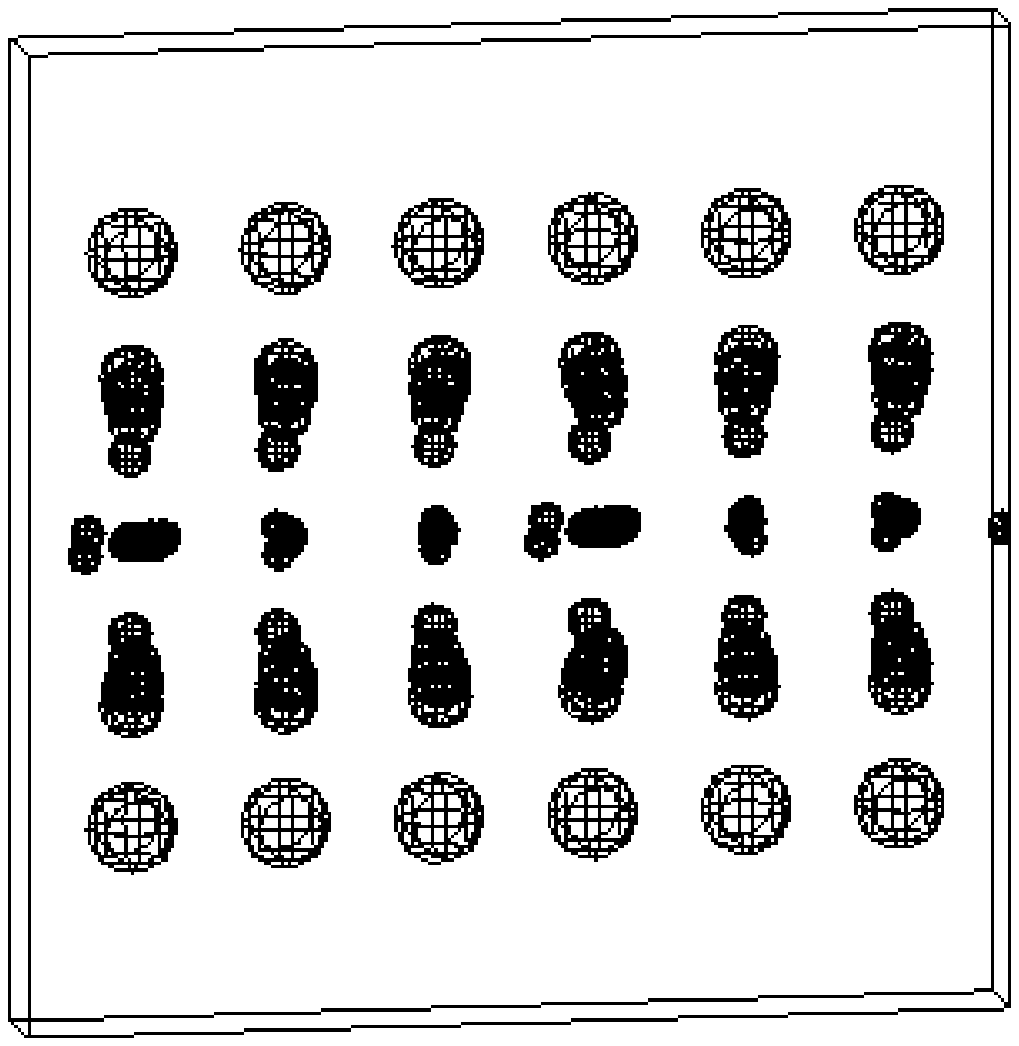}
\caption{Kernel plots for 3D Zel'dovich pancake simulations @
$a/a_c=2.5$.  We only display a thin slice $(x \in [0.4,0.6], y \in [0.5
0.52], z\in [0.4,0.6])$ out of unit volume simulations, rotated slightly
for clarity.  The $h=0.2$ isocontours are drawn as wire-mesh surfaces.
The left panel shows the ASPH simulation, and the right SPH.}
\label{3dpan001_ker.fig}
\end{figure}
We now revisit the single wavelength version of the problem in 3D with
$(k_x=0,k_y=0,k_z=1)$.  These 3D simulations are much lower resolution
than the 2D cases -- with only $32^3$ nodes in 3D as compared with
$64^2$ in 2D, we have effectively lowered the spatial resolution by a
factor of two and the mass resolution by a factor of 8.  In Figure
\ref{3dpan001_ker.fig} we present kernel plots of an ASPH and an SPH
simulation at $a/a_c=2.5$.  For clarity we have selected out only a single
plane of nodes in the central regions of the simulation to display.  The
surfaces are again $h=0.2$ isocontours -- note that in general ASPH kernels
are ellipsoidal in shape, while SPH are spherical.  It is evident that ASPH
retains the ability to recognize and adapt to the 1D nature of this
collapse, flattening out the kernels as they fall into the pancake caustic.
The SPH kernels, which are restricted to remain spherical, are not able to
adapt as well to the ongoing collapse.  By $a/a_c=2.5$, it is clear that
the SPH simulation is breaking the symmetry of the problem, as the SPH
nodes begin to slip by one another and become disorganized in the pancake
midplane.  The ASPH simulation maintains the physical symmetry of the
problem more effectively.

\begin{figure}[htbp]
\plottwo{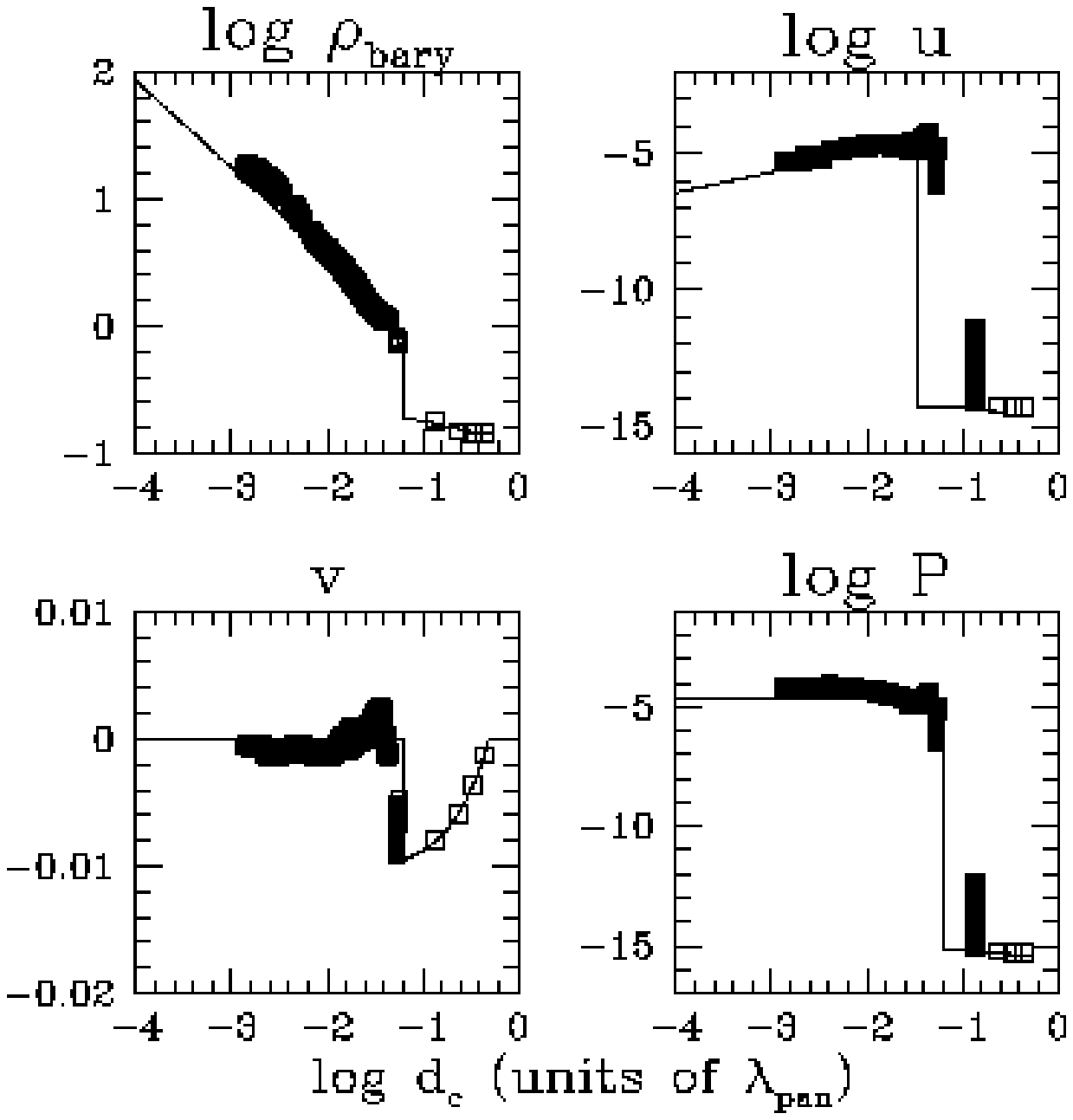}{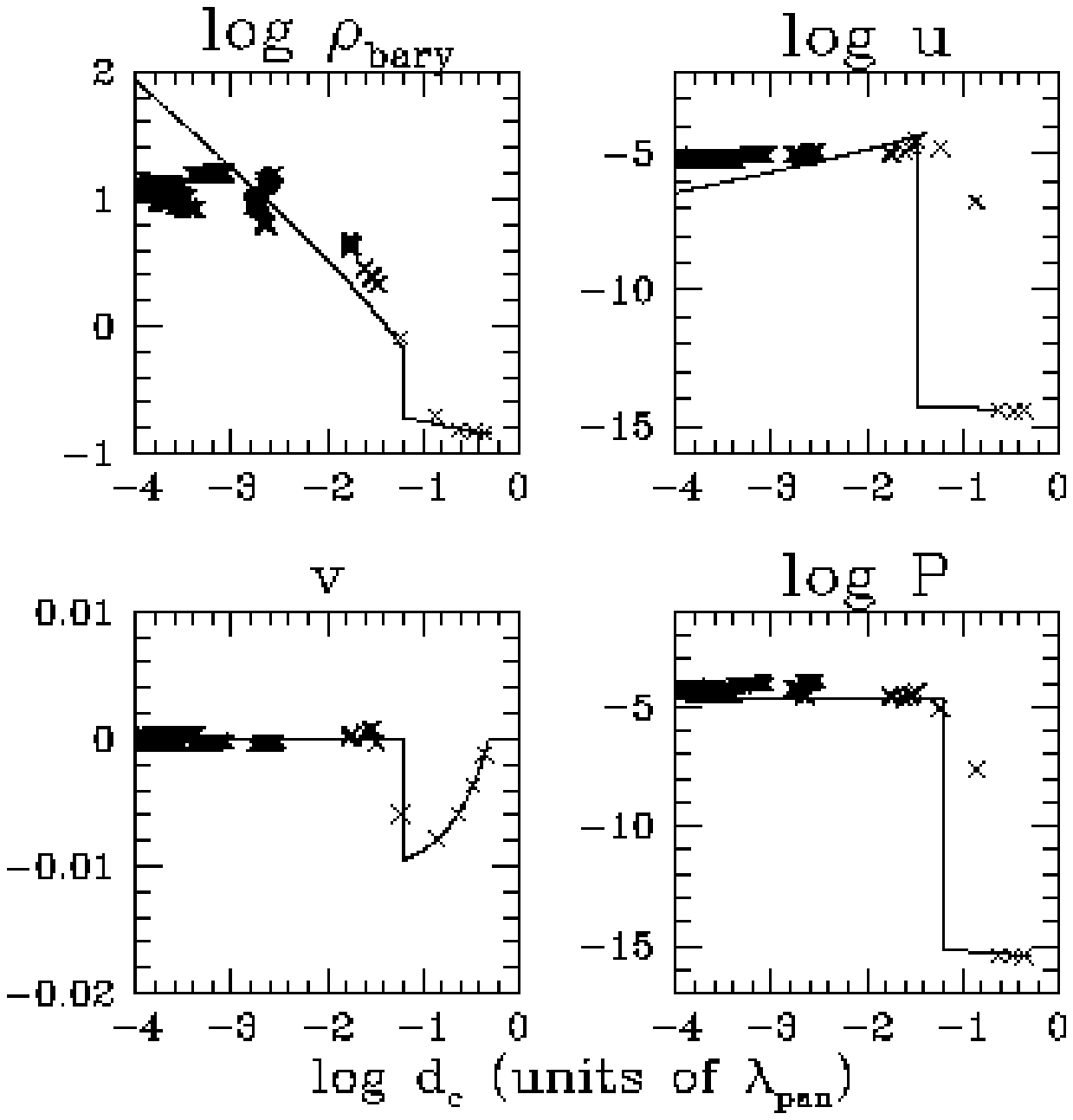}
\caption{Half-wavelength profiles of density $\rho$, specific thermal
energy $u$, velocity $v$, and pressure $P$ as a function of distance from
the pancake caustic $d_c$ for the $(k_x=0,k_y=0,k_z=1)$ 3D Zel'dovich
pancake simulations at $a/a_c=2.5$.  All quantities are converted to proper
coordinates and are expressed in units of the critical density, pancake
wavelength, and the Hubble time at the beginning of the simulation.  Solid
lines represent the analytical expectations.  The left-hand panel shows the
ASPH simulation, and the right SPH.}
\label{3dpan001_prof.fig}
\end{figure}

\begin{figure}[htbp]
\epsscale{0.5}
\plotone{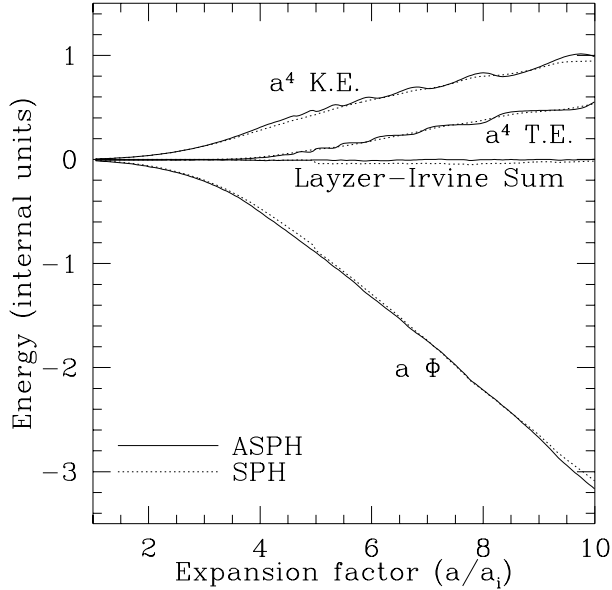}
\caption{Evolution of the global energies (kinetic, thermal, potential,
and Layzer-Irvine sum) for the $(k_x=0,k_y=0,k_z=1)$ 3D Zel'dovich pancake
simulations.}
\label{3dpan001_cons.fig}
\end{figure}
We can see the physical effects of these differences in the profiles
presented in Figure \ref{3dpan001_prof.fig}.  Comparing the density
profiles at $a/a_c=2.5$, the ASPH simulation follows the analytical profile
essentially as far the nodes can resolve (down to scales
$d_c/\Sub{\lambda}{pan} \sim 10^{-3}$), whereas the SPH simulation fails to
on scales of $d_c/\Sub{\lambda}{pan} \sim 10^{-2}$, becoming quite
disordered on smaller scales.  At all times the shock transition is both
more sharply defined and better localized under ASPH.  Finally in Figure
\ref{3dpan001_cons.fig} we present the evolution of the global energies
throughout these 3D simulations.  As in 2D, we find that both techniques
conserve energy to better than 1\%.

\subsubsection{Sedov Blastwave Test}
\label{Sedov.sec}
We will now turn to an entirely different class of test problems: that of
an intense explosion in a gas.  This problem possesses a set of well-known
similarity solutions (Sedov 1959).  We simulate this problem in both 2D and
3D, and compare the results to Sedov's solutions in the appropriate
geometry.  This problem represents a somewhat difficult case for Lagrangian
techniques such as (A)SPH, since the void is dynamically important.

The formal initial condition for this problem is to introduce an intense
point source of thermal energy into an initially pressureless, homogeneous
gas.  This immediately poses a problem for any (A)SPH formalism, as (A)SPH
cannot represent a discontinuous energy distribution.  Therefore, in order
to initialize this problem we distribute a thermal energy spike amongst a
small number of nodes in the gas, and then smooth this distribution.  This
results in an energy spike in the gas resembling the shape of our smoothing
kernel, as this is the closest (A)SPH can come to representing a delta
function.  The remaining nodes are initialized with a small, but finite,
internal energy in order to simulate an initially pressureless gas.  All
\Gt\ tensors are initialized as identical, spherical SPH tensors
appropriate for the undisturbed initial density $\rho_0$ and desired number
of significant neighbors per smoothing length $N_h$.  The major simulation
parameters we use for our Sedov blastwave simulations are summarized in
Table \ref{Sedov.tab}.

\paragraph{2D Sedov Blastwave Simulations:}
\begin{figure}[htbp]
\plottwo{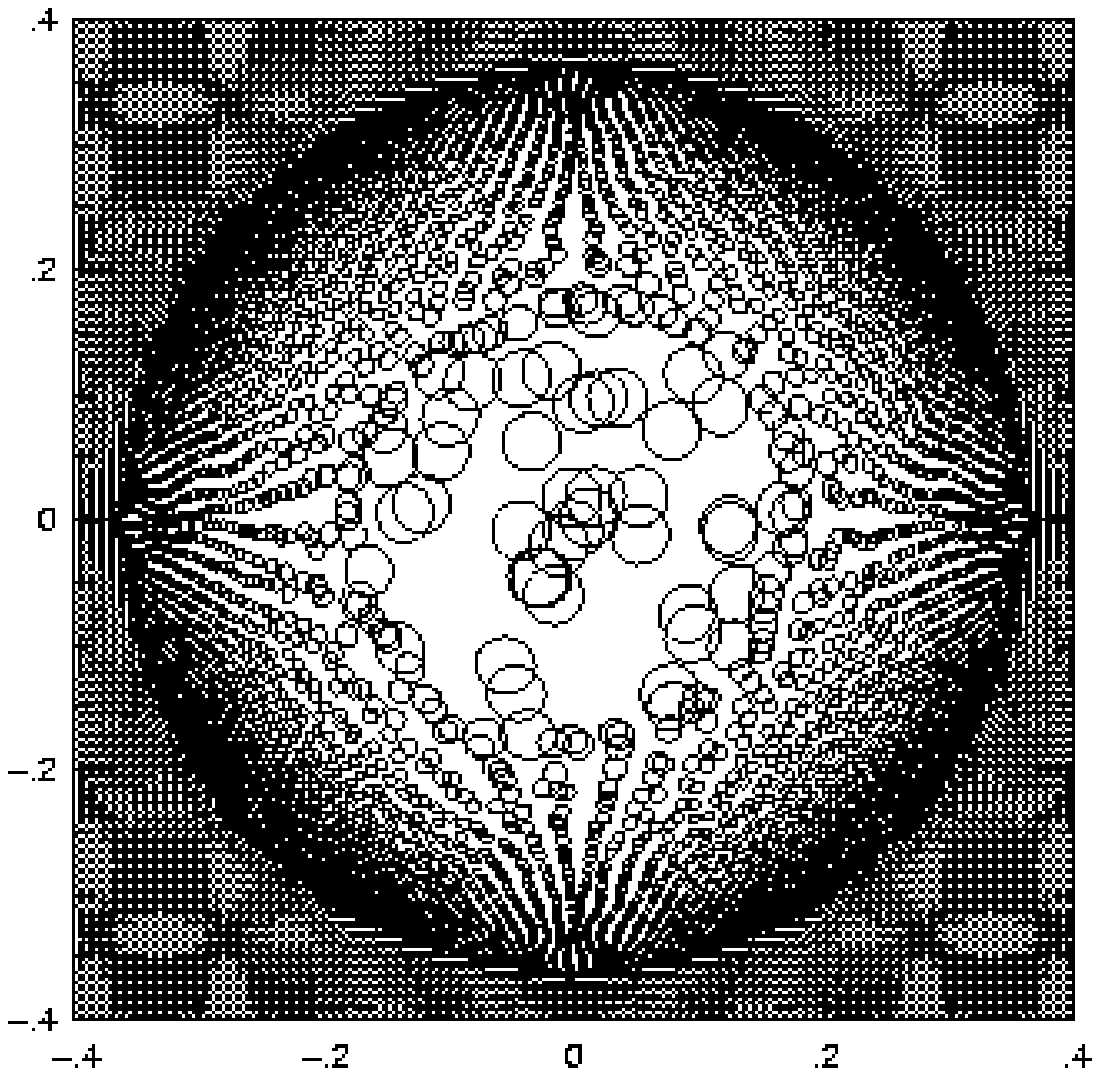}{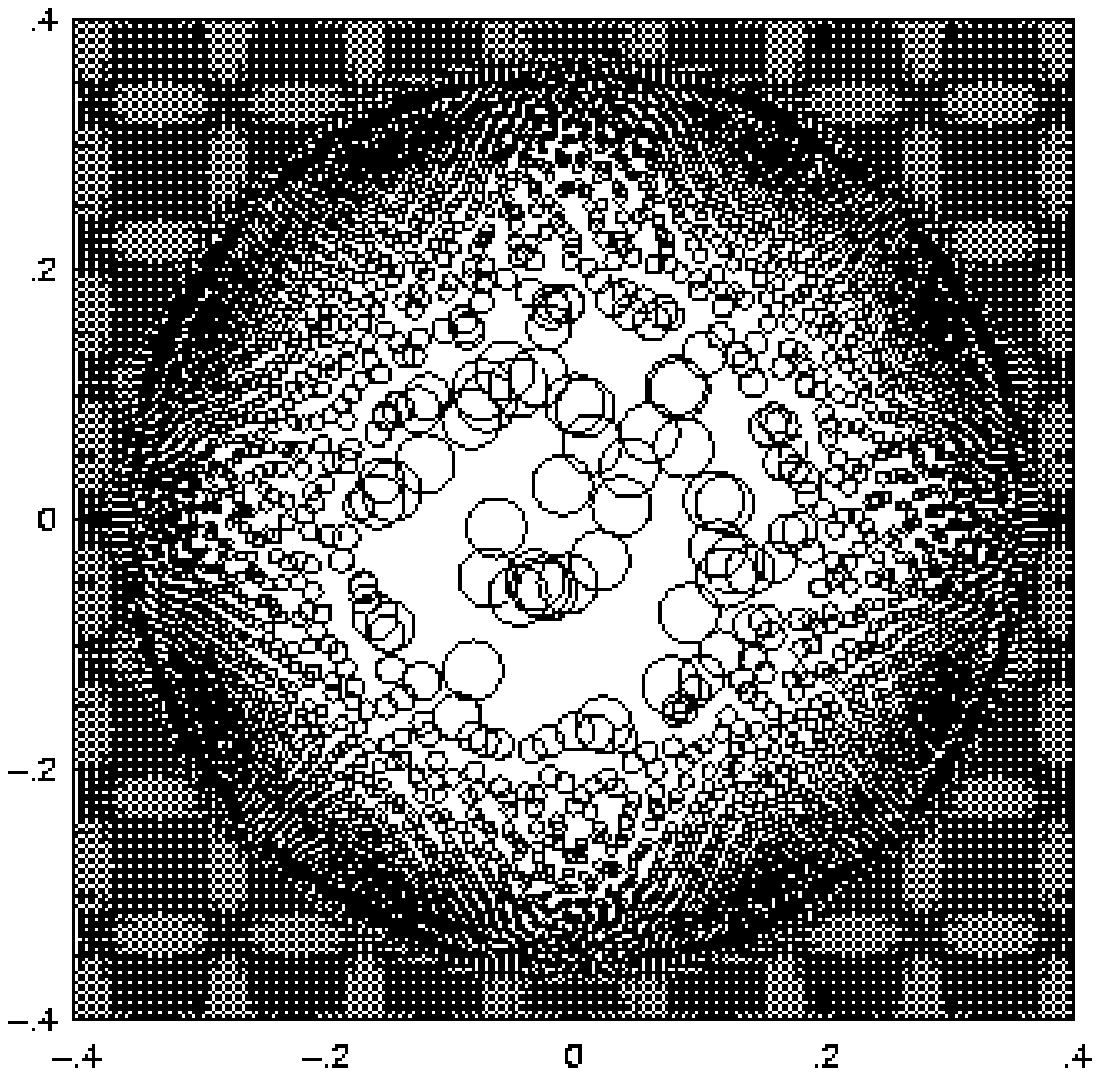}
\caption{Kernel plots for 2D Sedov blastwave simulations at time
$t=0.12$ for the cases of ASPH (left panel) and SPH (right panel).}
\label{Sed_ker.fig}
\end{figure}

\begin{figure}[htbp]
\plottwo{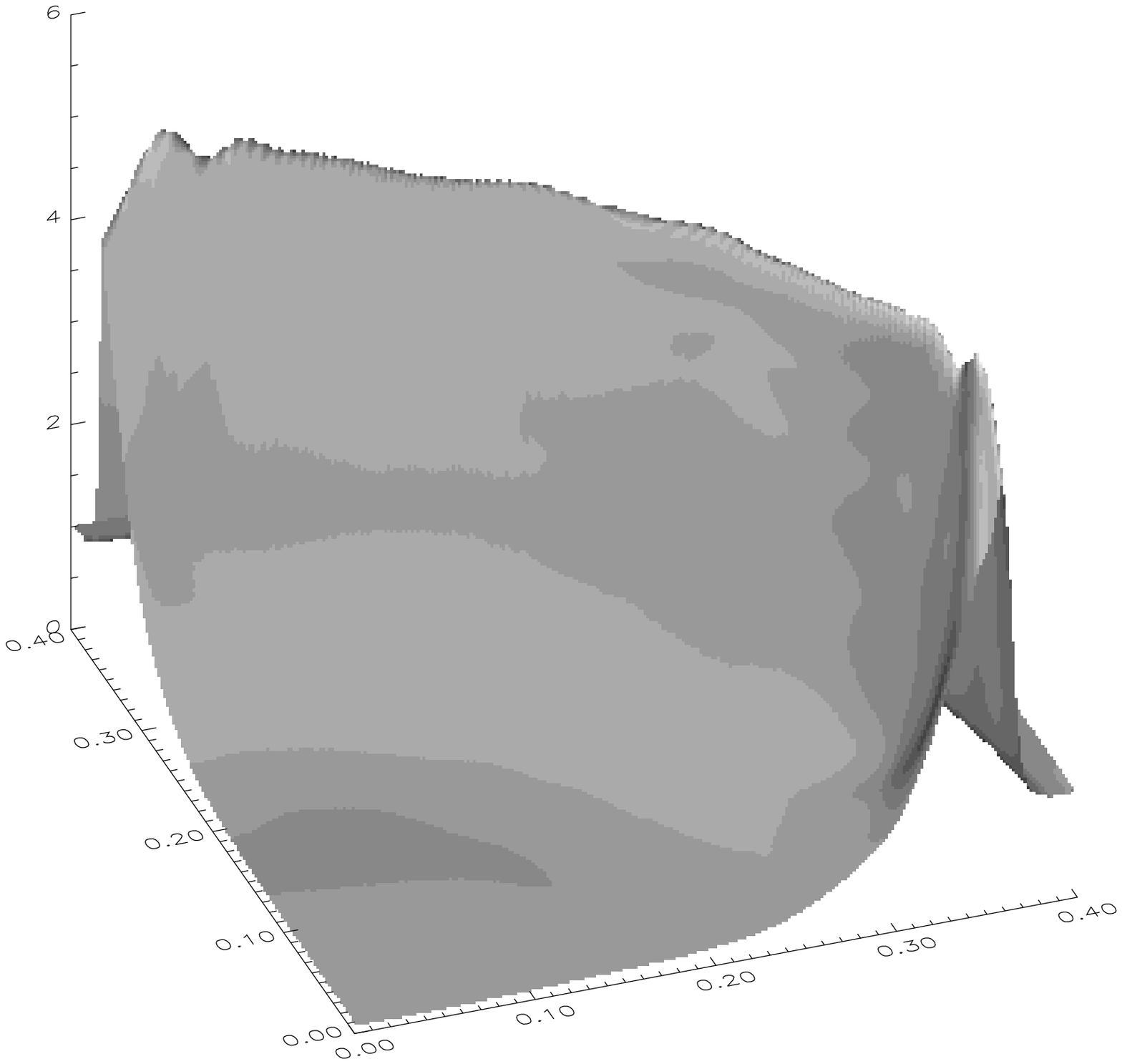}{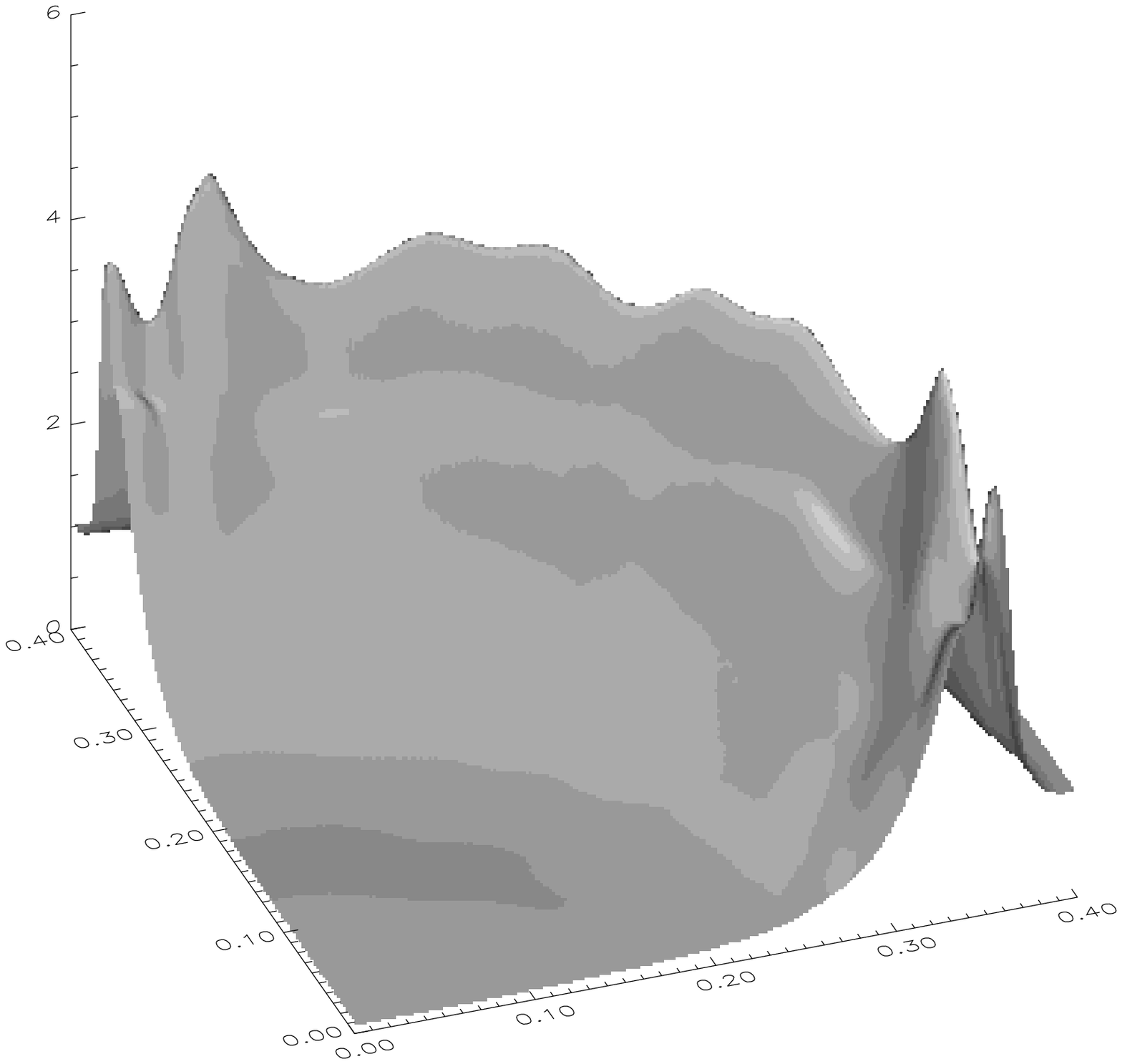}
\caption{Surface plots of the mass density field $\rho$ in one quadrant
of the expanding shockfront for the 2D Sedov blastwave simulations at
$t=0.12$.  The left panel shows the ASPH simulation, and the right SPH.}
\label{Sed_surf.fig}
\end{figure}
Figure \ref{Sed_ker.fig} presents kernel plots for an ASPH and an SPH
simulation of this problem at time $t=0.12$.  It is clear that the ASPH
model is more successful at producing a round, azimuthally symmetric
shockfront than the SPH case.  ASPH is able to produce a better shape for
the shock because the ASPH smoothing scale parallel to the shockfront is
not shrinking, and therefore communication between the nodes parallel to
the shockfront is maintained.  The effects of this difference are evident
in Figure \ref{Sed_surf.fig}, which shows surface plots of the density in
one quadrant of the blast for each of these runs.  It is clear that the SPH
density along the shock front shows greater fluctuations compared with the
ASPH runs, which maintains the most consistent symmetry.

\begin{figure}[htbp]
\plotone{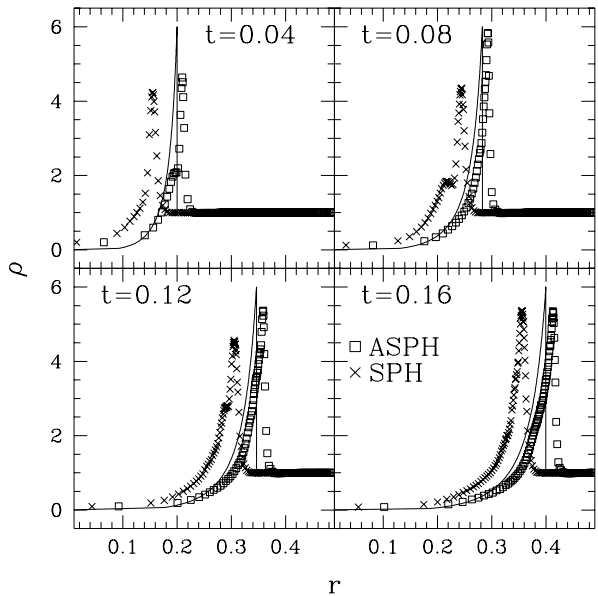}
\caption{Azimuthally averaged radial density profiles for ASPH (open
squares) and SPH (crosses) simulations of the 2D Sedov blastwave at times
$t$ = 0.04, 0.08, 0.12, and 0.16.  The SPH simulation is offset radially
for clarity -- the ASPH curves show the correct radial position.  Each
point represents the average of a radial bin containing 100 nodes.  The
solid lines show the analytic, cylindrical Sedov solution.}
\label{Sed_rhoprofs.fig}
\end{figure}

\begin{figure}[htbp]
\plotone{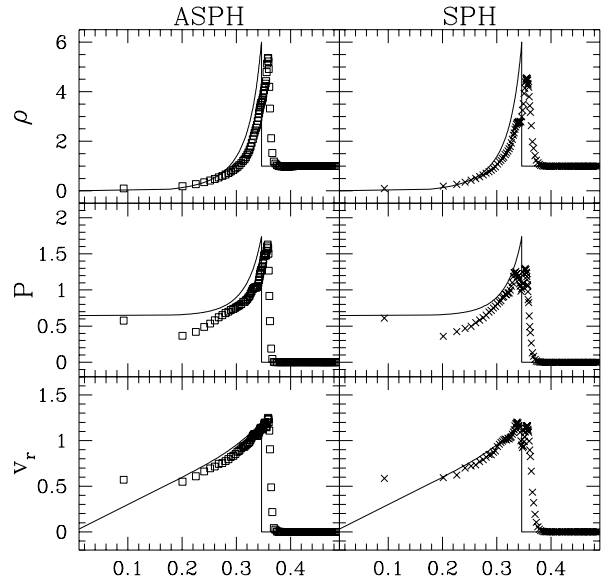}
\caption{Azimuthally averaged radial profiles of mass density $\rho(r)$,
pressure $P(r)$, and radial velocity $v_r(r)$ for the 2D Sedov blastwave
simulations at time $t=0.12$.  Shown are ASPH (open squares), SPH
(crosses), and the cylindrical Sedov solution (solid lines).  Each point
represents an average for a radial bin containing 100 nodes.}
\label{Sed_profs012.fig}
\end{figure}
Figure \ref{Sed_rhoprofs.fig} presents a time-series of azimuthally
averaged radial density profiles for these simulations, binned in
radial steps of equal numbers of particles.  Shown are times in the
interval $t \in [0.02,0.16]$.  The SPH run has been
offset radially for clarity.  Since we use $\gamma = 1.4$, the strong-shock
prediction for the density jump at the shock front is $\rho_2/\rho_1 =
(\gamma + 1)/(\gamma - 1) = 6$.  The ASPH runs converge to a peak shock
density $\rho_2 > 5$ fairly rapidly, while the SPH run does not achieve a
similar peak density until the end of the simulation, by which time the
shock has swept up most of the particles in the simulation.  The reason for
these differences is that the ASPH \Gt\ tensors are able to adapt to the
radial nature of the problem, allowing the radial smoothing scale to adjust
much more readily than SPH's isotropic approach allows.  We can gauge how
well ASPH's anisotropic smoothing is adapting to the predicted density jump
by examining how elliptical the kernels are becoming.  For the predicted
density jump of $\rho_2/\rho_1 = 6$, we would expect the ratio of the
shortest to longest axis of an ASPH node in the shock front to be $6^{-1}
\sim 0.167$, while we in fact find in the simulation $h_2/h_1 \sim 0.145$
at the shock front.  It is heartening that, as expected, the SPH results
converge with the ASPH as the number of nodes increases.  It should be
noted, however, that even though the SPH radial profiles do eventually
catch up with ASPH, SPH never achieves as round a shockfront.  Figure
\ref{Sed_profs012.fig} presents radial profiles of the radial velocity
$v_r$, mass density $\rho$, and pressure $P$ at $t=0.12$.  Also plotted is
the cylindrical Sedov solution, as the solid lines.  The break in the
pressure profile at $r \sim 0.2$ lies between the mass elements which mark
the edge of the initial thermal energy spike.

\begin{figure}[htbp]
\epsscale{0.5}
\plotone{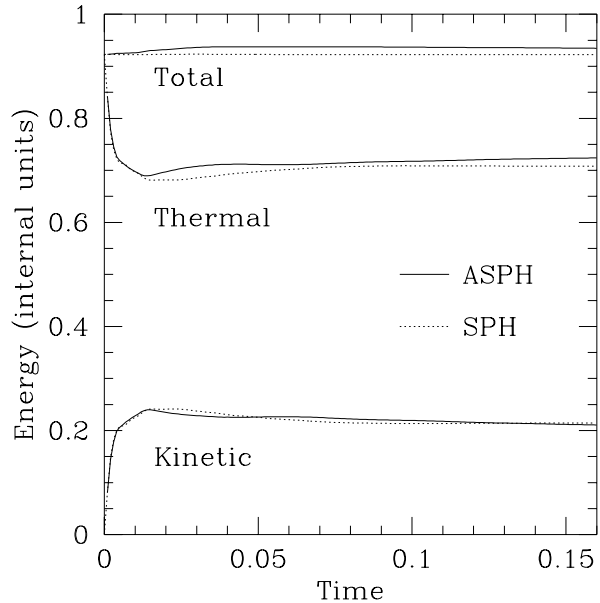}
\caption{Evolution of the global energies (kinetic, thermal, and total)
for the 2D Sedov blastwave simulations.}
\label{Sed_cons.fig}
\end{figure}
Finally, Figure \ref{Sed_cons.fig} presents the evolution of the global
energies for 2D Sedov models.  During the ASPH run the total energy
fluctuates peak to peak $\Delta E/E \sim 2\%$ the fluctuation of its
components, while SPH only suffers fluctuations of $\Delta E/E \sim 0.2\%$.
It is not unexpected for ASPH to have slightly worse energy conservation
than SPH, because ASPH introduces more degrees of freedom into the problem.
The difference is analogous to comparing SPH with a fixed smoothing scale
to SPH with variable smoothing.

\paragraph{3D Sedov Blastwave Simulations:}
We also examine 3D simulations of the Sedov blastwave problem under both
ASPH and SPH.  In this case we use $N=32^3$ (A)SPH nodes, which results in
a linear resolution of only one quarter what we have in 2D.  We can
therefore expect the 3D simulations to be much lower-resolution than the
previous 2D case.

\begin{figure}[htbp]
\plottwo{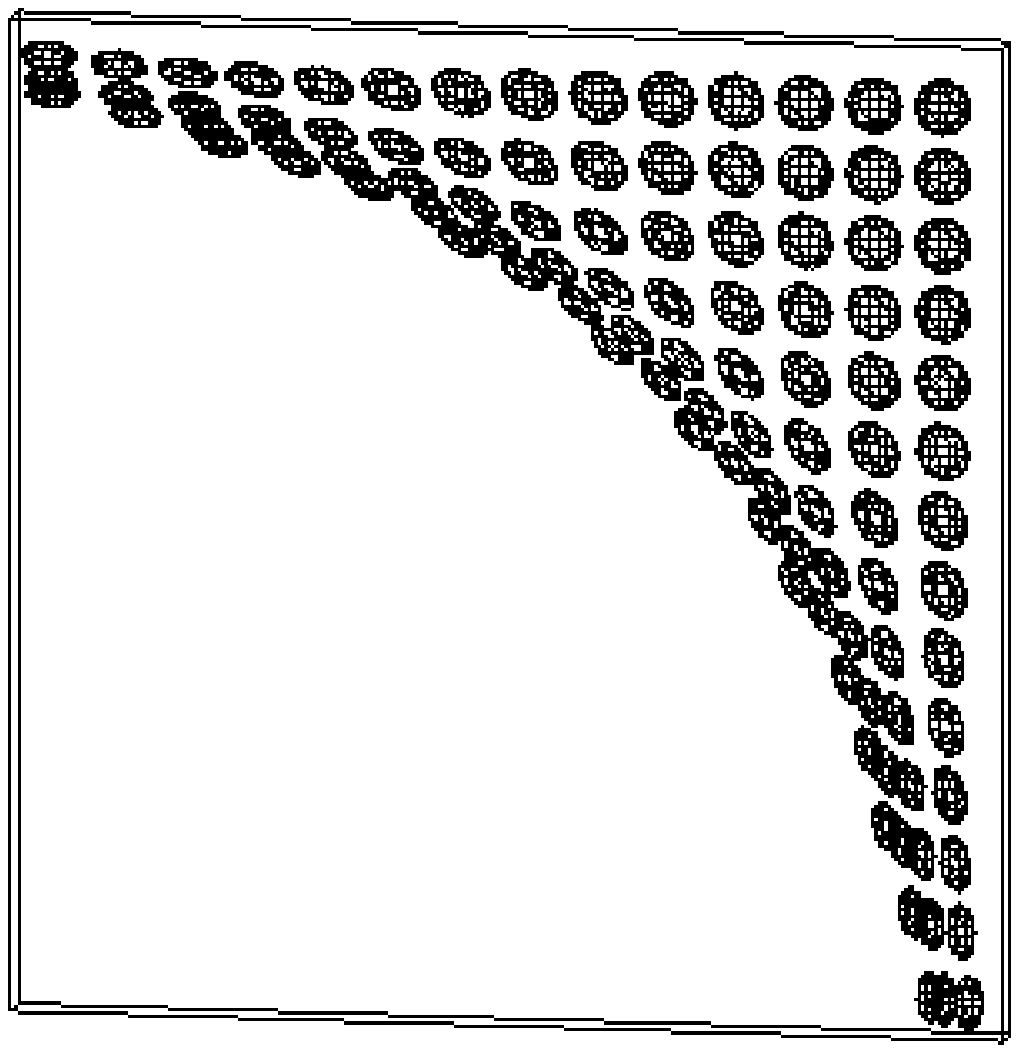}{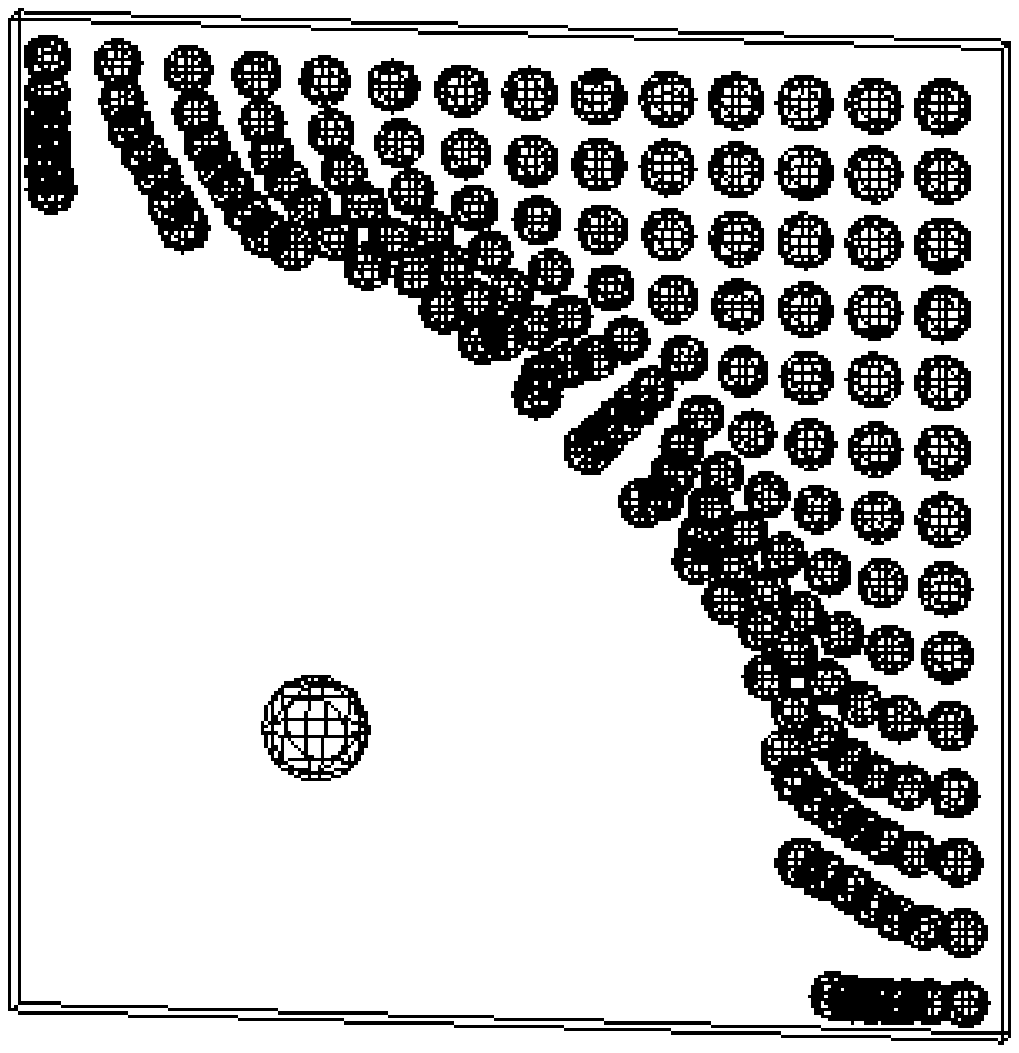}
\caption{Kernel plots for 3D Sedov blastwave simulations at time
$t=0.1$ for the cases of ASPH (left panel) and SPH (right panel).  These
figures show a small slice through the expanding blastwave for clarity: the
region plotted is $(x \in [0,0.45], y \in [0,0.45], z \in [0,0.02])$ out of
a unit volume.}
\label{3DSed_ker.fig}
\end{figure}
Figure \ref{3DSed_ker.fig} presents kernel plots for the ASPH and SPH 3D
Sedov blastwave simulations at time $t=0.1$.  We only plot a thin slice
through the volume (selecting out a single plane of nodes) for clarity.  It
is clear that the ASPH kernel shapes are indeed flattening radially as
expected, while the SPH nodes show relatively little change in the
expanding shockfront.  This difference is expected, since in 3D the ASPH
\Gt\ tensors are still able to approach the $\rho^{-1}$ evolution of the
smoothing scale in the radial direction, while the SPH smoothing scales are
constrained to evolve much more slowly as $\rho^{1/3}$.  Also, as is
evident in 2D, the ASPH nodes in 3D retain their ability to adapt to the
physical, spherical geometry of the problem well, avoiding the grid based
artifacts evident in the SPH node distribution.

\begin{figure}[htbp]
\plotone{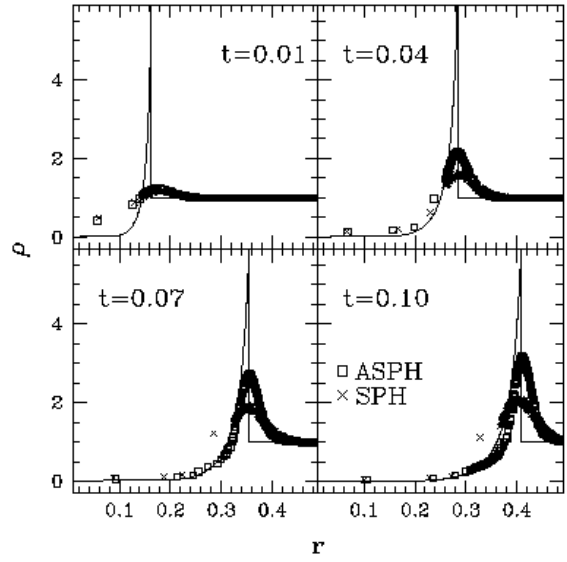}
\caption{Average radial density profiles for ASPH (open squares) and SPH
(crosses) 3D Sedov blastwave simulations at times $t$ = 0.01, 0.04, 0.07,
and 0.10.  Each point represents an average of a radial bin containing 50
nodes.  The solid lines show the Sedov solution.}
\label{3DSed_rhoprofs.fig}
\end{figure}

\begin{figure}[htbp]
\plotone{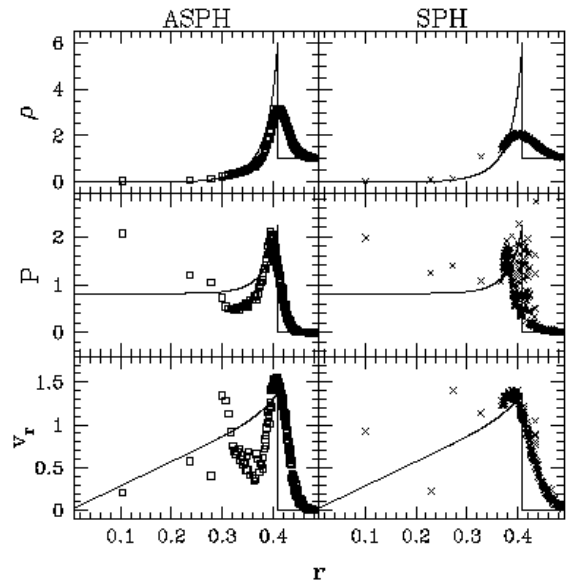}
\caption{Average radial profile of density $\rho(r)$, pressure $P(r)$,
and radial velocity $v_r(r)$ for 3D Sedov blastwave simulations at
$t=0.1$.  The squares represent the ASPH result, crosses SPH, and the solid
lines show the spherical Sedov solutions.}
\label{3DSed_profs010.fig}
\end{figure}

\begin{figure}[htbp]
\epsscale{0.5}
\plotone{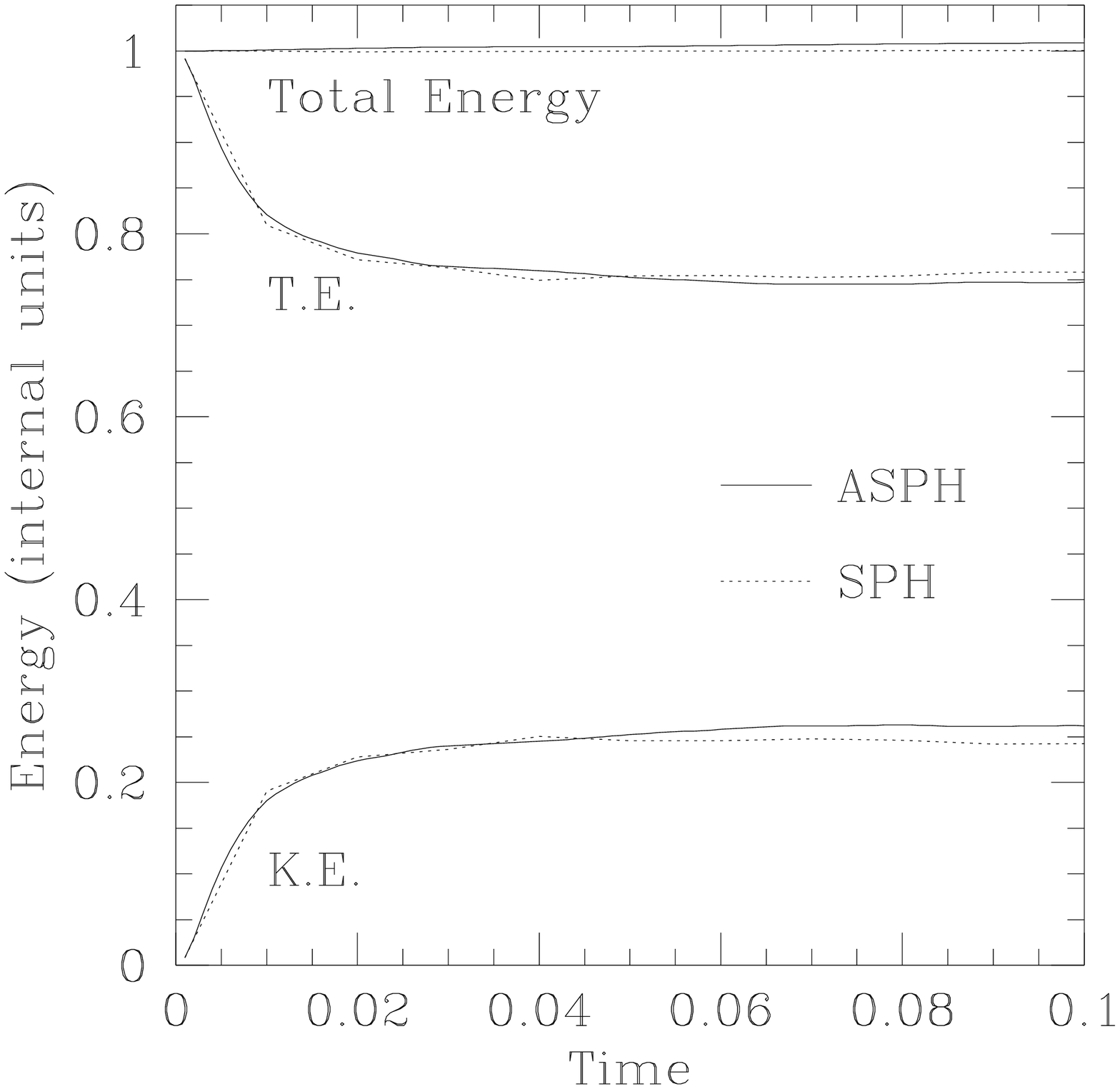}
\caption{Evolution of the global energies (kinetic, thermal, and total)
for the 3D Sedov blastwave simulations.}
\label{3DSed_cons.fig}
\end{figure}
Figure \ref{3DSed_rhoprofs.fig} presents a time sequence of radial profiles
for these simulations.  Each point represents an average of 50 particles in
a radial bin.  The solid lines show the Sedov similarity solution
predictions.  The lower resolution of these simulations in comparison with
the 2D simulations is quite evident, as both simulations fall well short
of the predicted density jump $\rho_2/\rho_1 = 6$.  If as in the 2D case
we examine the typical compression of an ASPH kernel in the shock front, we
find that typically the ratio of the shortest to longest axis is $h_3/h_1
\sim 0.4$, as compared with the theoretical prediction $6^{-1} \sim 0.167$.
Nevertheless, it is clear that ASPH is still able to resolve the density
jump more effectively than SPH, exceeding SPH's maximum density by $50\%$ at
$t=0.1$ and localizing the shock more precisely.  Figure
\ref{3DSed_profs010.fig} plots the radial profiles of the density,
pressure, and velocity at $t=0.1$ of both simulations against the
analytical solutions.  It is evident that that ASPH is better able to
reproduce the analytical profiles in all these quantities, though again the
lack of resolution due to paucity of particles hurts both simulations.
Finally, Figure \ref{3DSed_cons.fig} shows the evolution of the global
energies.  In this case ASPH conserves energy to better than $\Delta E/E
\sim 1\%$, while SPH conserves to $\Delta E/E \sim 0.1\%$.

\subsubsection{Riemann Shocktube Test}
\label{Shocktube.sec}
The Riemann shocktube is a well-known test problem to which SPH codes are
traditionally subjected (Monaghan \& Gingold 1983; HK89; Rasio \& Shapiro
1991).  This problem is also examined in Paper I, but only with a 1D code.
We reexamine it here since the ASPH formalism differs from that of Paper I,
and because the treatment here is 2D.  As this is such a well-known
problem, we will only briefly outline the setup.  The initial conditions
consist of placing two regions of gas, differing in density and pressure,
adjacent to one another across an interface.  This discontinuity will
collapse in a quasi-analytically known manner.  Because this problem
requires a fairly large number of nodes along the collapse dimension to get
reasonable results, we have chosen to initialize it in a slightly different
manner than previous problems.  In this case, our computational volume is a
rectangular strip (of aspect ratio 4:1) with the long axis aligned with the
direction of collapse.  This allows us to have more of our nodes in the
direction of interest, while still maintaining a 2D simulation.  We
should also note that since this simulation is periodic, we in fact have
two shocktubes evolving in our volume.  Therefore, even though we have 150
rows of particles along our $x$ dimension, we really only have 75 rows per
shocktube.  All \Gt\ tensors are initialized as circular SPH \Gt\ tensors
appropriate for the local density.  Table \ref{Shocktube.tab} presents the
major simulation parameters for this test.

\begin{figure}[htbp]
\plottwo{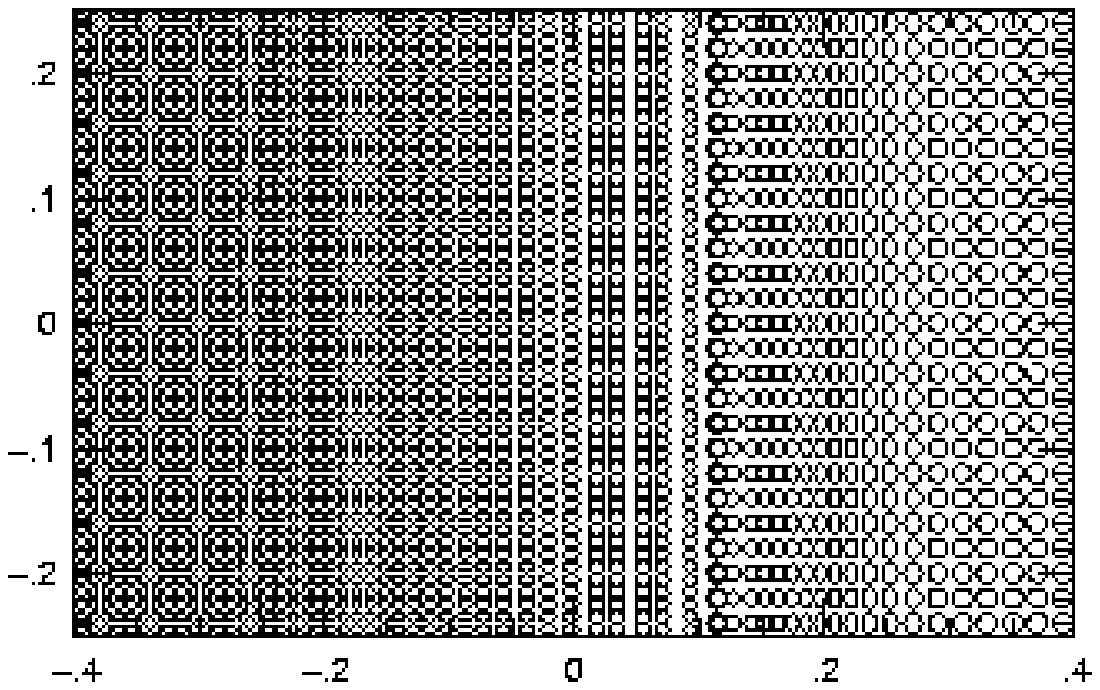}{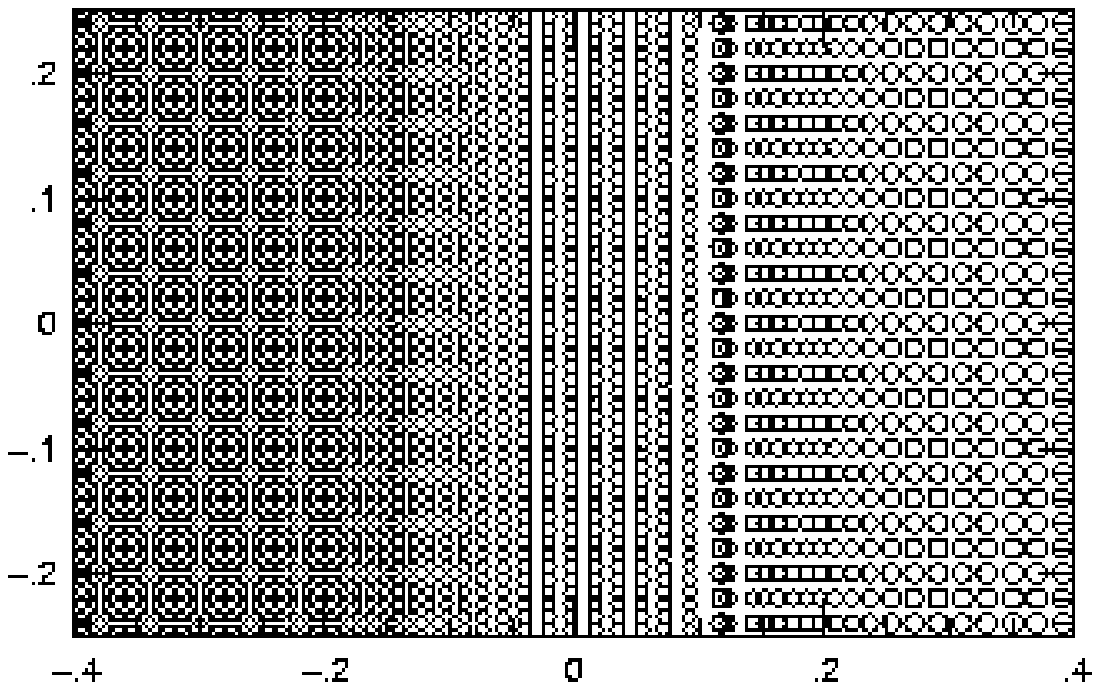}
\caption{Kernel plots for 2D Riemann shocktube simulations at $t=0.15$.
Shown are ASPH (left panel) and SPH (right panel) simulations.}
\label{Shocktube_ker.fig}
\end{figure}

\begin{figure}[htbp]
\plotone{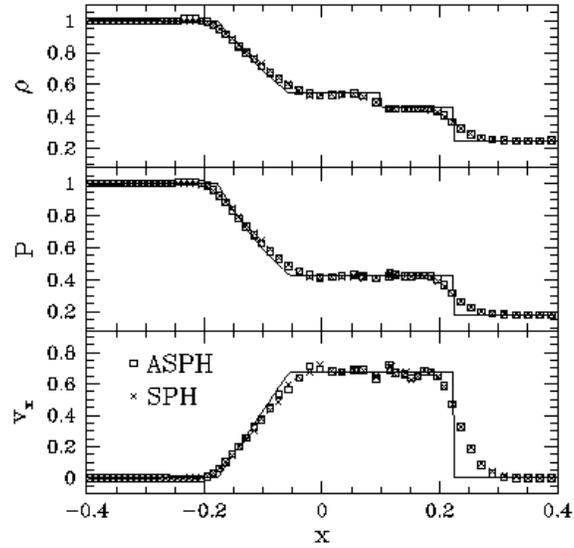}
\caption{Profiles of mass density $\rho(x)$, pressure $P(x)$, and
velocity $v_x(x)$ along the $x$ axis for the ASPH (squares) and SPH
(crosses) 2D Riemann shocktube simulations at $t = 0.15$.  Solid lines
show the analytical profiles.}
\label{Shocktube_state.fig}
\end{figure}
Figure \ref{Shocktube_ker.fig} shows kernel plots for a region including
one of the shockfronts at $t=0.15$, the end state of the simulations.
Note that the ASPH \Gt\ tensors are only slightly deformed by the
relatively gentle density evolution of this problem, with axis ratios in
the range $h_2/h_1 \in [0.63, 1.0]$, $\Interp{h_2/h_1} = 0.87$.  Figure
\ref{Shocktube_state.fig} shows the $x$ profiles for the mass density
$\rho$, pressure $P$, and velocity $v_x$ for the regions shown in Figure
\ref{Shocktube_ker.fig}, along with the analytic solutions for these
quantities.  We plot all nodes in this region, so each point plotted
actually represents many overlapping points as we project in the $y$
direction through the system.  There is little evidence for symmetry
breaking, as the points overlap nicely.  ASPH and SPH appear to solve this
problem equivalently, with little distinction between the two.  This is not
surprising, as the density evolution is fairly gentle and therefore there
is little need for large dynamic range in the resolution scale.  

\subsubsection{Two Interacting Blast Waves}
\label{DoubleBlast.sec}
We will now discuss a test problem popularized by Woodward \& Colella
(1984), which has become a fairly standard hydrodynamic test.  This problem
involves multiple interactions of strong shocks, rarefaction waves, and
contact discontinuities, and is in general a rather stringent test.  The
initial conditions are discussed in detail in Woodward (1982) and Woodward
\& Colella (1984), so we only summarize here.  A $\gamma = 1.4$ gas is
initialized at unit density ($\rho = 1$) in a unit length.  Three regions
of differing pressure are established according to
\beq
  \label{DBP.eq}
  P(x) = \left\{ \begin{array}{l@{\quad}l}
    1000, & x \in [0,0.1]; \\
    0.01, & x \in [0.1,0.9]; \\
    100, & x \in [0.9,1].
  \end{array} \right.
\eeq
The system is evolved within reflecting boundary conditions.  Note that
these initial conditions result in two strong blast-waves (of Mach numbers
$\sim 170$ and $\sim 51$ respectively) propagating toward one another
through the low pressure gas, as well as two rarefaction waves moving
backwards through the hot gas.  In addition to Woodward \& Colella's
comparisons, this problem has also been investigated under 1D SPH by
Steinmetz \& M\"{u}ller (1993) and in 2D under the free Lagrangian method
by Whitehurst (1995).  In order to mimic the reflecting boundary
conditions, we use a periodic volume with the initial conditions of
equation (\ref{DBP.eq}) mirrored about $x=0$.  Additionally, as in the
previous Riemann shocktube this problem requires a fair degree of linear
resolution in the direction of evolution, so we evolve the system in a
rectangular strip of gas.  Even using an exceedingly thin strip of gas we
still only have marginal resolution along the direction of interest (256
nodes in the cases shown here).  Table \ref{DB.tab} summarizes our major
simulation parameters for this test.

\begin{figure}[htbp]
\plottwo{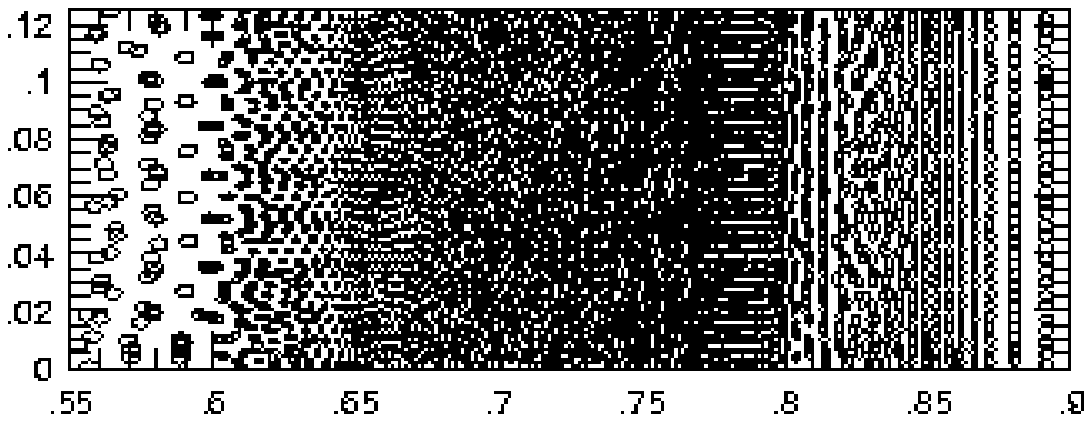}{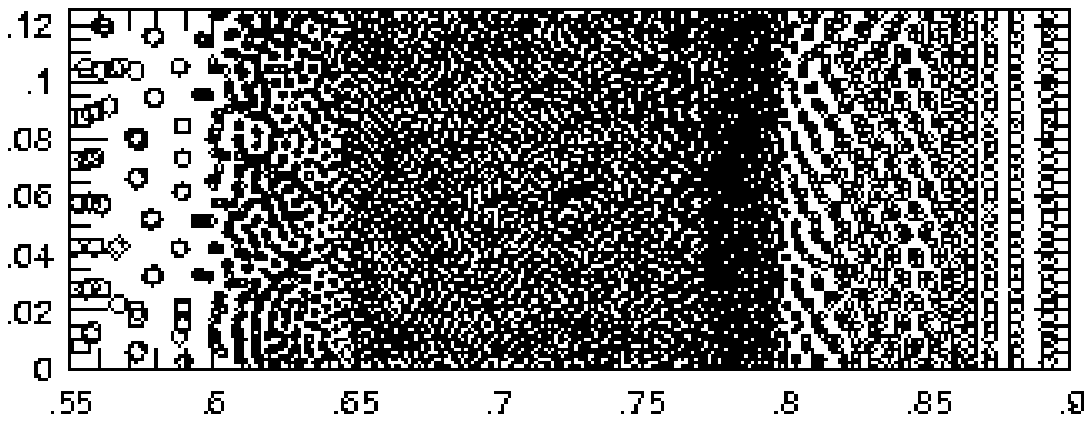}
\caption{Kernel plots for the 2D Woodward double blastwave simulations
at $t=0.038$ in the sub-region ($x \in [0.55,0.9], y \in [0,0.125]$).
Shown are ASPH (left panel) and SPH (right panel).}
\label{DB_ker.fig}
\end{figure}

\begin{figure}[htbp]
\figurenum{22a}
\epsscale{0.8}
\plotone{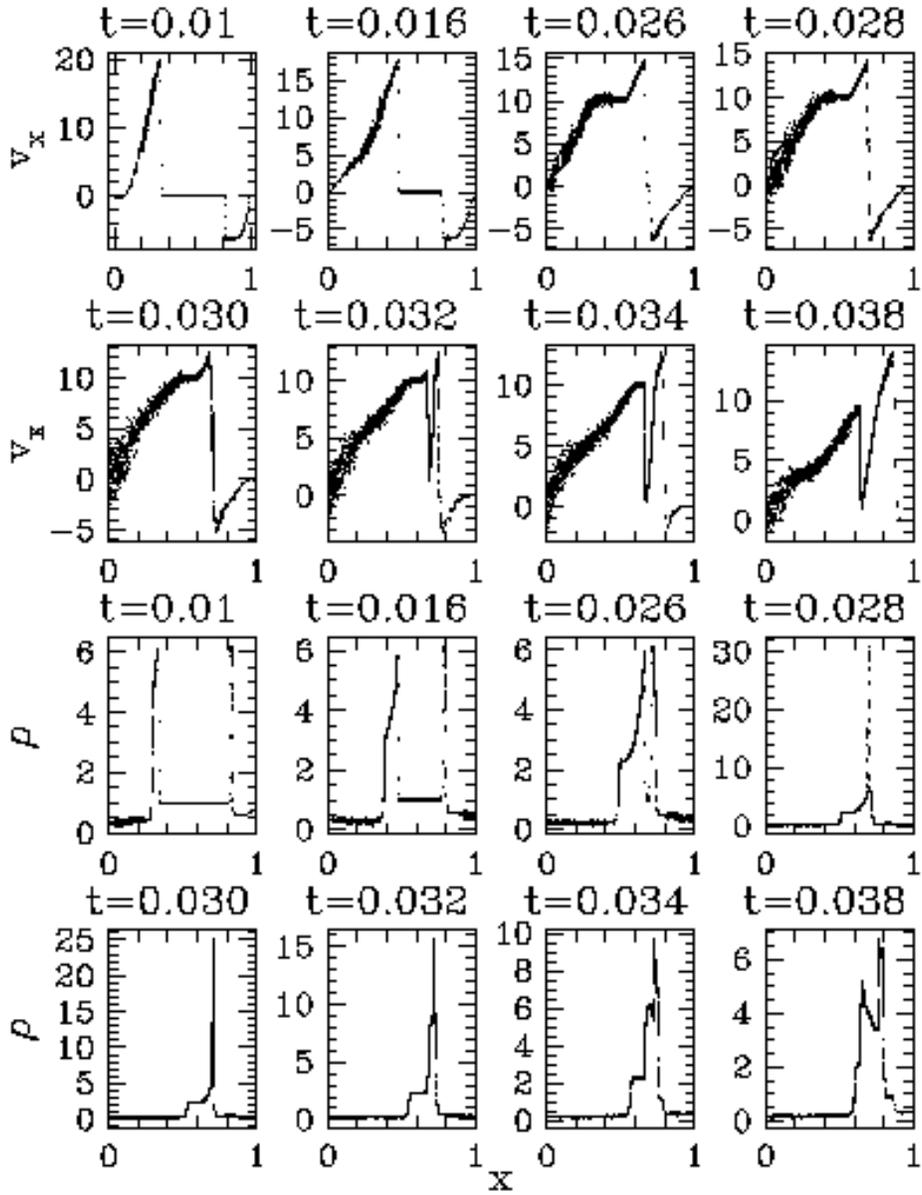}
\caption{Velocity ($v_x$) and mass density ($\rho$) profiles plotted
against $x$ for the ASPH simulation of the 2D Woodward double blastwave
problem.  Shown are profiles for times $t$ = 0.01, 0.016, 0.026, 0.028,
0.030, 0.032, 0.034, and 0.038, which are chosen for direct comparison to
the corresponding figures in Woodward \& Colella (1984) and Steinmetz \&
M\"{u}ller (1993).}
\label{DB_state.fig}
\end{figure}

\begin{figure}[htbp]
\figurenum{22b}
\epsscale{0.8}
\plotone{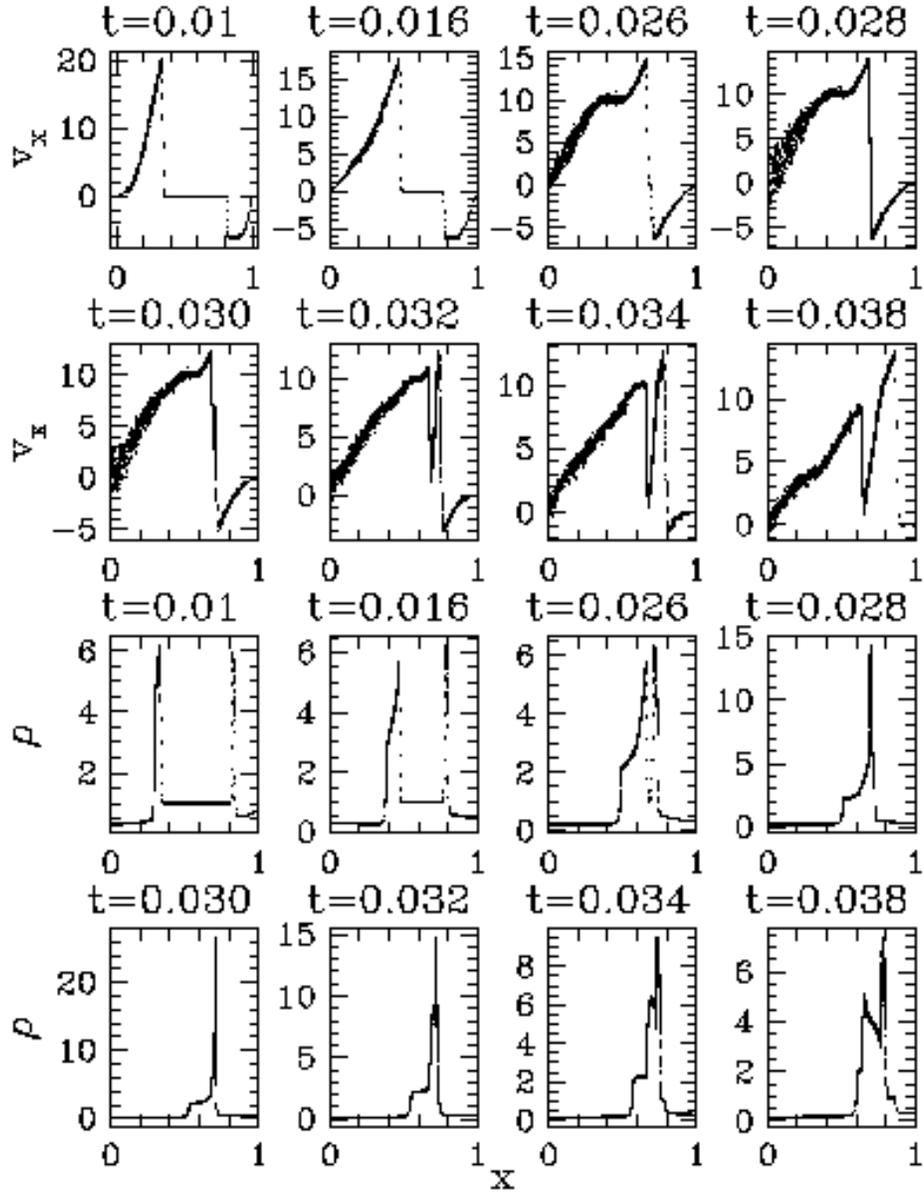}
\caption{Velocity ($v_x$) and mass density ($\rho$) profiles plotted
against $x$ for the SPH simulation of the 2D Woodward double blastwave
problem.  Panels selected and arranged as in Figure
\protect\ref{DB_state.fig}.}
\end{figure}
\stepcounter{figure}

\begin{figure}[htbp]
\epsscale{0.5}
\plotone{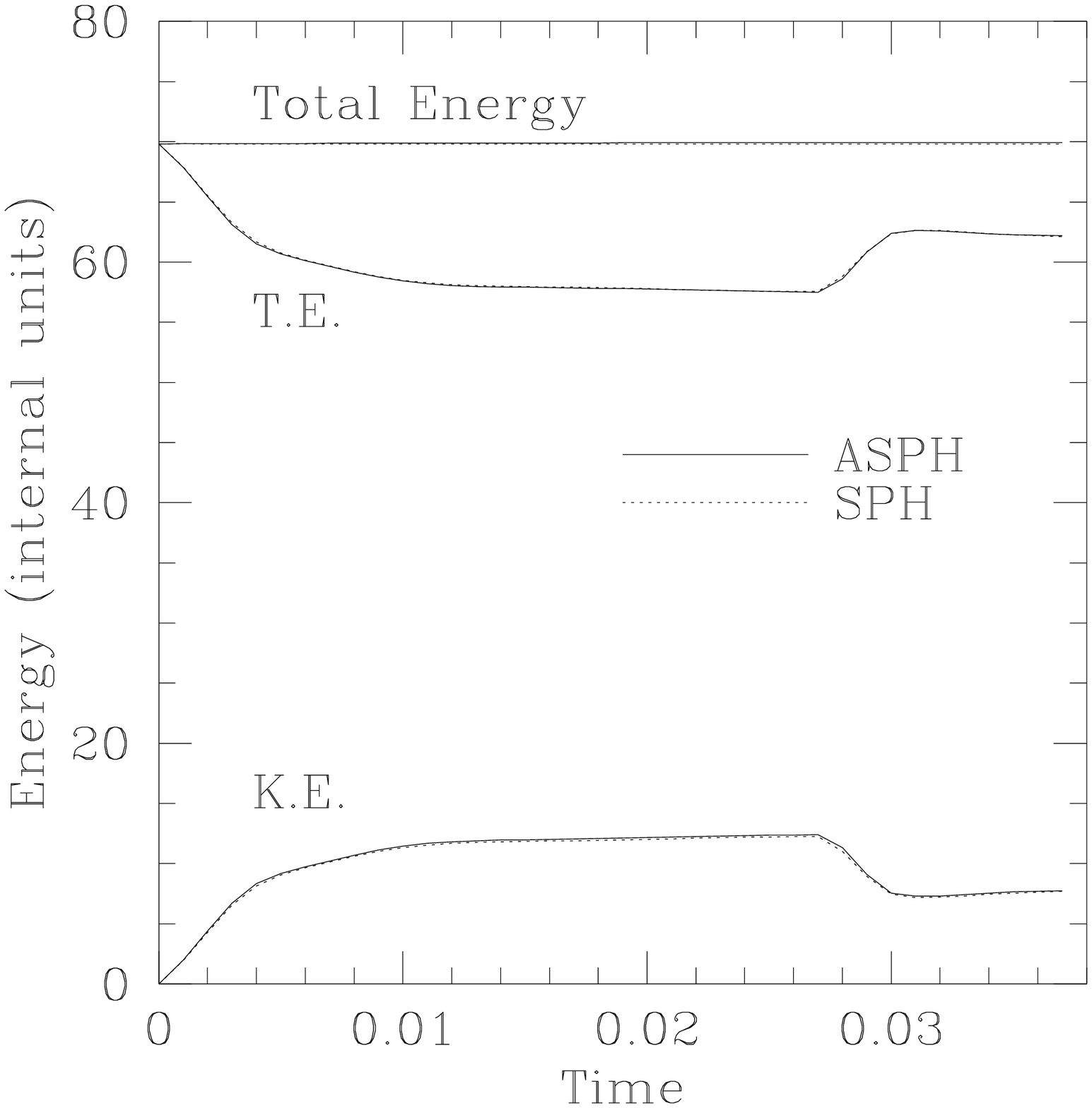}
\caption{Evolution of the global energies (kinetic, thermal, and total)
for the 2D Woodward double blastwave simulations.}
\label{DB_cons.fig}
\end{figure}
Figure \ref{DB_ker.fig} shows kernel plots for the region $(x,y) \in
([0.55,0.9], [0,0.125])$ at the end state $t=0.038$ of these simulations.
In accordance with the 1D nature of this problem, the ASPH \Gt\ tensors
adapt primarily in the $x$ direction, and remain marginally better ordered
(closer to the initial lattice) than the SPH case.  Figures
\ref{DB_state.fig}a \& b show the velocity and mass density profiles for
these simulations at the same times slices shown in both Woodward \&
Colella (1984) and Steinmetz \& M\"{u}ller (1993), in order to facilitate
comparison.  Both ASPH and SPH represent this problem reasonably well,
though note that at time $t=0.028$ (just as the two shockfronts collide)
the ASPH simulation resolves a much higher density spike (of order
$\rho_{peak} \sim 30$), as compared with SPH (which finds $\rho_{peak} \sim
14$).  The predicted value for this density jump is 24, so ASPH slightly
overshoots.  The higher ASPH value is a direct result of ASPH's ability to
achieve superior spatial resolution in this small region.  We also note
that in the regions where there is evident disorder in these profiles
(particularly behind the rarefaction wave moving through the initially
$P=1000$ gas) the system is hitting the maximum smoothing scale we can
allow due to the narrow width of the simulation, so some fraction of this
scatter is in fact likely an artifact.  In general, as found by M\"{u}ller
\& Steinmetz (1993), both SPH and ASPH seem capable of solving this
problem, despite the commonly held belief that SPH cannot deal with strong
shock phenomena.

\subsection{2D Test Cases}
\label{2dtests.sec}
\begin{figure}[htbp]
\plotone{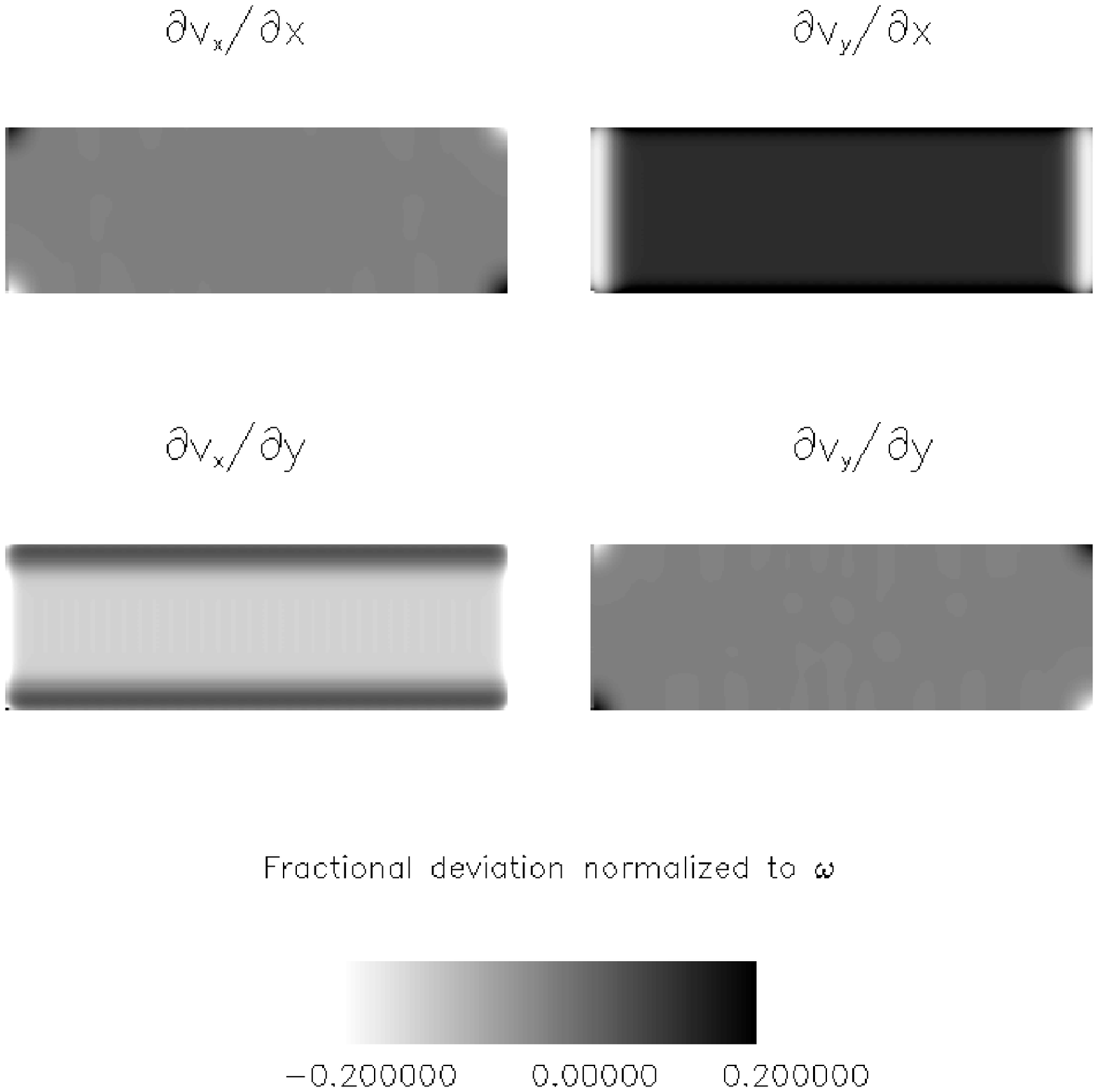}
\caption{Gray-scale images of the SPH estimated components
$\Interp{\partial v_\alpha/\partial x_\beta}$ for a bar in solid-body
rotation with angular velocity $\omega$.  These quantities formally should
be constant throughout the bar.}
\label{dvdx_tv.fig}
\end{figure}
Each of the previous test cases represent idealized situations with no
``handedness'' to the problem (\ie\ they are physically 1D).
Additionally, due to the physical symmetry of these 1D problems, the
\Gt\ tensors are able to align themselves such that all significant
interactions take place along one of their major axes.  In this section
we will investigate problems which break these symmetries and are truly
2D, concentrating on rotational test cases in order to investigate the
question of angular momentum conservation under ASPH (discussed in \S
\ref{SHASPH.sec}).  Specifically we will examine a gas disk subjected to
an external Keplerian potential (\S \ref{KepDisk.sec}) and a
self-gravitating disk undergoing radial collapse with rotation (\S
\ref{ColDisk.sec}).

Before we begin this discussion, though, we should address why we do not
examine an obvious test case for angular momentum.  The simplest example of
a problem with a well-defined angular momentum is a rigidly rotating bar,
something with an equation of state like a solid so that it should simply
maintain solid body rotation.  There is, however, an unfortunate flaw with
such a deceptively simple system, which is that it possesses an edge.
(A)SPH is derived assuming that there are no distinct surfaces or edges
present in a modeled system, but rather that all quantities vary smoothly.
(A)SPH sampled averages, such as defined by equations
(\ref{Wint.eq})-(\ref{Norm.eq}), will be in error near an edge.  Of particular
concern for ASPH is the fact that the estimates for the spatial gradients
of the velocity field $\Interp{\partial v_\alpha/\partial x_\beta}$
(eq. [\ref{sigker.eq}]) will be in error near any edges, which ensures
that the evolution of the \Gt\ tensor will also be incorrect.  Figure
\ref{dvdx_tv.fig} presents images of the SPH estimates of the elements of
$\Interp{\partial v_\alpha/\partial x_\beta}$ for a 2D bar rigorously in
solid body rotation.  If these estimates were correct, then the images
should be uniform throughout.  This figure clearly shows that, as expected,
the SPH estimates are in error near the edges, by as much as $50\%$.  So long
as this is the case, the \Gt\ tensors near the edges of the system will not
be evolved correctly, resulting in a false torque about the surface, and
thereby violating angular momentum conservation.  We have in fact found
this to be the case for ASPH simulations of such systems, and the magnitude
of the angular momentum violation scales as the number of surface nodes in
the system.  It is interesting to note that SPH cannot deal with surfaces
any better than ASPH (see, \eg, Monaghan 1992), but because by construction
SPH cannot violate global angular momentum conservation, such tests are
meaningless.  This is simply an example that global conservation alone does
not guarantee a successful simulation.  Fisher \& Owen (1997) are currently
investigating a modified technique based upon SPH designed to account and
correct for such problems.

In order to investigate the angular momentum issue, it is necessary to
study a non-periodic system, which requires the density fall to zero at
some point.  For this reason, we have investigated rotating disk systems in
this paper, in an effort to have the density fall off smoothly to an edge.
The fact that there is still an edge is somewhat worrisome, but by having
the density fall off smoothly we hope to moderate this problem.
Additionally, these sort of systems are more representative of the sorts of
rotating problems we are likely to encounter in cosmological structure
formation scenarios, such as if we were to model a disk galaxy.

\subsubsection{Pseudo-Keplerian Disks}
\label{KepDisk.sec}
Our first rotational test is a type of Keplerian disk, modified to include
pressure support.  Since this is not a standard test case, we will describe
the initial conditions in some detail.  A gas disk is created in rotational
and pressure balance with a fixed, external gravitational potential
produced by a theoretical point mass at the disk's center.  The initial
radial density and pressure profiles of the disk are chosen arbitrarily.
The disk is not self-gravitating.  In order to make this problem
computationally feasible, it is necessary to use a softened potential for
the point mass
\beq
  \Phi(r) = -\frac{G M}{(r^2 + r_c^2)^{1/2}} \quad \Rightarrow \quad 
  {\bf g}(r) = -\frac{G M {\bf r}}{(r^2 + r_c^2)^{3/2}},
\eeq
where $M$ is the mass of the gravitating point mass and $r_c$ is the
softening core radius.  Note that the nodes in this gravitational potential
are treated as point masses in 3D, not as infinite rods (as was the case
for the Zel'dovich pancake simulations).  We are now simulating gas confined
to a plane.

The forms we choose for the initial radial mass density and pressure
profiles are
\beq
  \label{Kdiskrho.eq}
  \rho(r) = \rho_0 \lp 1 - \frac{r^2}{r_d^2} \rp,
\eeq
\beq
  P(r) = K \rho^2(r) = K \rho_0^2 \lp 1 - \frac{r^2}{r_d^2} \rp^2,
\eeq
where $\rho_0$ is the central gas density, $r_d$ is the outer radius of
the gas disk, and $K$ is a constant which sets the amount of pressure
support.  Based on these choices, the balancing rotational velocity of
the gas is
\beq
  \label{Kdiskvt.eq}
  v_t^2(r) = \frac{G M r^2}{(r^2 + r_c^2)^{3/2}} - 4 K \rho_0 
             \frac{r^2}{r_d^2},
\eeq
where $v_t(r)$ is the supporting circular (or tangential) velocity.
Equation (\ref{Kdiskvt.eq}) limits the possible range of the pressure
constant $K$ to
\beq
  K \in \left[ 0, \; \frac{G M r_d^2}{4 \rho_0 (r_c^2 + r_d^2)^{3/2}} \right].
\eeq
We generically choose the largest possible value for $K$ in order to
maximize the amount of pressure support.  Note that maximizing $K$ in
this fashion sets $v_t(r_d) = 0$.

There is no physical motivation for choosing this problem to study.
Rather, this test is designed solely for the purpose of testing ASPH.
Since there is no preferred resolution direction, ASPH has no real
advantage under this problem.  Therefore ASPH's extra degrees of freedom,
rather than offering a real advantage, can only lead to trouble.  For $r >
r_c$, this system possesses a strong radial velocity shear, which will tend
to elongate the ASPH \Gt\ tensors into the flow.  This is a shear field
that cannot be represented under the first-order treatment our \Gt\
evolution derivation is based upon.  The problem can be understood by
considering the evolution of a hypothetical fluid element in such a
Keplerian potential: for $r > r_c$ a true fluid element will distend and
eventually be infinitely sheared around an arc of the disk.  This creates a
situation which our first-order derivation of \Gt\ and its evolution cannot
ideally adapt to (an ellipse cannot be distorted to follow such a curving
arc and remain an ellipse).  Therefore our evolution equations for the \Gt\
tensor field must fail at some level for this case.  Additionally, the
imposed point mass potential is very strongly centralized, whereas our
initial density profile is flat-topped.  This situation is unstable, and
although the system is born in radial balance, we can expect as it evolves
it will rapidly deviate from the initial conditions.  These effects will
tend to force the ASPH \Gt\ tensor field to evolve under conditions it
cannot simply, nor ideally, adapt to.  Finally, our use of an external
central force as the binding global potential will rigorously ensure
angular momentum conservation for the gravitational interactions.  Any
angular momentum errors incurred are therefore due to either numerical
problems or ASPH.

In order to initialize the (A)SPH node positions, we first select candidate
positions quasi-randomly based on the Sobol sequence.  These potential
positions are subjected to a Monte-Carlo acceptance/rejection scheme, with
a probability distribution appropriate to match the density profile of
equation (\ref{Kdiskrho.eq}).  Once a node position is selected, the
specific thermal energy is uniquely identified by the required pressure
profile in combination with the theoretical density.  The \Gt\ tensors are
initialized as SPH \Gt\ tensors with determinants $|\Gt|$ scaled
appropriately for the theoretical local density.  The system is evolved
until it is seen to settle into an apparently equilibrium distribution.
Finally, in order to obtain meaningful measurements of the angular
momentum, these simulations are evolved in a non-periodic computational
volume.  Table \ref{Kep.tab} summarizes our input numerical parameters for
these simulations.

\begin{figure}[htbp]
\plottwo{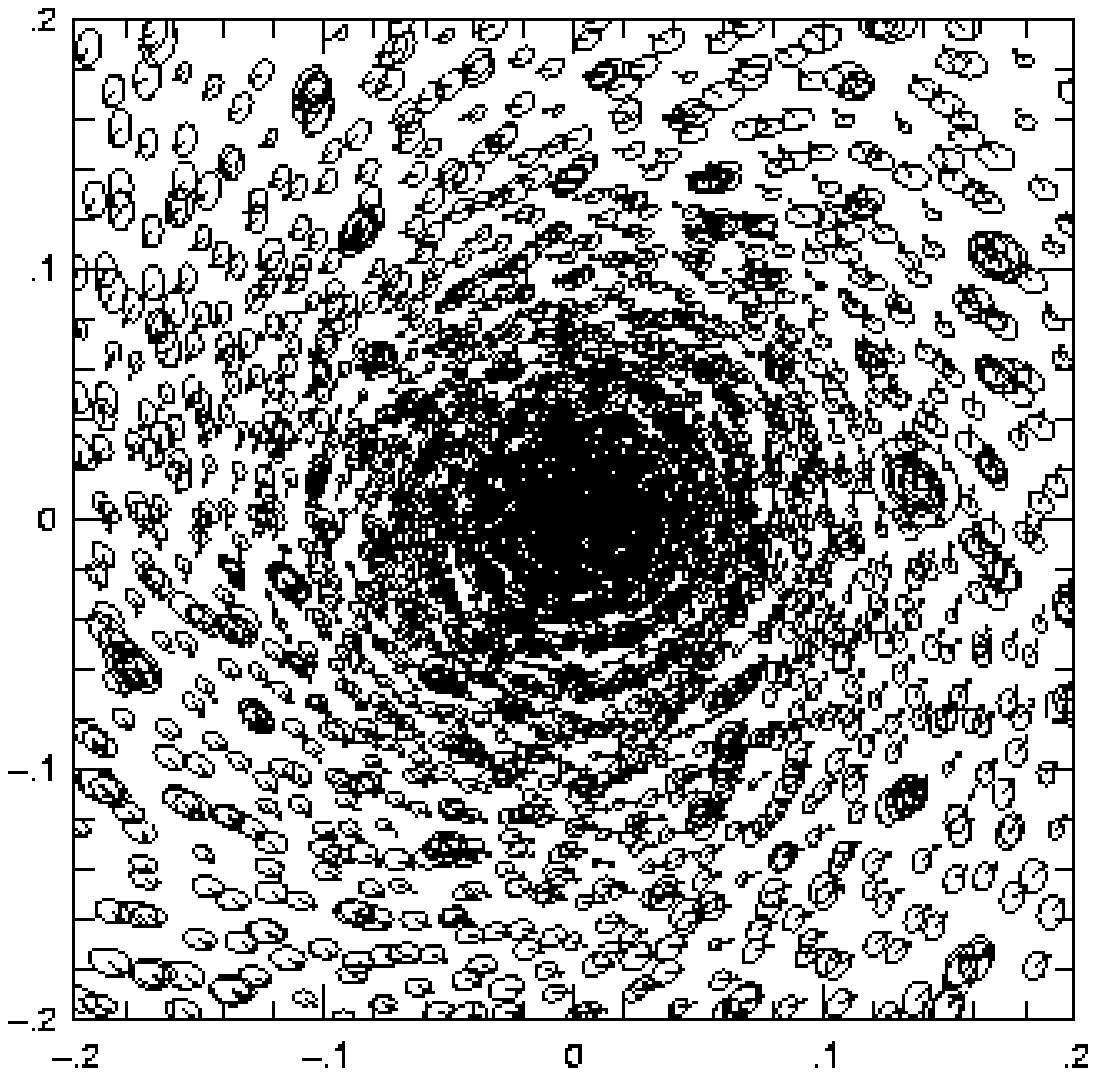}{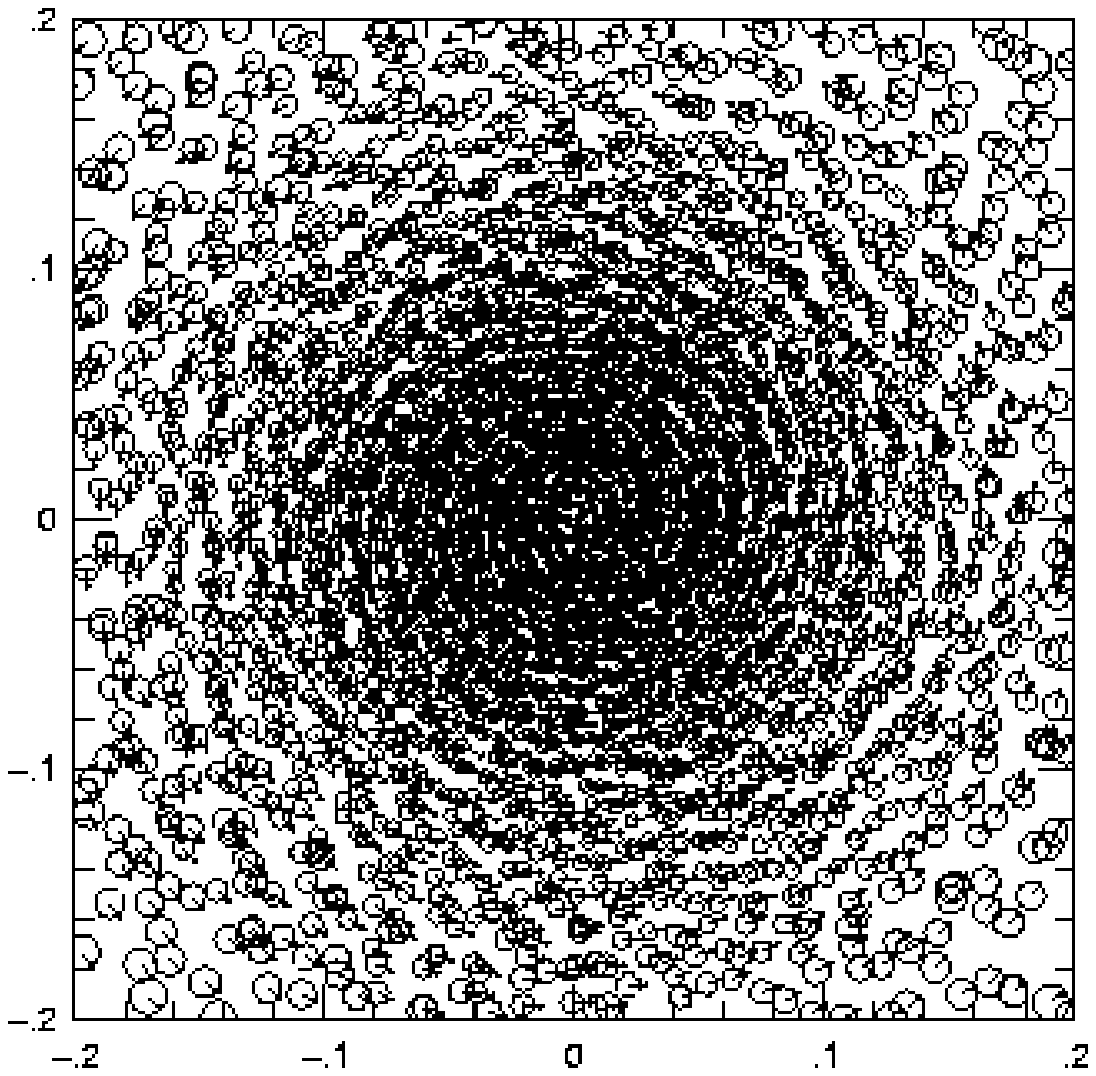}
\caption{Kernel plots for the 2D Pseudo-Keplerian Disk simulations at
$t=3.0$ for the cases of ASPH (left panel) and SPH (right panel).  Arrows
are drawn for each node indicating the direction and magnitude of their
velocity.}
\label{Kep_ker.fig}
\end{figure}
Figure \ref{Kep_ker.fig} shows kernel plots of the central core regions for
both the ASPH and SPH simulations at $t=3.0$.  In order to emphasize the
steady rotation field, we also plot arrows for each node indicating both
the direction and relative magnitude of the velocity.  Though grossly
similar, there are interesting differences in these two simulations.  It
appears that the core of the ASPH simulation is relatively denser or more
collapsed as compared with SPH.  The ASPH kernels seem to be elongated into
the direction of rotation, and there is an apparent trend for the ASPH
kernels' ellipticity to increase with radius.  The geometry of the \Gt\
tensors ranges from $h_2/h_1 \in [0.5, 1.0]$, $\Interp{h_2/h_1} = 0.75$.
This behaviour is understandable in terms of our smoothing algorithm.  At
all radii the rotational shearing field tends to elongate the \Gt\ tensors
into the direction of rotation.  At ``small'' radii ($r/h \sim 1$), the
requirement that each node sample several neighbors implies that each sees
neighboring nodes which are significantly further around the arc of
rotation, and therefore will be trying to elongate their \Gt\ tensors in
noticeably different directions.  When the \Gt\ tensor field is smoothed,
averaging over many different $\Gt_i$'s elongated in different directions
will result in an average round shape.  As we move out to larger radii
($r/h \gg 1$), this averaging process will progressively sample more and
more ``local'' conditions.  In this case, each ASPH node sees itself
embedded within a coherently shearing field, with neighbors elongating in a
similar manner.  Averaging in this case retains the elliptical shape,
though limits how elliptical each element can become.  This sort of
behaviour is desirable.  When an ASPH node is in a region where on scales
of $h$ there are conflicting signals dictating the evolution of \Gt, the
``safest'' choice is for \Gt\ to adopt a round shape, emulating SPH.  This
reflects the fact that the evolution of the \Gt\ tensor is based upon
approximating the local velocity field by the first-order argument
${\bf v}({\bf r} + d{\bf r}) \approx {\bf v}({\bf r}) + \sigt ~d{\bf r}$.
If this approximation is invalid, then the evolution equations for \Gt\
break down.  Only when there is a clear, unambiguous signal dictating the
evolution of \Gt\ should it be allowed to deviate from this ``safe''
choice.

\begin{figure}[htbp]
\plotone{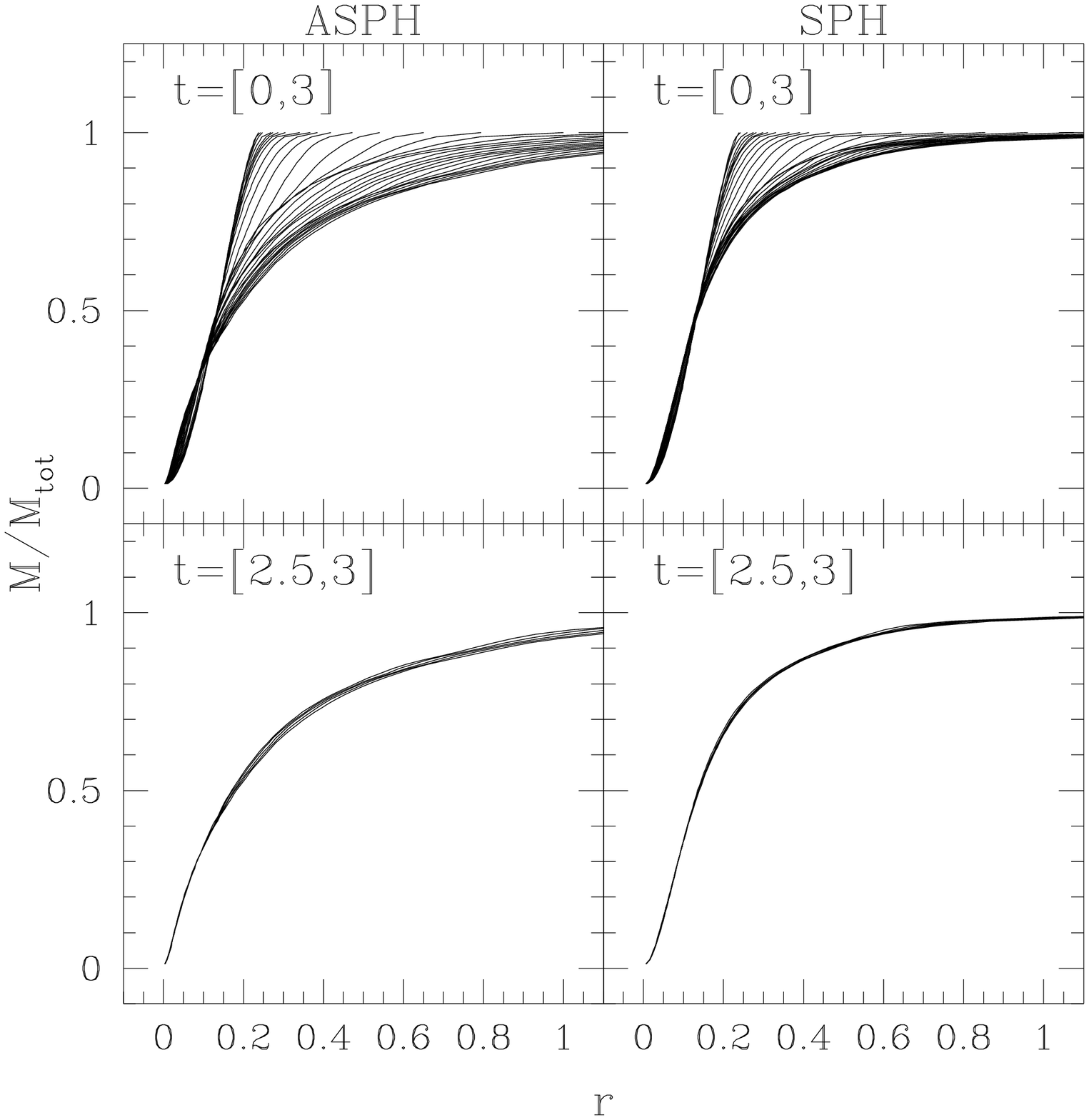}
\caption{Azimuthally averaged radial mass profiles for the 2D
Pseudo-Keplerian Disk simulations.  Each curve represent the fraction of
the mass of the disk contained within the radius $r$ at a particular time.
The top panels show the function $M(r)$ for times varying from the
beginning of the simulation to the end ($t \in [0,3]$).  The bottom panels
show only the last 6 measurements of $M(r)$, in the time range $t \in
[2.5,3]$.}
\label{Kep_Mr.fig}
\end{figure}

\begin{figure}[htbp]
\plotone{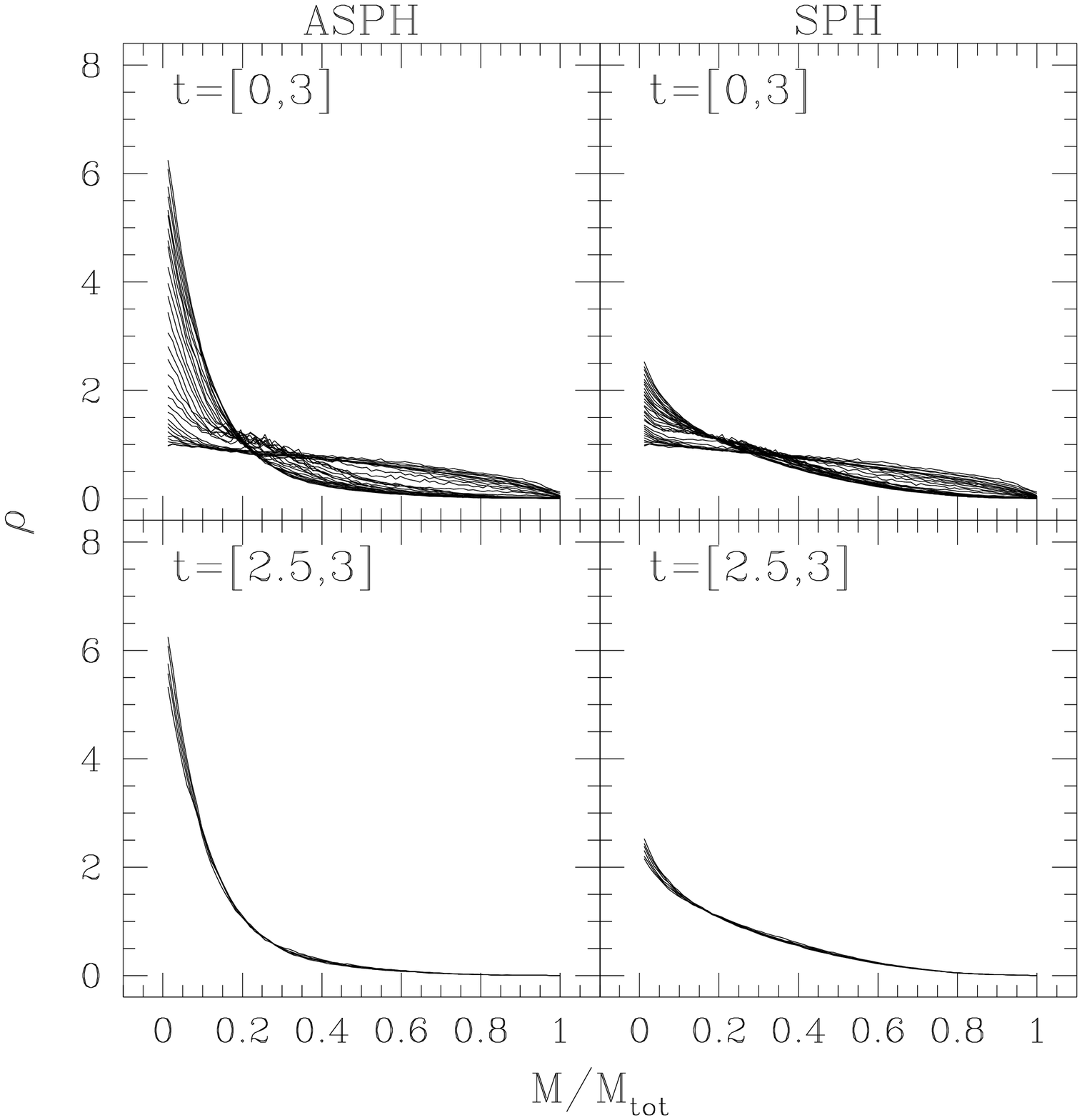}
\caption{Azimuthally averaged radial density profiles for the 2D
Pseudo-Keplerian Disk simulations, expressed as a function of the enclosed
mass fraction $\rho(\protect\Sub{M}{interior})$.  The top panels show the
function $\rho(\protect\Sub{M}{interior})$ for times varying from $t \in [0,3]$,
while the bottom panels only show the last few measurements at times $t \in
[2.5,3]$.}
\label{Kep_rhoM.fig}
\end{figure}
In order to examine these different mass distributions in a more
quantitative fashion, Figures \ref{Kep_Mr.fig} and \ref{Kep_rhoM.fig} show
the distribution of $M(r)$ and $\rho(M)$ respectively, measured radially
outward from the center of the potential.  Note that these two figures are
plotted subtly differently: Figure \ref{Kep_Mr.fig} shows the mass fraction
as a function of radius $M(r)$/\Sub{M}{tot}, whereas Figure
\ref{Kep_rhoM.fig} shows the density distribution as a function of the
radially enclosed mass fraction $\rho(\Sub{M}{interior}/\Sub{M}{tot})$.
Each figure plots many curves: the upper panels show the evolution of
these functions from the beginning to the end of the simulations, while
the lower panels only show the final few measurements.  It is evident that
in the beginning the mass distribution evolves fairly steadily up to a
point, and then settles into an equilibrium (or at least slowly evolving)
state.  In Figure \ref{Kep_Mr.fig} we can see that the total ASPH mass
distribution is slightly more diffuse than SPH, indicating more mass has
been thrown to the outer regions of the disk or become unbound.  Figure
\ref{Kep_rhoM.fig} also shows that the core of the ASPH disk is more dense
than the SPH by a little more than a factor of two, confirming the visual
impression of the kernel plots.

\begin{figure}[htbp]
\plotone{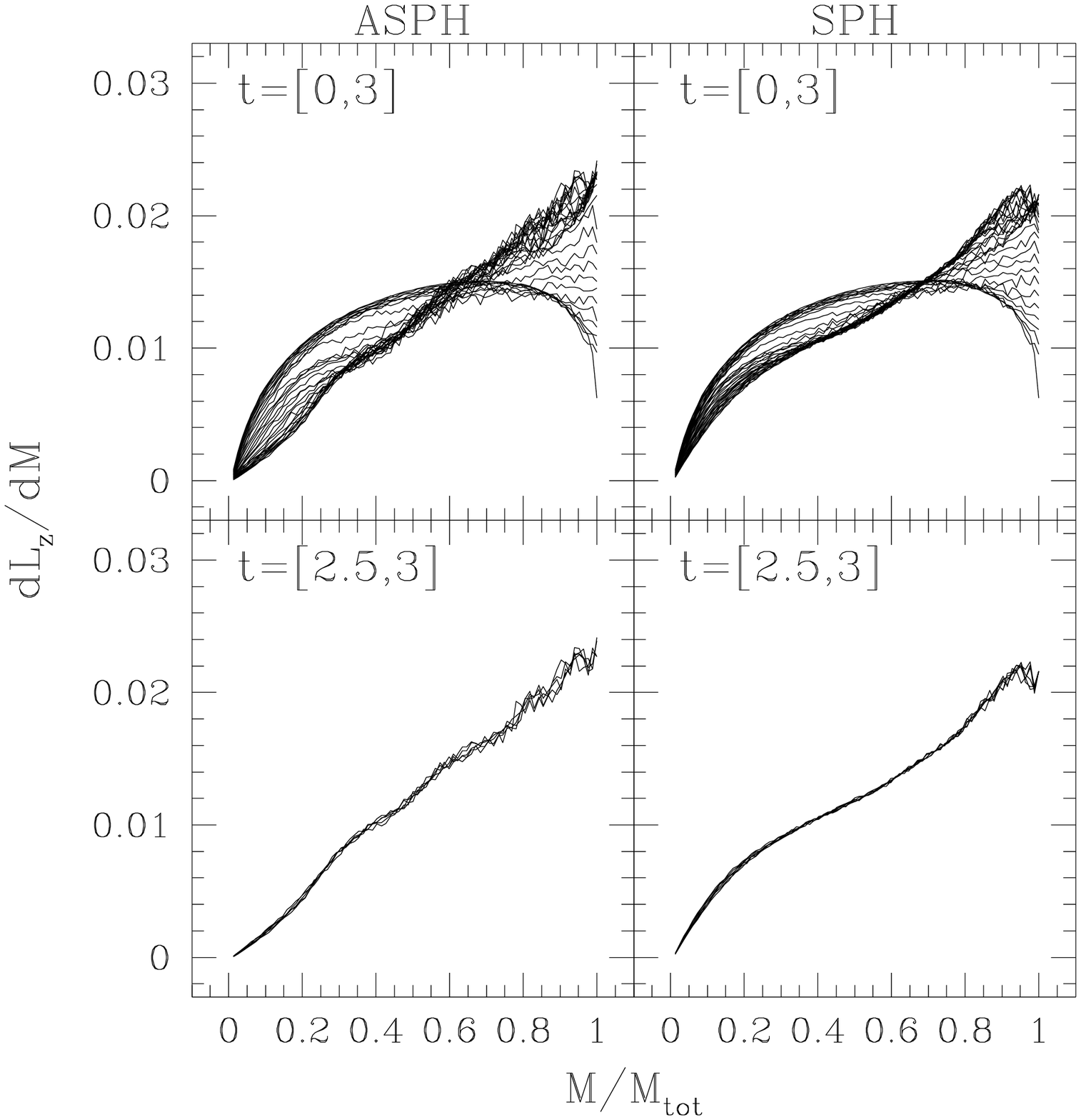}
\caption{Azimuthally averaged radial measurements of angular momentum
$\Delta L_z$ as a function of the enclosed mass for the 2D
Pseudo-Keplerian Disk simulations.  These curves represent a radial
measurement of $dL_z/dM$, where $M$ is interpreted as the radially enclosed
mass.  The top panels show the function $dL_z/dM$ for times varying from $t
\in [0,3]$, while the bottom panels only show the last few measurements at
times $t \in [2.5,3]$.}
\label{Kep_dLzdM.fig}
\end{figure}

\begin{figure}[htbp]
\plotone{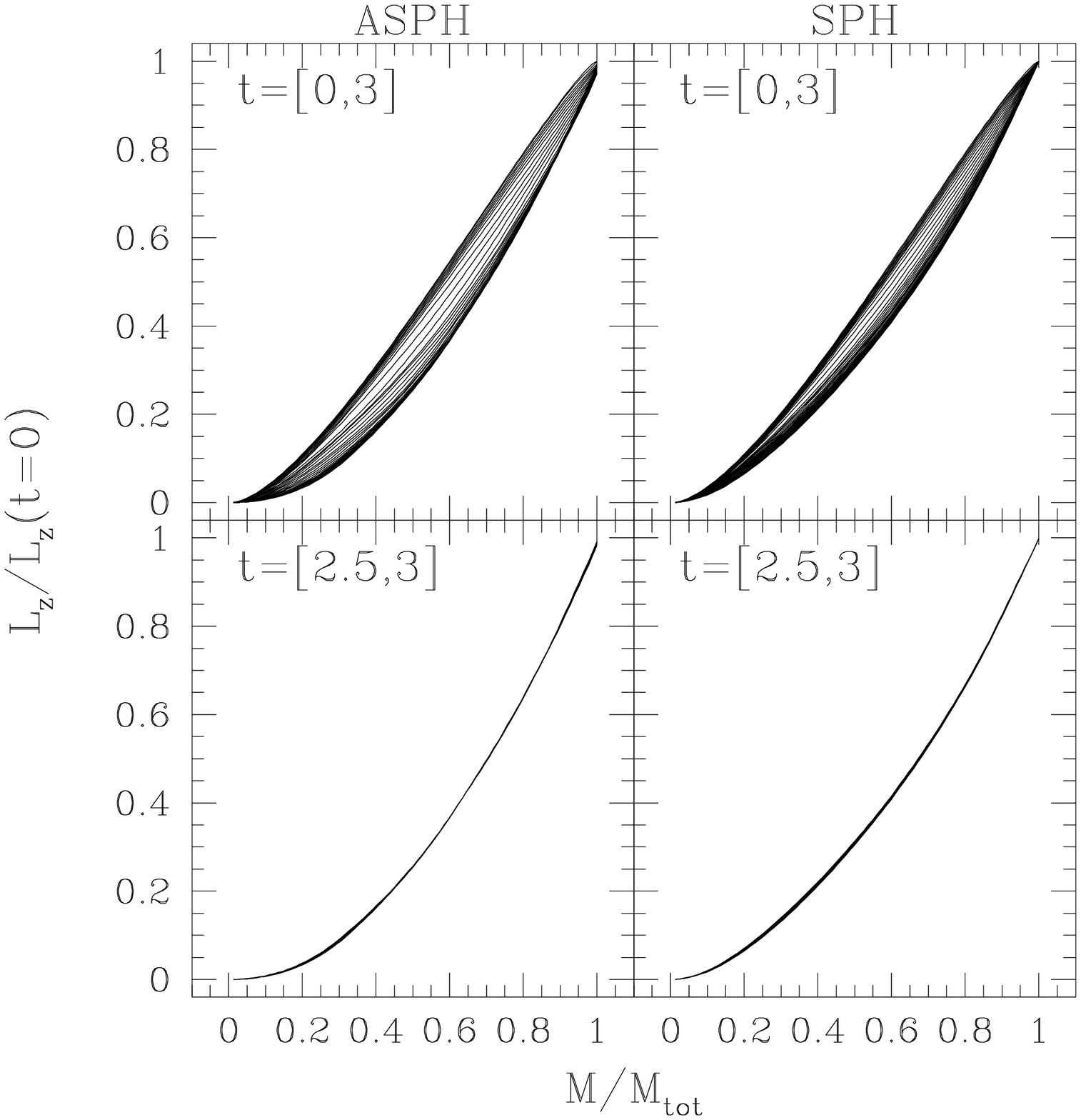}
\caption{The total enclosed angular momentum $L_z$ as a function of the
enclosed mass $M$ for the 2D Pseudo-Keplerian Disk simulations, as
measured radially from the center of the disk.  These curves of $L_z(M)$
represent the integration $\int_0^M (dL_z/dM) dM$ of the function
$dL_z/dM(M)$ presented in Figure \protect\ref{Kep_dLzdM.fig}.}
\label{Kep_LzM.fig}
\end{figure}

\begin{figure}[htbp]
\plottwo{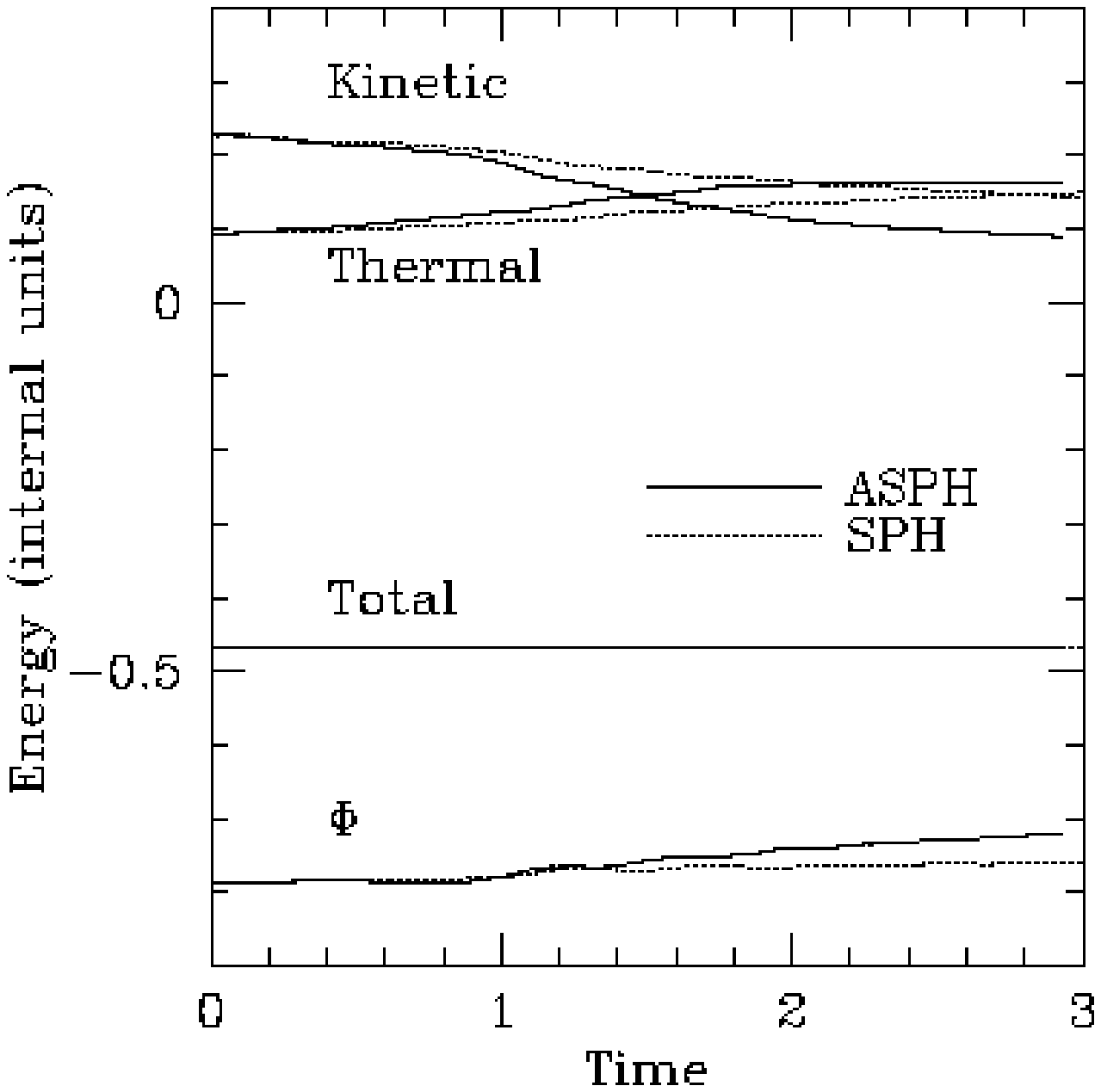}{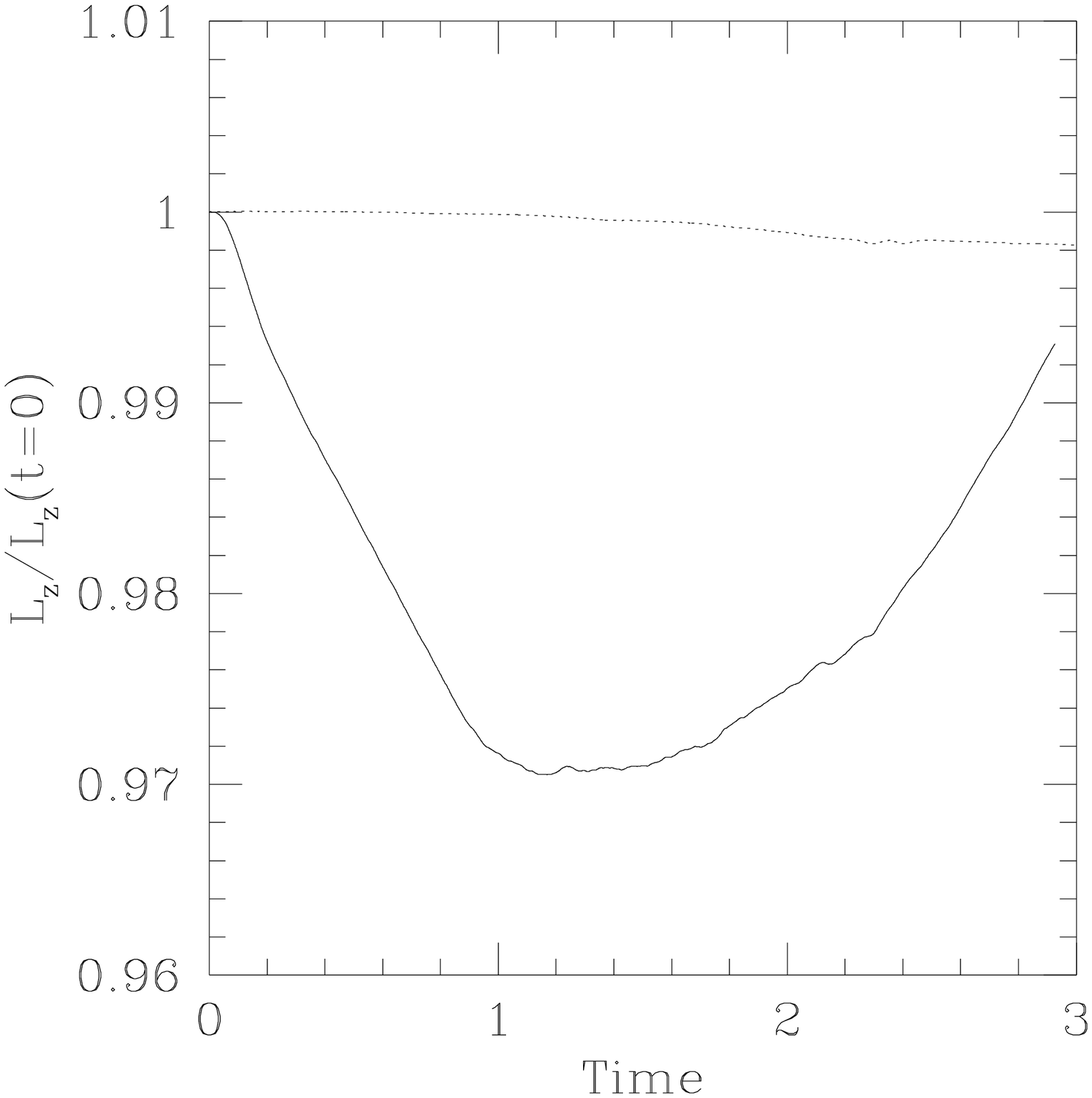}
\caption{Evolution of the global energies (left panel: kinetic, thermal,
potential, and total) and the total angular momentum $L_z$ (right panel)
for the 2D Pseudo-Keplerian Disk simulations.}
\label{Kep_cons.fig}
\end{figure}
Our primary interest in this problem is the evolution of the angular
momentum.  Figure \ref{Kep_dLzdM.fig} shows the function
$dL_z/dM(M)$, which is the amount of angular momentum contained in
shells as a function of the total mass interior to that shell.  Figure
\ref{Kep_LzM.fig} shows the integration of this function $L_z(M) = \int_0^M
(dL_z/dM) dM$, which represents the total angular momentum contained
within the mass fraction $M$.  Since this problem should settle to a
condition of axisymmetry, we expect that the radial angular momentum
distribution should settle to a steady state.  It is evident from Figure
\ref{Kep_dLzdM.fig} that the radial distribution of $dL_z/dM$ initially
changes rather rapidly, and then settles for both simulations to a fairly
steady configuration.  The ASPH and SPH distributions of angular momentum
are somewhat different, however.  ASPH settles into a nearly linear
distribution of $dL_z/dM$ with respect to increasing mass, while SPH
characteristically maintains more angular momentum in the core region.
These differences are expected since ASPH finds a denser core, and
therefore must lose more angular momentum from the core mass.
Nevertheless, both simulations do converge to equilibrium, rotating
configurations.  Figure \ref{Kep_cons.fig} shows the evolution of
the global energies and angular momentum.  SPH of course conserves global
angular momentum nearly exactly (by construction), while ASPH suffers an
overall fluctuation $\Delta L_z/L_z \sim 3\%$.

Since we do not have an analytical expectation for this problem, it is
difficult to know which technique better represents the ``true'' solution.
However, we can state that ASPH appears to conserve the global angular
momentum reasonably, and also does not seem to suffer dramatic local
transport problems (as evidenced by the fact that the local radial angular
momentum distribution establishes an equilibrium state).  It is worth
noting that smoothing the \Gt\ tensors is critical for this problem.  In
all of the previous tests we found that smoothing \Gt\ has relatively
little impact on the outcome.  However, in this case, without smoothing the
\Gt\ tensors rapidly become extremely distorted in the shearing velocity
field, which leads to poor angular momentum conservation (angular momentum
losses $\Delta L_z/L_z \sim 100\%$ in a single rotation or two).  These
extreme distortions in the individual \Gt\ tensors are unphysical, since
they are trying to track neighboring ASPH nodes around the arc of rotation,
which they cannot do.  Smoothing the \Gt\ tensors limits this process and
keeps the distortion manageable, though there may ultimately prove to be a
better solution.

\subsubsection{Collapsing Disk with Rotation}
\label{ColDisk.sec}
The Pseudo-Keplerian disk simulation offers encouraging evidence that ASPH
can solve rotational problems while maintaining reasonable conservation of
the angular momentum.  However, it is not convincing in and of itself.
Although the Pseudo-Keplerian disk does undergo significant radial
evolution, this evolution is relatively gentle in time, and does not
involve strongly distorting the \Gt\ tensors themselves.  This is mainly
due to the fact that the only strong signal to drive the evolution of the
\Gt\ tensors is the shearing rotational field.  We now wish to propose a
more stringent test of angular momentum conservation, in which we can
expect significant evolution of the \Gt\ tensors right down into the core,
at least initially.  In order to accomplish this, we simulate a modified
form of a standard Maclaurin disk (Binney \& Tremaine 1987), which is an
analytically tractable class of rotating hydrodynamic disks.  The
modification we make to this problem is to take away the majority of its
pressure support, making the system unstable to collapse.  When the
timescale for collapse is significantly shorter than the rotational
timescale, the collapse process will dominate the \Gt\ tensor evolution.
Additionally, a rotational problem undergoing rapid and violent collapse
presents a difficult problem both in terms of the transport and
conservation of angular momentum, making this an overall much more
difficult test.

As the Maclaurin disk is a well-documented problem, we will only briefly
outline this setup here, emphasizing our modification of the pressure
term.  The Maclaurin disks represent a class of gas disks which have
radial density, pressure, and circular velocity profiles
\beq
  \label{Cdiskrho.eq}
  \rho(r) = \rho_0 \lp 1 - \frac{r^2}{r_d^2} \rp^{1/2},
\eeq
\beq
  P(r) = f K \rho^3(r),
\eeq
\beq
  v_t(r) = \Omega r,
\eeq
\beq
  \Omega \in [0, \Omega_0], \quad \quad 
  \Omega_0^2 = \frac{\pi^2 G \rho_0}{2 r_d},
\eeq
where $\rho_0$ is the central density, $r_d$ the disk radius, $K$ a
constant normalizing the pressure, $f \in [0,1]$ a multiplicative fudge
factor, $\Omega$ the frequency of rotation, and $\Omega_0$ the natural
frequency of the system.  Our modification is the introduction of $f$,
which allows us to tweak the fraction of the required pressure support
actually introduced into the system.  If we set $f = 1$ we recover the
traditional Maclaurin disk problem, for which solid body rotation
provides radial balance.  For $f < 1$, the pressure support is
inadequate and the disk becomes unstable to collapse.

As with the Pseudo-Keplerian disk, this simulation is performed in a
non-periodic computational volume in order make measurements of the global
angular momentum meaningful.  We solve for the self-gravity utilizing a
non-periodic PM code.  The gravity calculation treats the nodes as points
in 3D, so we are again considering gas confined to a plane in a 3D space
rather than infinite parallel rods.  In order to seed the initial positions
of the (A)SPH nodes to match the density profile of equation
(\ref{Cdiskrho.eq}), we select candidate positions using a Sobol sequence
which are subjected to a monte-carlo acceptance/rejection criterion.  All
\Gt\ tensors are initialized as round SPH \Gt\ tensors, scaled
appropriately for the local theoretical density.  Table
\ref{Cdisk.tab} summarizes our major numerical parameters for these
simulations.

\begin{figure}[htbp]
\begin{center}
\begin{minipage}[t]{0.9\hsize}
\plottwo{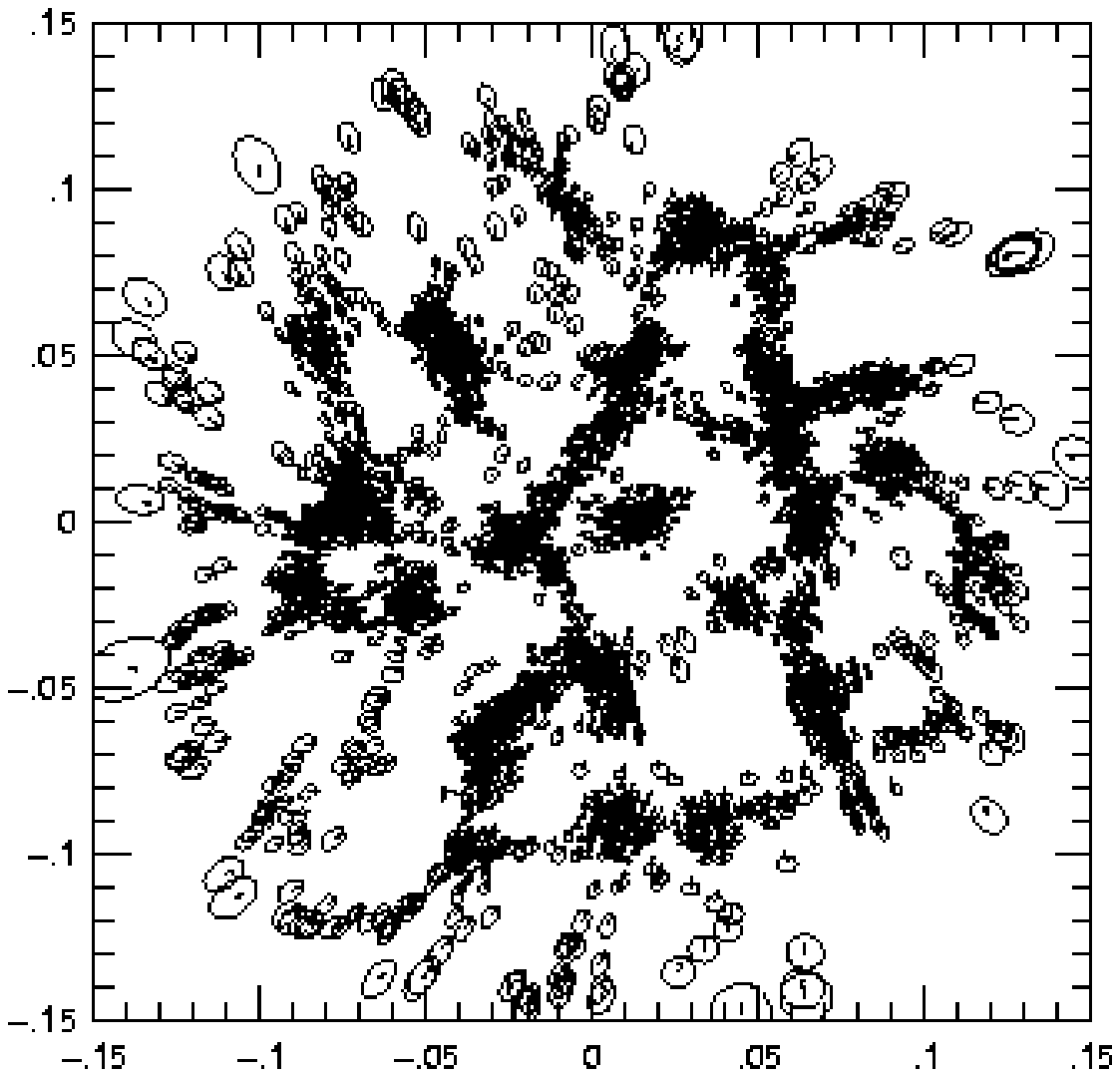}{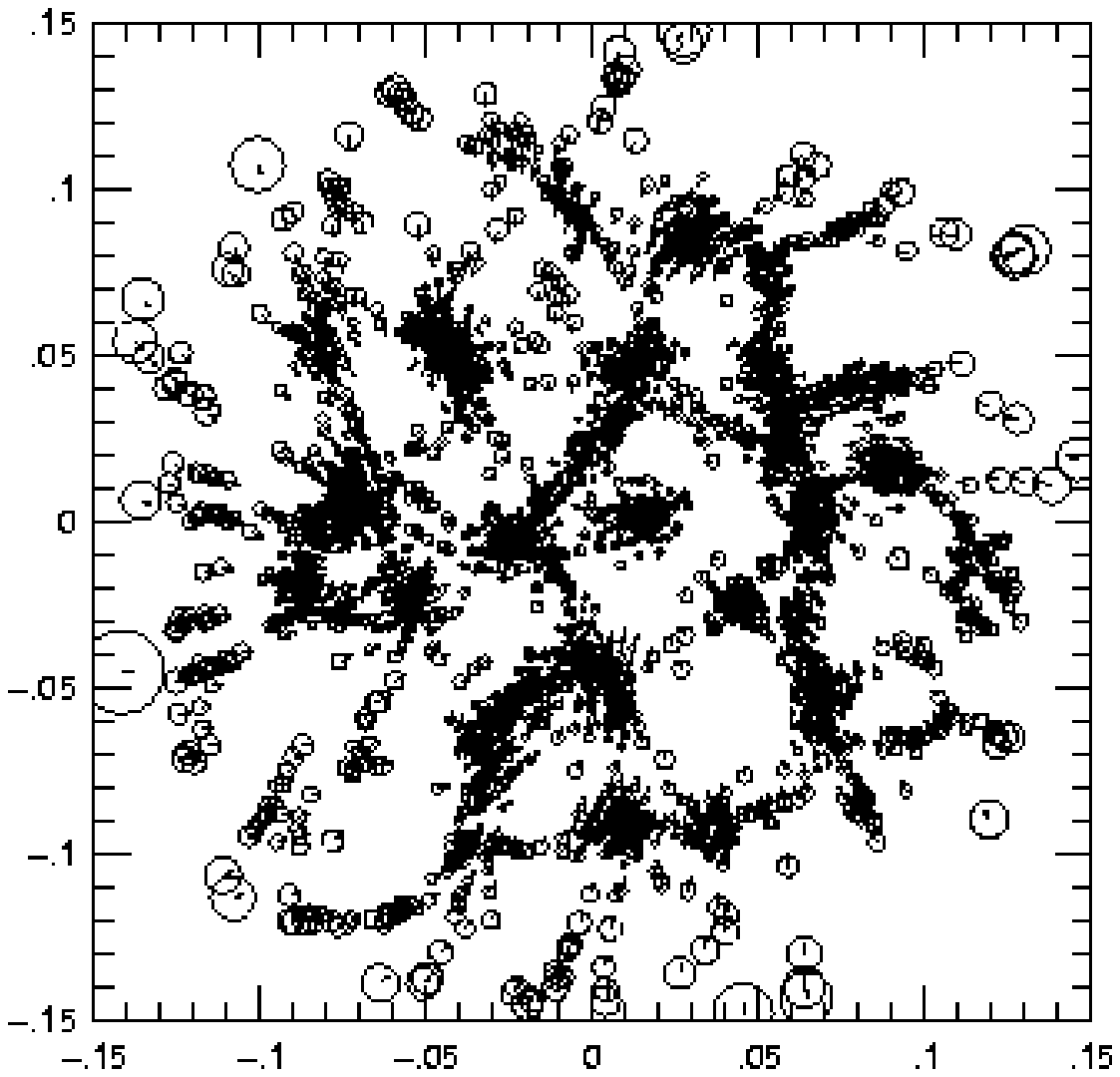}
\end{minipage}
\begin{minipage}{0.9\hsize}
\plottwo{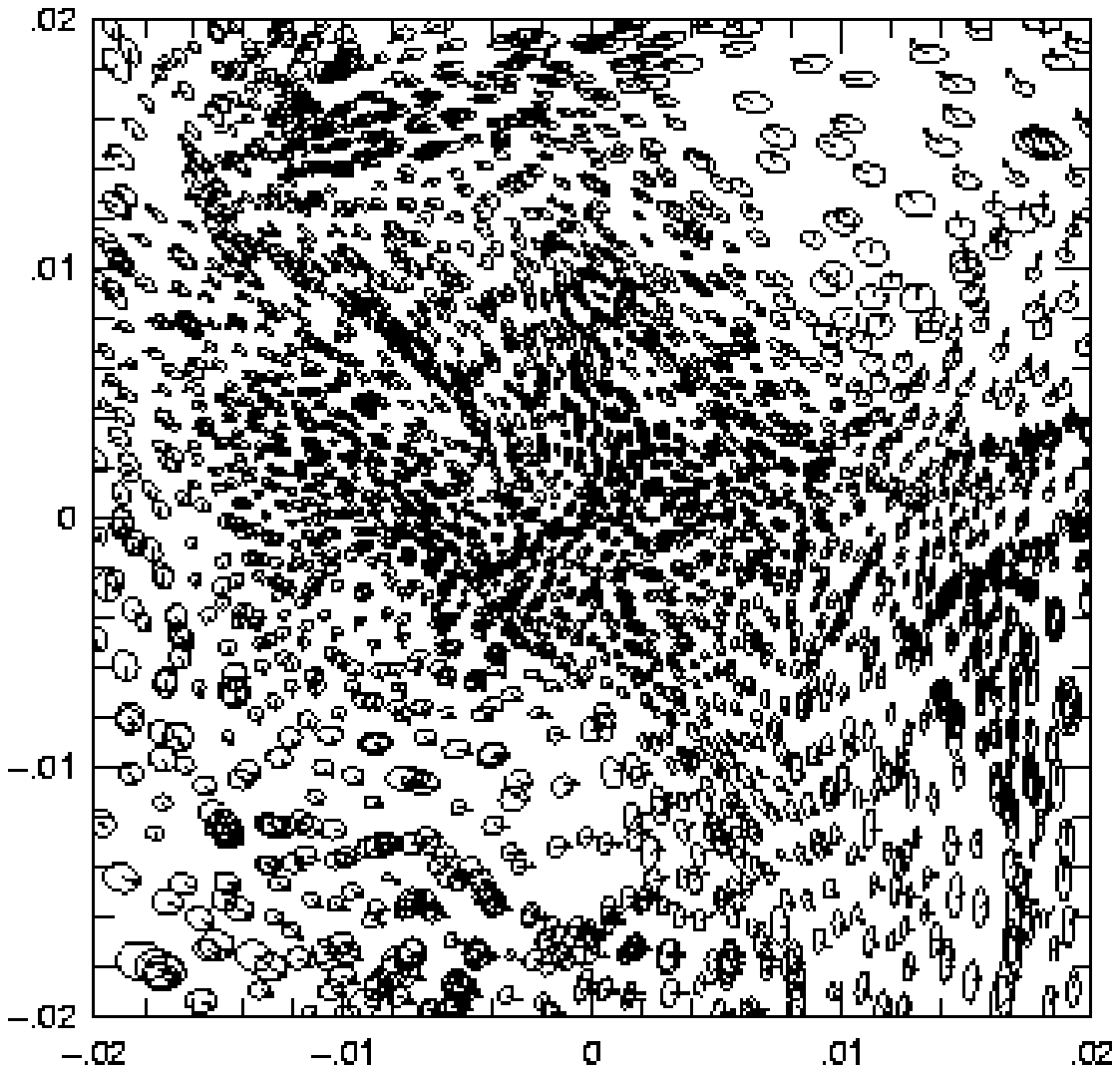}{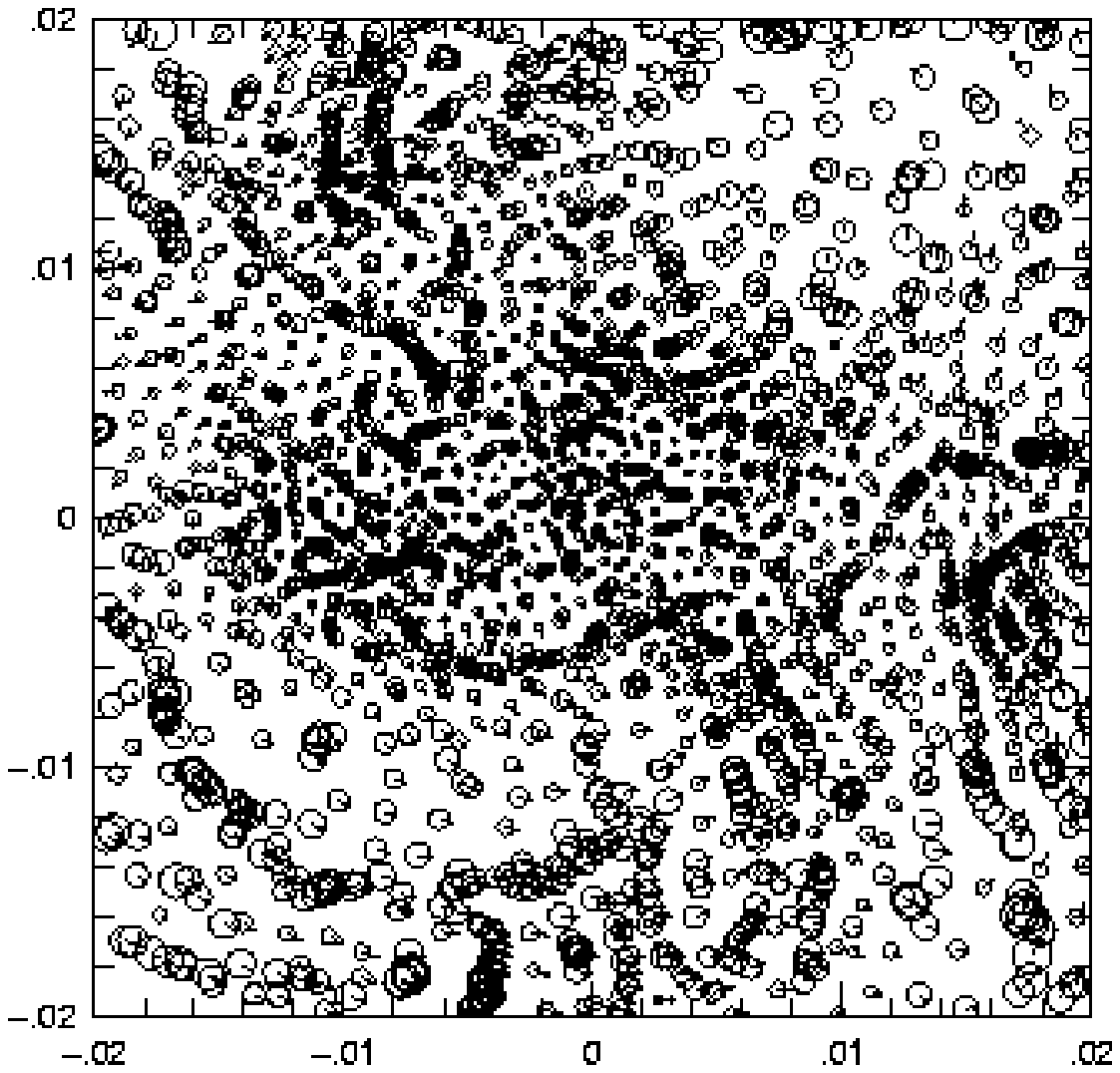}
\end{minipage}
\begin{minipage}{0.9\hsize}
\plottwo{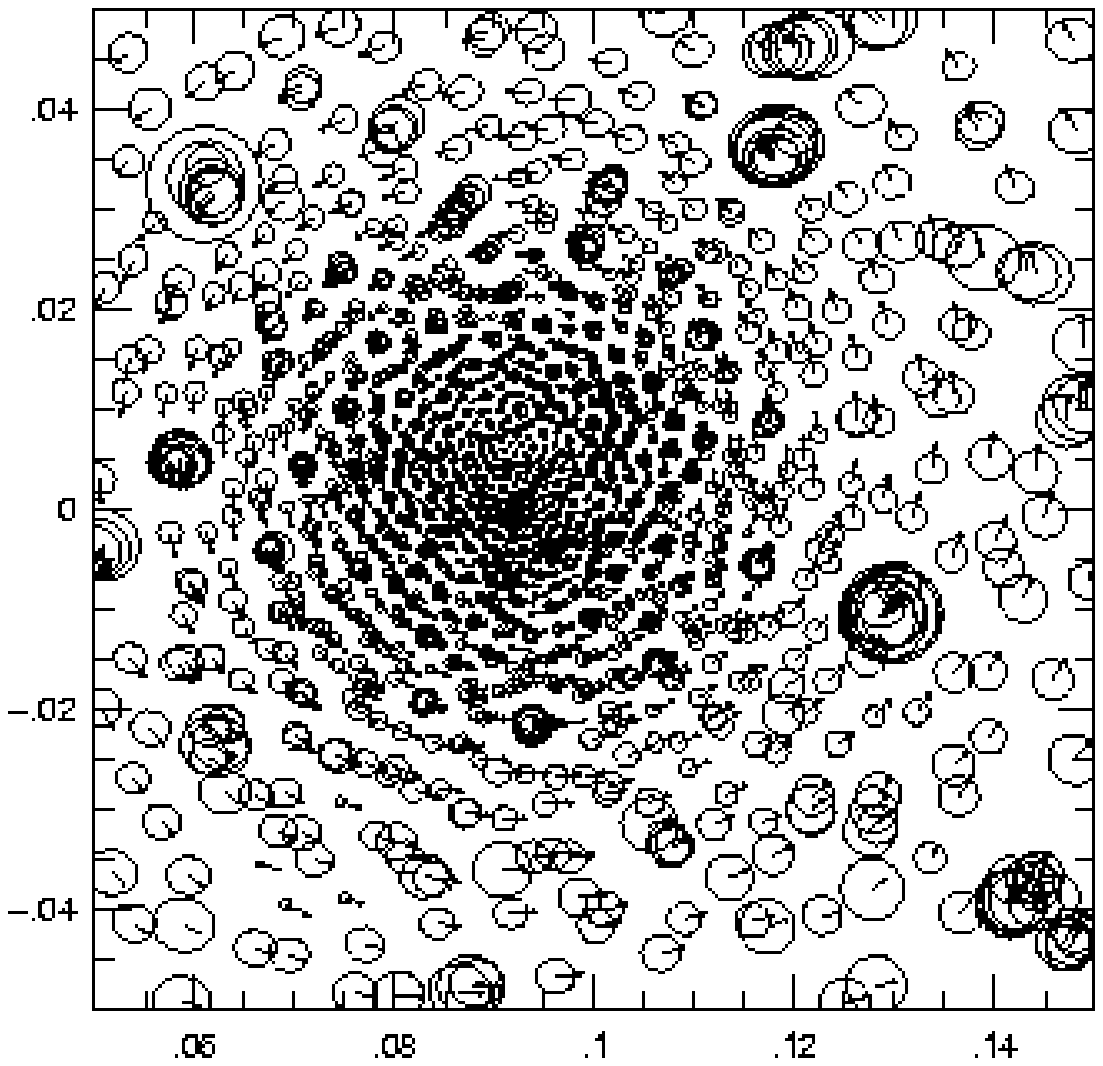}{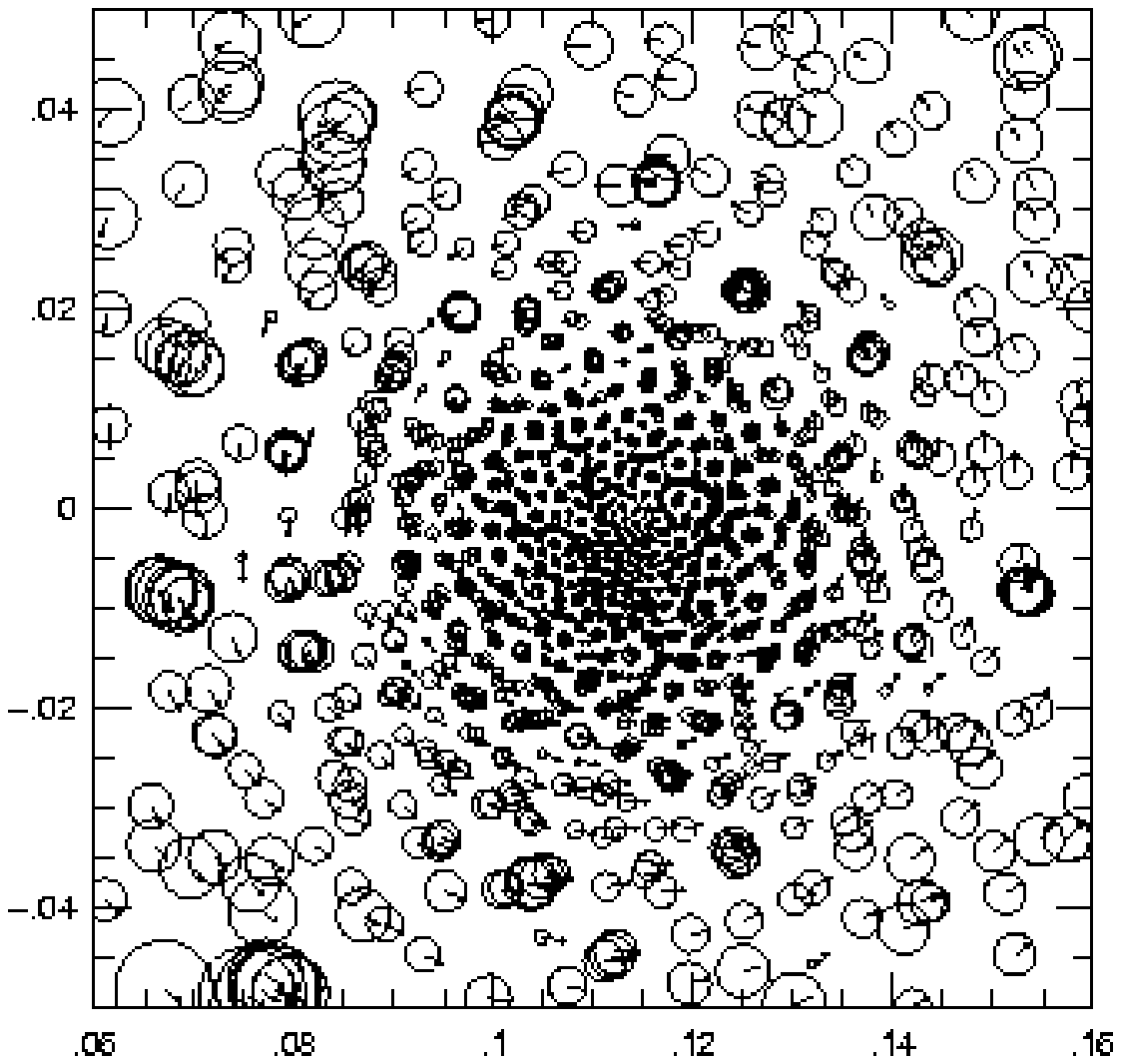}
\end{minipage}
\end{center}
\caption{Kernel plots for the 2D collapsing disk simulations.  Panels are
arranged with time increasing down columns: $t=0.08$
(pre-collapse) in the upper row, $t=0.1$ (maximum collapse) in the middle
row, and $t=0.5$ (post-collapse) in the lower row.  The ASPH simulation is
on the left, and SPH on the right.  The arrows associated with each node
indicate the direction and magnitude of the velocity.}
\label{Col_ker.fig}
\end{figure}
The evolution of this system is quite complex.  With $f=0.001$, the system
is highly unstable, undergoing immediate, radical collapse.  As the overall
radial collapse proceeds, the disk fragments into a collection of filaments
due to the gravitational amplification of perturbations in the noisy
initial density field.  During this filamentary stage, the ASPH \Gt\
tensors become quite elliptical, aligning themselves along the filaments.
By $t = 0.1$ the system reaches maximum collapse, and this filamentary
structure is destroyed.  After this time, we find that the system settles
into a two-phase structure.  There is a hot, dense, rapidly rotating core
surrounded by a diffuse, slowly expanding halo.  Figure \ref{Col_ker.fig}
presents kernel plots for the ASPH and SPH simulations at times $t = 0.08,
0.1, 0.5$, corresponding to the pre-collapse, maximal-collapse, and
post-collapse regimes.  We also plot the velocity fields on these kernel
plots as vector fields.  At $t=0.08$ (top panels), the pre-collapse
filamentary structure is evident.  During this stage, the ASPH simulation
exhibits \Gt\ tensors with ellipticities in the range $h_2/h_1 \in
[0.28,0.99]$, $\Interp{h_2/h_1} = 0.64$.  Examining the middle row of
panels at $t=0.1$ (the initial point of maximum collapse), there is an
evident clumpiness in the distribution of nodes with a large range of
smoothing scales.  The velocity field demonstrates a rather complex
structure, though an overall counterclockwise sense of rotation is
maintained, reflecting our input angular momentum.  Note that the ASPH
simulation shows markedly elliptical kernels in this collapsing core
region, with axis ratios in the range $h_2/h_1 \in [0.21, 0.98]$,
$\Interp{h_2/h_1} = 0.59$.  Despite this, the particle positions and
overall fluid flow corresponds well between the ASPH and SPH runs at this
time.  Finally, at $t=0.5$ in the bottom row of panels, we find that the
system settles into orderly rotation about a dense core.  The ASPH
simulation exhibits basically round \Gt\ tensors in the core (with overall
\Gt\ geometries in the range $h_2/h_1\in [0.46, 1.0]$, $\Interp{h_2/h_1} =
0.92$) reminiscent of the Pseudo-Keplerian Disk, which is appropriate as
there is no preferred resolution direction.

\begin{figure}[htbp]
\plotone{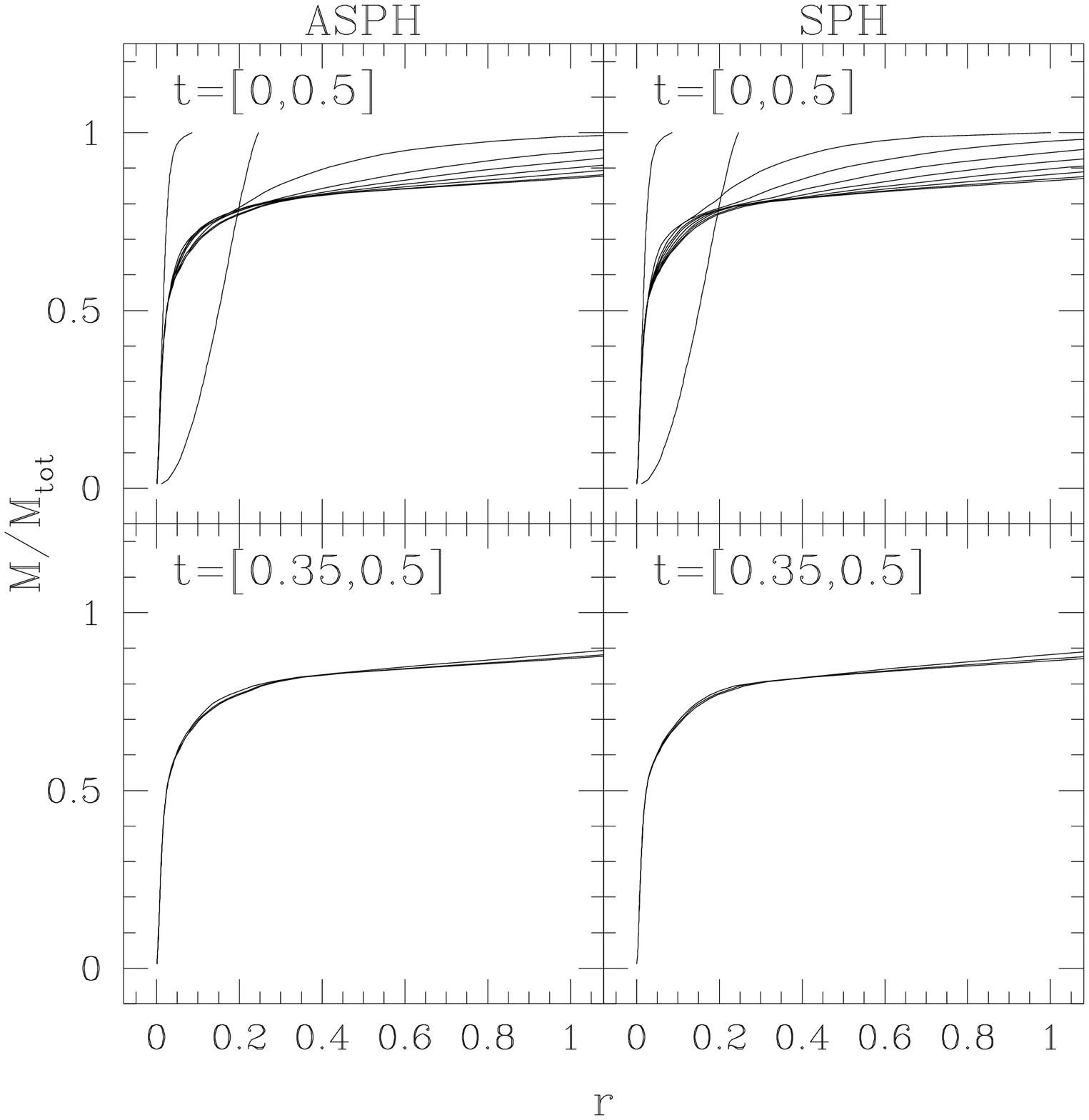}
\caption{Azimuthally averaged radial mass profiles for the 2D
collapsing disk simulations.  Each curve represent the fraction of
the mass of the disk contained within the radius $r$.  The top panels show
the function $M(r)$ for times varying from the beginning of the simulation
to the end ($t \in [0,0.5]$).  The bottom panels show only the last
measurements of $M(r)$, in the time range $t \in [0.35,0.5]$.}
\label{Col_Mr.fig}
\end{figure}

\begin{figure}[htbp]
\plotone{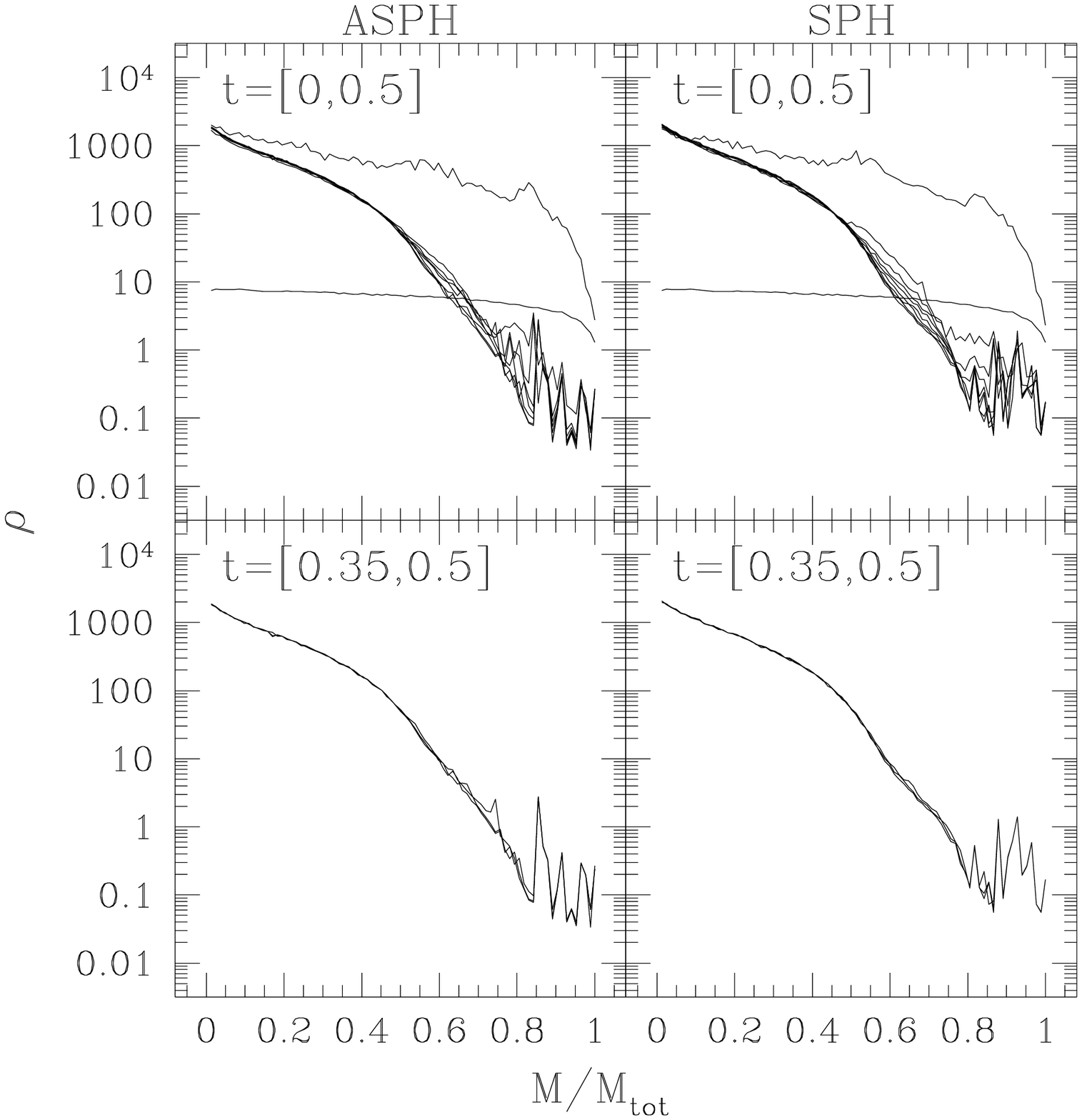}
\caption{Azimuthally averaged radial density profiles for the 2D
collapsing disk simulations, expressed as a function of the enclosed mass
fraction $\rho(\protect\Sub{M}{interior})$.  The top panels show the
function $\rho(M_{interior})$ for times varying from $t \in [0,0.5]$, while
the bottom panels only show the last few measurements at times $t \in
[0.35,0.5]$.}
\label{Col_rhoM.fig}
\end{figure}
Figures \ref{Col_Mr.fig} and \ref{Col_rhoM.fig} show the evolution of the
functions $M(r)$ and $\rho(M)$, analogously to Figures \ref{Kep_Mr.fig} and
\ref{Kep_rhoM.fig} for the Pseudo-Keplerian disk simulations.  In this case,
though, we measure the radial coordinate out from the center of mass of the
central disk, since this is a self-gravitating system and therefore the
minimum of the potential moves with the center of mass.  It is clear that
the system rapidly forms a concentrated central disk containing $\sim 70\%$
of the mass following the maximum collapse.  The radial density gradient of
this central disk is also quite large, varying from $\rho \sim 10^3$ in the
central regions to $\rho \sim 10^{-1}$ at the edge.  The ASPH and SPH
simulations both seem to agree well on this mass distribution.

\begin{figure}[htbp]
\plotone{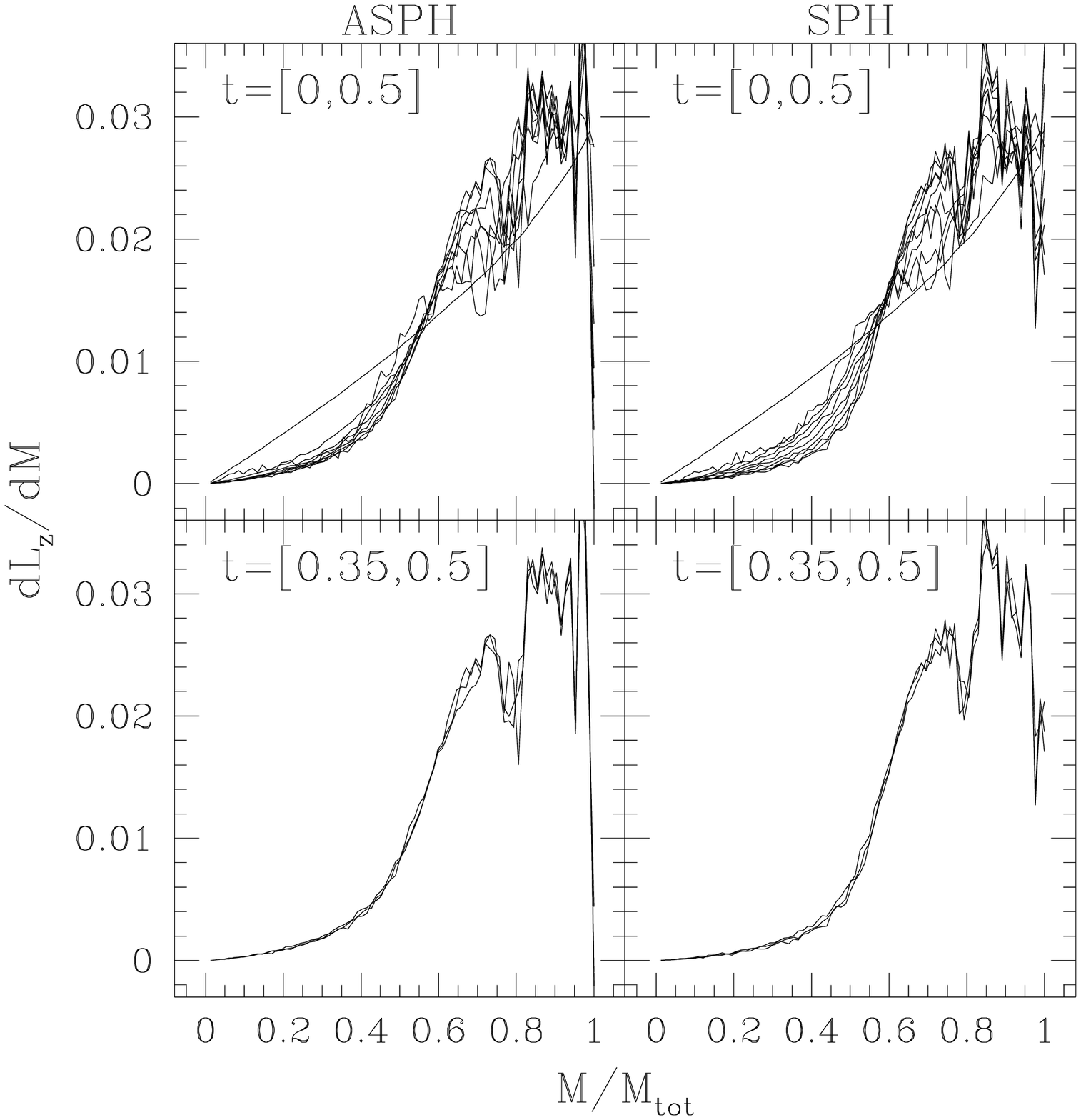}
\caption{Azimuthally averaged radial measurements of angular momentum
$\Delta L_z$ as a function of the enclosed mass for the 2D collapsing disk
simulations.  These curves represent a radial measurement of $dL_z/dM$,
where $M$ is interpreted as the radially enclosed mass.  The top panels
show the function $dL_z/dM$ for times varying from $t \in [0,0.5]$, while
the bottom panels only show the last few measurements at times $t \in
[0.35,0.5]$.}
\label{Col_dLzdM.fig}
\end{figure}

\begin{figure}[htbp]
\plotone{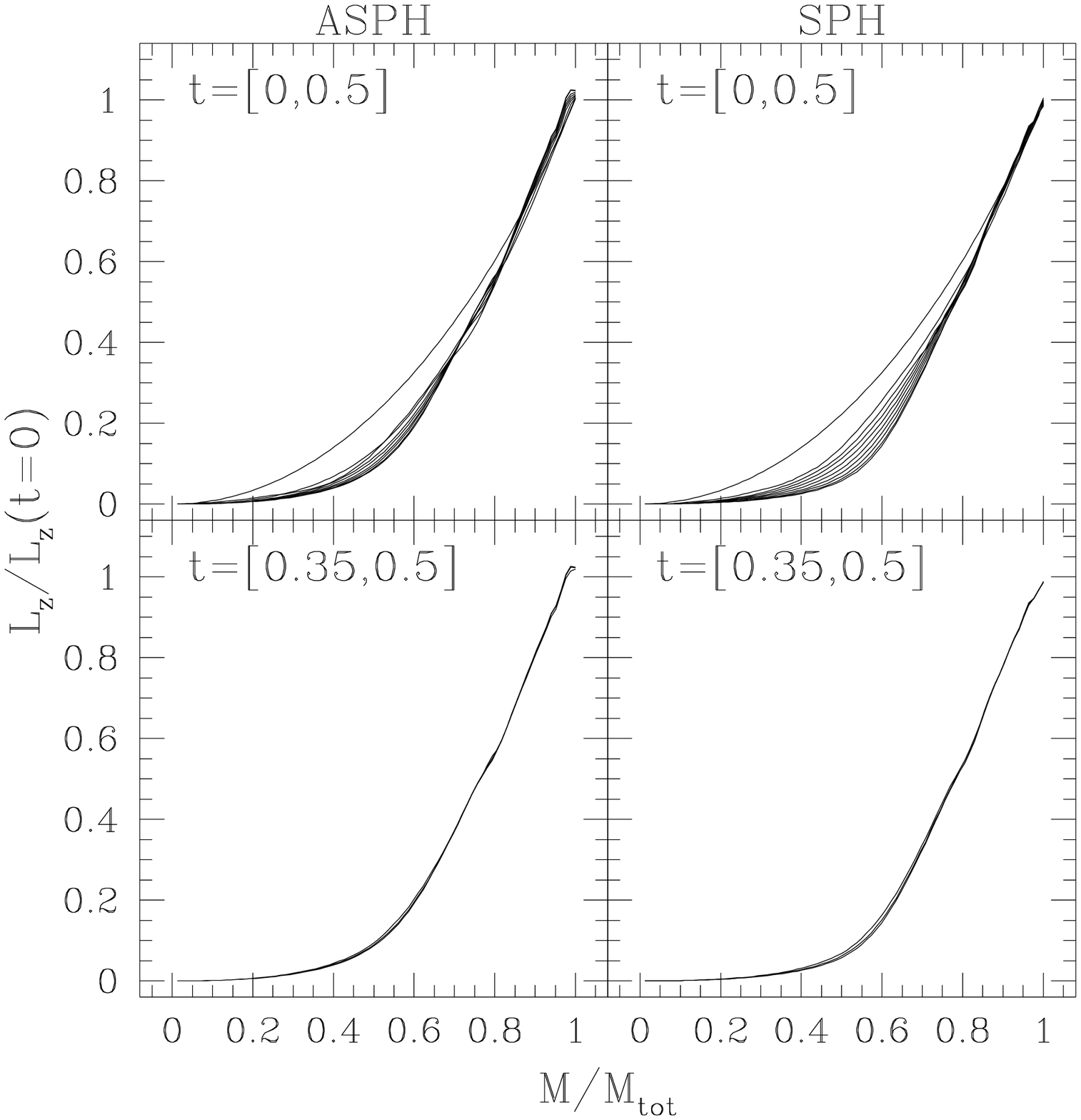}
\caption{The total enclosed angular momentum $L_z$ as a function of the
enclosed mass $M$ for the 2D collapsing disk simulations, as measured
radially from the center of the disk.  The curves of $L_z(M)$ represent the
integration $\int_0^M (dL_z/dM) dM$ of the function $dL_z/dM(M)$ presented in
Figure \protect\ref{Col_dLzdM.fig}.}
\label{Col_LzM.fig}
\end{figure}
Figure \ref{Col_dLzdM.fig} shows the radial distributions of
$dL_z/dM(M)$ throughout the simulations.  It is evident that once
the systems achieve equilibrium, the central $70\%$ of the mass (representing
the collapsed disk) maintains a steady distribution of angular momentum.
However, outside the collapsed disk this distribution becomes quite noisy
for the hot, diffuse gas.  It is also curious to note that it appears the
ASPH simulation maintains slightly more angular momentum in the disk as
compared with SPH.  This is also evident in the cumulative distribution of
$L_z(M) = \int_0^M (dL_z/dM) dM$ in Figure \ref{Col_LzM.fig}.  This
trend is the opposite that found in the previous Pseudo-Keplerian disk,
where SPH maintains a larger core angular momentum than ASPH.  Regardless,
it is clear that both ASPH and SPH settle to equilibrium distributions of
the local angular momentum once the system settles and axisymmetry is
achieved.

\begin{figure}[htbp]
\plottwo{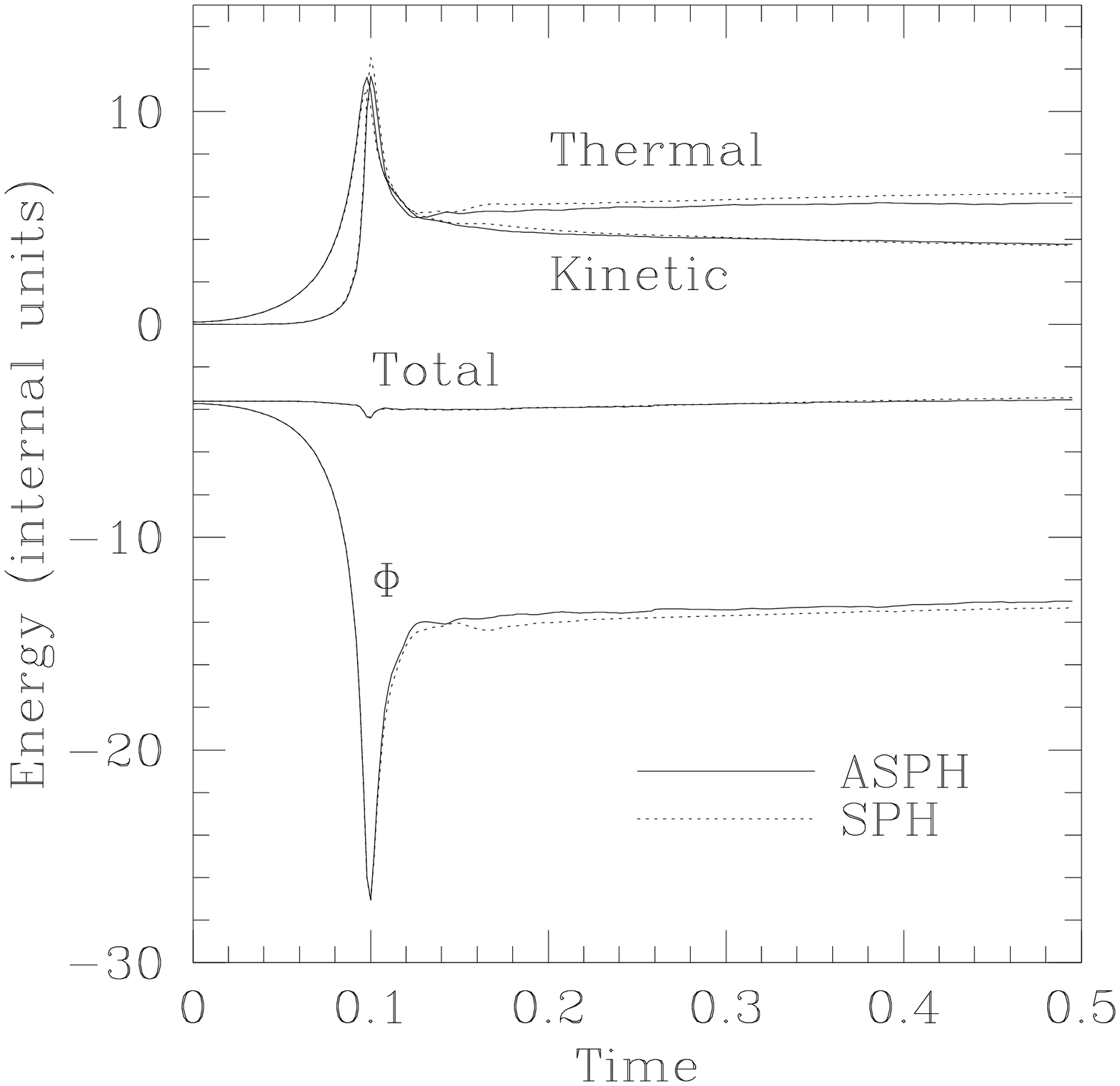}{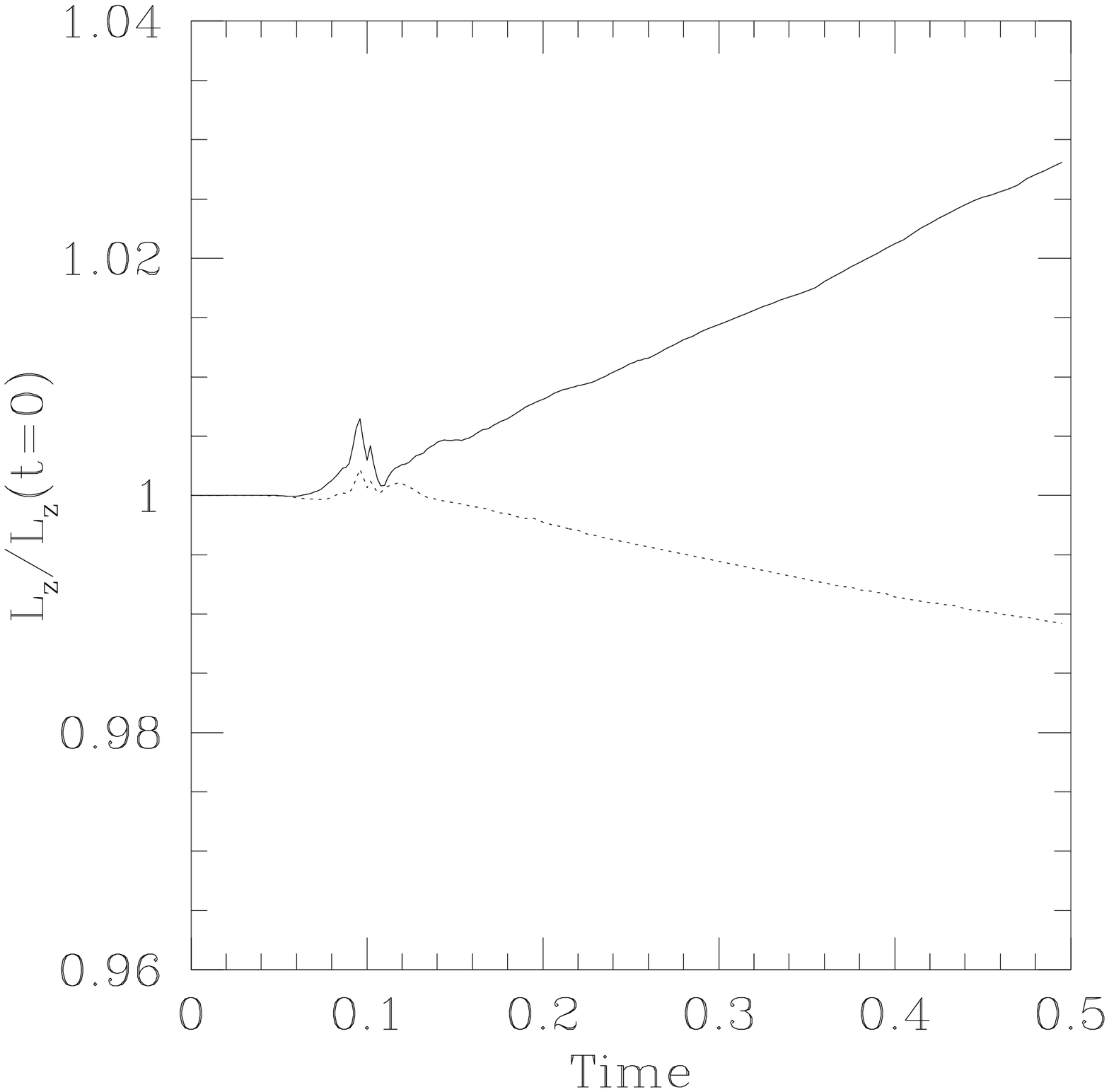}
\caption{Evolution of the global energies (left panel: kinetic, thermal,
potential, and total) and the total angular momentum $L_z$ (right panel)
for the 2D collapsing disk simulations.}
\label{Col_cons.fig}
\end{figure}
Finally, in Figure \ref{Col_cons.fig} we present the time evolution of the
global energies and angular momentum.  It is apparent that at time $t=0.1$
the system undergoes maximal collapse, at which time all the energies
suffer a dramatic spike.  After this time the energies achieve equilibrium
values.  The global angular momentum for the ASPH simulation varies by
$\Delta L_z/L_z \sim 3\%$ throughout the simulation, while SPH varies by
$\Delta L_z/L_z \sim 1\%$.  This error is likely due to the PM code, which
does not conserve angular momentum precisely.  We therefore conclude that
in this test ASPH again conserves angular momentum reasonably, both locally
and globally.

\section{Summary and Conclusions}
\label{Disc.sec}
A new version of SPH, which we call Adaptive SPH, or ASPH, which replaces
the isotropic smoothing of standard SPH by anisotropic smoothing, is
described in detail here, along with the results of test problems and
comparisons with standard SPH. This is Paper II of a pair of methodology
papers describing ASPH and its application to cosmological gas dynamics. In
Paper I, we demonstrated the relative shortcomings of SPH compared to ASPH
whenever the flow problem to be simulated involves highly anisotropic
volume changes, such as occur commonly in cosmic structure formation,
involving gravitational collapse and strong shocks. Paper I described the
ASPH algorithm in detail, including a mathematical prescription for
evolving the ellipsoidal smoothing kernels based upon the idea that each
ellipsoid rotates and deforms with the flow so as to track the Lagrangian
evolution of infinitesimal fluid elements centered on each SPH particle. In
Paper I, this ASPH algorithm was implemented in 2D and tested against a
variety of problems like the Riemann shock and cosmological pancake
problems, in comparison with standard SPH.  In addition to anisotropic
smoothing, the ASPH method described in Paper I included an algorithm
designed to reduce the spurious effects of artificial viscous heating of
gas undergoing supersonic compression far from a shock, by selective
inclusion of artificial viscosity in the energy equation only for particles
which are close to being overtaken by shocks. The combined effects of
anisotropic smoothing and this suppression of viscous heating away from
shocks was shown in Paper I to result in a substantial increase in spatial
resolving power at fixed particle number, for problems like cosmological
pancake collapse, generic to cosmic structure formation.

In Paper II, we have described further developments of the ASPH method, further
tests, and an alternative mathematical prescription for the evolution of 
the anisotropic smoothing kernels. The latter is based upon a linear 
transformation of spatial coordinates for each particle to one in which the
mean spacing of particles in the vicinity of that particle appears to be
isotropic. Paper II has described the implementation of this ASPH algorithm
in both 2D {\it and} 3D. Several refinements of the ASPH scheme described
in Paper I are described here, such as the use of separate time-steps for
separate particles and a new scheme for nearest-neighbor assignment, which
improve the efficiency of the method, and the use of ellipsoidal 
smoothing kernels whose axis length and orientations are themselves 
``smoothed'' over nearest neighbors, which enhances the stability of the
ASPH method. An alternative to the artificial viscosity suppression algorithm
of Paper I is also introduced, involving the use of a spatially compact kernel
just for artificial viscosity, which improves the overall energy conservation
of the ASPH method, albeit at the cost of decreasing somewhat the quality
of the comparison between the ASPH and exact solutions
for the Zel'dovich pancake problem. However, even with this
more conservative algorithm in Paper II for artificial viscosity than
was utilized in Paper I, the improvement of the ability of ASPH to resolve
pancake shocks and postshock profiles over that of standard SPH is substantial.
Tests of the ASPH method in Paper II also go beyond those in Paper I to show
that the advantages of ASPH over SPH at fixed particle number extend even to
problems other than anisotropic gravitational collapse, as in the Zel'dovich
pancake problem, to include situations like the classical problem of a 
point explosion in a uniform gas, described by the Sedov similarity solution.
The conclusion of this set of tests is that
ASPH achieves a significant improvement relative to SPH in a much wider range
of circumstances than those involving anisotropic gravitational collapse during cosmic structure formation which initially motivated us to develop ASPH.

%%%%%%%%%%%%%%%%%%%%%%%%%%%%%%%%%%%%%%%%%%%%%%%%%%%%%%%%%%%%%%%%%%%%%%%%%%%%%%%%
It is worth discussing in detail how the resolving power of these
techniques is defined, in order to make the sorts of comparisons discussed
above clear.  Since ASPH and SPH are Lagrangian techniques, their
respective resolutions are most naturally described in terms of a mass
resolution, essentially given by the mass per smoothing volume.  For the
same number of particles per simulation volume, the higher spatial
resolving power of ASPH over SPH ensures that the true mass resolution of
ASPH is closer to the mass per smoothing volume than is that of SPH, since
the evolution of the properties of the mass in each smoothing volume is
more accurately accomplished by ASPH.  This is a consequence of the fact
that the interpolation amongst nearest neighbor particles is more accurate
when proper account is taken of the mean interparticle spacing along
different directions.  In SPH, however, the failure of the length resolution 
to adapt in direction to follow the anisotropy of volume changes causes the
interpolation amongst particles to degrade the spatial resolution in a
direction-dependent way, which causes the true mass resolution of the
method to be worse than the nominal mass per smoothing volume.  Quantifying
the degree to which ASPH improves both the spatial and mass resolution in
this sense is difficult to do in a problem-independent way.  Instead, one
must appeal to specific application problems to assess the relative
resolving powers of SPH and ASPH.

For the case of cosmological pancake collapse, one can compare the ASPH and
SPH results and assess how many SPH particles would be required to get the
shock jump and postshock fluid variable profiles to be within some
fractional error of the exact solution.  This was done in Paper I, for
example, by using 1D versions of the SPH and ASPH methods which enabled us
to vary the number of particles per dimension over a very much wider range
than would have been possible if we had used a true 3D code, and yet serves
to indicate the relative number of particles required per pancake
wavelength per column along the direction of collapse to resolve the flow
adequately.  In that case, it was found that an order of magnitude more
particles per dimension are required in SPH than ASPH, in order to ensure
that a randomly oriented 1D pancake is as well-resolved by the SPH results
as by the ASPH results.  This means that, for a 3D calculation of a
randomly oriented pancake, the total number of SPH particles per simulation
volume must be as much as 1000 times greater than for ASPH.  (The results
in this paper are not quite so dramatic since we only consider 2D
comparisons here.)  This result is for the case with no radiative cooling.
If radiative cooling is included, this ratio is even larger.  This is a
necessary condition for the SPH code to do as well as the ASPH code at
resolving a 1D pancake collapse. The latter is a fundamental and generic
structure which forms during cosmological structure formation under quite
general circumstances (see, for example, the introduction of Valinia \etal\
1997 for a review of this latter point and references).  As such, it is
realistic to use this problem-specific comparison of SPH and ASPH as a
general indication of how many more particles of SPH are required to do as
well as ASPH at resolving cosmological structure formation.

It is, of course, true that in using SPH with more particles to achieve the
equivalent spatial resolution as ASPH with fewer particles, 
one is improving the mass resolution of the SPH calculations
at the same time.  It is difficult to quantify that improvement, however,
since the true mass resolution is worse than the mass per smoothing volume
wherever the isotropic interpolations of SPH encounter anisotropic volume
changes, as they inevitably must in a wide range of problems, including
cosmological structure formation.  A side benefit of this increase in the
number of particles is that the initial conditions for the gas, which for
cosmological structure simulations are usually close to uniform with small
amplitude density fluctuations, can include a larger range of a wavenumbers
for the initial perturbations.  This is analogous to the effect of
comparing an Eulerian grid-based method with adaptive mesh refinement (AMR)
to one without, where the AMR method achieves the same high resolution as
the non-AMR method but with many fewer initial grid-cells and a smaller
total number of grid-cells, even after the highest level of refinement has
occurred.  In the AMR case, the resolution of the method in space improves
during the simulation as mesh refinement is added, while in the non-AMR
case the mesh must be refined everywhere and at all times to the full
extent of the maximum local refinement added by the AMR code in order to
match the AMR code's resolving power.  In that case the AMR code, while
much more efficient, can never recover the extra dynamic range of the
initial conditions which the more expensive, non-AMR code gains as a side
benefit of its having a uniformly higher number of cells than the AMR code
at the initial time-slice. Nevertheless, it is well-established that the
AMR modification of Eulerian grid-based hydrodynamics methods is a powerful
enhancement of the resolving power of the Eulerian method at fixed
computational resources.  A similar statement can be made for the effect of
ASPH as an improvement of the resolving power of the SPH method, for a wide
range of astrophysically interesting flows.

In this context, it is important to point out that the range of problems
for which the ASPH method gives better results than standard SPH is much
wider than just that of the Zel'dovich pancake problem (or the general
gravitational growth of cosmological density fluctuations, which often
lead to highly anisotropic collapse problems such as pancake and filament
formation).  Our results here for the Sedov blastwaves clearly demonstrate
that ASPH has strong advantages over SPH even for problems which do not
involve pancake collapse or any other anisotropic gravitational
compression.  It is more difficult to quantify the increase in the number
of SPH particles which would be necessary in order to yield results as
close to the exact solution as ASPH, however, simply because our 2D and
especially our 3D results indicate that this problem requires a very large
number of particles in order to yield a close match to the exact solution,
even for ASPH.  As a result, it is not yet possible computationally to
handle a large enough number of SPH particles to demonstrate how large a
number is required to match the success of the ASPH results.
%%%%%%%%%%%%%%%%%%%%%%%%%%%%%%%%%%%%%%%%%%%%%%%%%%%%%%%%%%%%%%%%%%%%%%%%%%%%%%%%

We 
have also investigated the potential shortcomings of the ASPH technique, the
most serious of which is that while SPH is rigorously guaranteed to
conserve angular momentum globally, 
ASPH is not.  Through a set of 2D rotational
problems we show that, with reasonable precautions enforced, ASPH 
does globally conserve
angular momentum satisfactorily (to a level of order a few percent) over
many dynamical times, for the challenging types of rotating gas problems 
involving shearing and collapsing disks which we have devised to test it.

In addition to the global conservation of the angular momentum, we have also
considered the possibility that ASPH could result in erroneous local
transport of the angular momentum due to false interparticle torques on the
ellipsoidal kernels.  We have tested
this hypothesis by examining rotating systems
which eventually settle to axisymmetric configurations, since inviscid,
axisymmetric systems cannot physically transport angular momentum radially and
therefore any transport measured would have to be a numerical effect.  We
find that once our test problems settle to axisymmetric configurations {\em
both} ASPH and SPH cease to transport the angular momentum, despite the
continued presence of elliptical kernels in the ASPH models.  Thus, for the
2D rotating systems considered here, ASPH does not result in either
significant violations of the global conservation of angular momentum, nor
spurious transport of the local angular momentum.  We are encouraged by
these test results to pursue the issue of angular momentum conservation by
the ASPH method further in future papers, with particular emphasis on 3D
rotating problems.

In general ASPH appears to be quite promising for a number of applications
where SPH has been previously used.  ASPH offers the ability to maximize
the spatial resolution for a given number of computational nodes, at only a
modest computational penalty compared to standard SPH.  This can be
understood by examining the mechanics of an ASPH implementation vs.\ SPH.
Where SPH involves dividing a scalar distance by a smoothing scale, ASPH
requires a matrix multiplication, incurring a computational penalty of a
few operations.  Determining the evolution of the \Gt\ tensor relies solely
upon evaluating $\sigma_{\alpha \beta} = \partial v_\alpha/\partial
x_\beta$, the trace of which is required by SPH regardless.  This quantity
can be determined in the same loop over neighbors which calculates the
normal SPH dynamical equations, and therefore the computational penalty is
again light.  The majority of the increased computational burden for ASPH
is due to the necessity of smaller timesteps and therefore more integration
cycles, required by stability criteria for higher resolutions (see appendix
\ref{timestep.app}).  This is a penalty which will always be necessitated
by higher resolution, though, and should not reflect disfavorably on ASPH.
For this reason it is critical to devise efficient integration schemes for
use with large scale ASPH simulations, such as the asynchronous algorithm
described in appendix \ref{integrator.app}, which allows each ASPH node to
evolve at its own timescale, semi-independently of the simulation as a
whole.

\acknowledgements 
The authors (particularly JMO) would like to thank Mike Fisher and Mathias
Steinmetz for many useful conversations and observations, Richard Pogge and
David Weinberg for their support and help during this work, and many
members of Ohio State's Astronomy department for both useful comments and
allowing me to use their workstations for extended computations.  JMO
acknowledges the support of NASA grant NAG5-2882, an Ohio State GSARA award
for travel funds, and the hospitality of the Max Planck Institut f\"{u}r
Astrophysik in Garching.  PRS and HM acknowledge the support of NASA grant
NAG5-2785 and NSF grant ASC 9504046.  PRS is grateful to the Institute for
Theoretical Physics at the University of California at Santa Barbara for
its hospitality in June 1995, during the 1995 Astrophysics Workshop on
Galaxy Formation, during part of this work.  We also acknowledge the
support of the Ohio Supercomputer Center, the RZG Computing Center at
Garching, and the use of the University of Texas High Performance Computing
Facility.

\begin{appendix}
\section{(A)SPH Dynamical Equations}
\label{Dyneqs.app}
In this appendix we briefly describe the (A)SPH dynamical equations,
with the exception of the evolution of the \Gt\ tensor which is
treated in detail in Appendix \ref{Genderiv.app}.  In \S
\ref{propdyneqs.app} we present the set of proper coordinate, time
dependent equations we use to evolve non-cosmological simulations -- \S
\ref{codyneqs.app} presents the set of comoving dynamical equations,
expressed in terms of a power of the cosmological expansion factor, rather
than time, as the independent variable.  We express all dynamical equations
in terms of the normalized position vector $\veta$, such that they are
equally applicable to SPH and ASPH.
%The reader who wishes a
%more thorough derivation of these equations should consult the review
%articles on SPH by either Monaghan (1992) or Benz (1990).

\subsection{(A)SPH Equations in Proper Space}
\label{propdyneqs.app}
The (A)SPH dynamical equations are based upon solving discretized forms of
the Lagrangian conservation equations at the positions of each (A)SPH node.
We use the following forms of the Lagrangian conservation equations
\beq
  \label{cont.eq}
  \frac{D\rho}{Dt} = -\rho \grad \cdot {\bf v},
\eeq
\beq
  \label{mom.eq}
  \frac{D{\bf v}}{Dt} = -\frac{\grad P}{\rho} + {\bf g},
\eeq
\beq
  \label{therm.eq}
  \frac {Du}{Dt} = -\frac{P}{\rho} \grad \cdot {\bf v}
                   + \frac{1}{\rho} \left[ \Gamma(T) - \Lambda(T) \right],
\eeq
which represent the conservation laws for mass, momentum, and energy.  Note
we make use of the convention that the Lagrangian derivative is $D/Dt
\equiv \partial/\partial t + ({\bf v} \cdot \grad)$.  We define $\rho$ as
the mass density, ${\bf v}$ velocity, $P$ pressure, ${\bf g}$ gravitational
acceleration, $u$ specific thermal energy, $T$ temperature, $\Gamma(T)$ the
photoionization heating rate, and $\Lambda(T)$ the radiative cooling rate.
These forms of the conservation equations account for pressure and
gravitational forces, as well as the lowest order influences of radiation.
We neglect the influence of viscous drag, heat conduction, and radiation
pressure.  These other effects could in principle be included at the
penalty of complicating the equations.  However, throughout this work we
have chosen to follow a similar philosophy to that of Evrard (Evrard 1988;
Evrard \etal\ 1994) in keeping the dynamical system relatively ``clean'',
avoiding the introduction of too many complicating parameters.  The
physical processes accounted for here are thought to be the dominant
processes in the formation of cosmological structures such as large
galaxies (Rees \& Ostriker 1977; Silk 1977; White \& Rees 1978).

We use the equation of state appropriate for an ideal gas, which yields
the relations
\beq
  \label{idealP.eq}
  P = (\gamma - 1) \rho u,
\eeq
\beq
  \label{idealT.eq}
  k T = (\gamma - 1) \mu m_p u,
\eeq
with $\gamma \equiv c_P/c_V$ the ratio of the specific heats, $k$
Boltzmann's constant, $\mu$ the average atomic weight (in atomic units),
and $m_p$ the mass of a proton.

Granted these forms of the conservation equations and an equation of
state, the evolution equations for a given node $i$ in proper
coordinates are
\beq
  \label{Sphpos.eq}
  \frac{D {\bf r}_i}{Dt} = {\bf v}_i,
\eeq
\begin{eqnarray}
  \label{Sphmom.eq}
  \frac{D {\bf v}_i}{Dt} 
    &=& -\sum_j m_j \left[ \lp \frac{P_i}{\rho_i^2} + \frac{P_j}{\rho_j^2} 
        \rp \gWij + \Pi_{ij} \gWPij \right] + {\bf g}_i \\
    &=& -\sum_j m_j \left[ (\gamma - 1) \lp \frac{u_i}{\rho_i} + 
        \frac{u_j}{\rho_j} \rp \gWij + \Pi_{ij} \gWPij \right] + {\bf g}_i,
	\nonumber
\end{eqnarray}
\begin{eqnarray}
  \label{Sphtherm.eq}
  \frac{Du_i}{Dt} &=& \frac{P_i}{\rho_i^2} \sum_j m_j \vij
                      \cdot \gWij \, + \, \frac{1}{2} \sum_j 
                      m_j \Pi_{ij} \vij \cdot \gWPij 
                      + \: \frac{\Gamma(T_i) - \Lambda(T_i)}{\rho_i} \\
                  &=& \sum_j m_j \vij \cdot \left[ (\gamma - 1) 
                      \frac{u_i}{\rho_i} \gWij + \frac{1}{2} \Pi_{ij} 
                      \gWPij \right]
                      + \: \frac{\Gamma(T_i) - \Lambda(T_i)}{\rho_i},
                      \nonumber
\end{eqnarray}
\beq
  \label{Sphcont.eq}
  \frac{D \rho_i}{Dt} = \sum_j m_j \vij \cdot \gWij,
\eeq
\beq
  \label{Sphdens.eq}
  \rho_i = \sum_j m_j \Wij.
\eeq
Note that there is no direct assumption about the dimensionality of the
problem in these expressions.  This exemplifies one of the nice properties
of (A)SPH based techniques, in that they are easily implemented in any
number of dimensions.  It is only in the evolution of the smoothing scales
that the dimensionality of the problem becomes important, as is discussed
in the body of the paper (\S \ref{DhsDt.sec} and \S \ref{DGDt.Sec}) and in
detail in appendix \ref{Genderiv.app}.

Inspection of the above set of equations reveals that equations
(\ref{Sphcont.eq}) and (\ref{Sphdens.eq}) are redundant.  Either are
capable of specifying the density.  If we choose equation
(\ref{Sphcont.eq}), then the density becomes another parameter to be
evolved from a given set of initial conditions.  This approach has the
disadvantage of not ensuring absolute mass conservation.  If we instead
choose to use the summation approach of equation (\ref{Sphdens.eq}), then
mass conservation is rigorously enforced.  We nominally use the summation
technique to update the density.  However, our code utilizes an
asynchronous integration algorithm, which requires intermediate estimates
of the mass density.  For this reason we also calculate equation
(\ref{Sphcont.eq}) for use in making these estimates, and therefore include
it here.

The term $\Pi_{ij}$ in the momentum and energy equations (eqs.
[\ref{Sphmom.eq}] and [\ref{Sphtherm.eq}]) represents an artificial
viscosity, which is present to account for shock conditions in the gas.
Without this artificial viscosity, (A)SPH is insufficiently dissipative to
prevent interpenetration of the nodes, and shock conditions will be poorly
represented.  We adopt a standard SPH form of the artificial viscosity due
to Monaghan \& Gingold (1983), which is defined on a pair-by-pair basis as
\beq
  \label{Sphvisc.eq}
  \Pi_i = \left\{ \begin{array}{l@{\quad}l}
                  \rho_i^{-1} (-\alpha_\Pi c_i \mu_i + \beta_\Pi \mu_i^2),
                  & ({\bf v}_i - {\bf v}_j) \cdot
                    ({\bf r}_i - {\bf r}_j) 
                 = \vij \cdot \rij < 0; \\
                0, & \mbox{otherwise},
                  \end{array} \right.
\eeq
\beq
  \label{Sphmu.eq}
  \mu_i = \frac{({\bf v}_i - {\bf v}_j) \cdot ({\bf r}_i - {\bf r}_j)}
               {h_i \left( \frac{|{\bf r}_i - {\bf r}_j|^2}
	       {h_i^2} + \zeta^2 \right)} 
        = \frac{\vij \cdot \veta_i}{\veta_i \cdot \veta_i + 
          \zeta^2},
\eeq
where $\alpha_\Pi$ and $\beta_\Pi$ are numerical constants of order unity,
$c_i^2 \equiv (dP/d\rho)_S = \gamma P_i/\rho_i$ the sound speed, and
$\zeta$ a numerical factor required to avoid divergences.  Note that this
artificial viscosity is easily represented as a function of $\veta$,
so that it may be directly implemented under ASPH.  This form of artificial
viscosity represents a hybrid of the usual Von Neumann-Richtmyer artificial
viscosity $\Pi = \beta_\Pi \rho l^2 (\grad \cdot {\bf v})^2$ and a bulk
viscosity $\Pi = -\alpha_\Pi \rho l c_s \grad \cdot {\bf v}$ (where $l$ is
the characteristic resolution scale for the shock), and has units of
$P/\rho^2$.  In order to maintain the symmetry of the SPH equations we must
use a symmetrized version of the artificial viscosity, which can be defined
analogously to equation (\ref{Wij.eq}) as $\Pi_{ij} = (\Pi_i + \Pi_j)/2$.

The Monaghan-Gingold formulation of the artificial viscosity outlined is
very successful at reproducing shock conditions in SPH simulations, but has
the unfortunate drawback of producing a great deal of artificial shear
viscosity.  This can cause spurious transport of angular momentum in
rotating systems, and is therefore of concern to us in this paper since we
are concerned with testing the angular momentum properties of ASPH.  In
order to deal with this problem, we implement a multiplicative correction
factor $f$ suggested by Balsara (1995)
\beq
  \tilde{\Pi}_{ij} = \Pi_{ij} \frac{f_i + f_j}{2},
\eeq
\beq
  \label{balsara.eq}
  f_i = \frac{|\Interp{\grad \cdot {\bf v}}_i|}
             {|\Interp{\grad \cdot {\bf v}}_i| + 
              |\Interp{\grad \times {\bf v}}_i| + 0.0001 c_i/h_i}.
\eeq
Note that in a shear-free flow ($\grad \cdot {\bf v} \neq 0$, $\grad \times
{\bf v} = 0$), $f = 1$ and we recover the normal Monaghan-Gingold form for
$\Pi_{ij}$.  However, in a pure shear flow ($\grad \cdot {\bf v} = 0$,
$\grad \times {\bf v} \neq 0$), $f = 0$ and the artificial viscosity is
suppressed.  Clearly this correction term cannot completely fix the problem
of artificial shear viscosity (such as in a combined compressive, shearing
flow), but at least it moderates the problem.

Artificial viscosity is a necessary evil for many hydrodynamic techniques,
and ideally should be used as little as possible.  The artificial viscosity
is present solely to account for shock conditions and ensure that (A)SPH
nodes do not interpenetrate.  In the definition of equation
(\ref{Sphvisc.eq}) $\Pi$ is restricted to be active solely for convergent
flows.  While a convergent flow is a minimal requirement for the presence
of a shock, not all convergent flows necessarily indicate shock conditions.
Overuse of $\Pi$ can lead to spurious heating of the system, resulting in
poor resolution of shocks.  This issue is of great concern for
gravitational collapse simulations, which is precisely the sort of
situation we are concerned with modeling in a gravitationally dominated
cosmological structure simulation.  In an effort to overcome these
shortcomings, we utilize a distinct interpolation kernel with the
artificial viscosity (\WPij\ vs.\ \Wij, see eqs. [\ref{Sphmom.eq}] and
[\ref{Sphtherm.eq}]).  In concert with ASPH's finer resolution, a judicious
choice for \WPij\ allows significant improvement in the resolution of
shocks.  This approach reduces the spurious preheating problem endemic to
the standard SPH implementation of the artificial viscosity, allowing the
capture of the shock jump over a few nodes.  This is discussed in \S
\ref{AV.sec}.  The Zel'dovich pancake test case in \S \ref{Zeldovich.sec} is
one example of the improvement that can be gained by this approach.

For the case of SPH, these forms of the dynamical equations rigorously
ensure the conservation of mass (with the summation approach to density),
linear momentum (due to the symmetrization of the momentum equation), and
angular momentum (because all node interactions are radially symmetric on a
pair-by-pair basis) -- energy is only conserved to second-order.  It is
worth noting, though, that some of these conservation properties are
somewhat artificially enforced.  For instance, we allow a variable
smoothing scale $h({\bf r})$, and therefore formally the expression for the
gradient of the kernel should include $\grad h$ terms.  These terms are
almost universally neglected by practitioners of SPH, as they must be
numerically estimated and it can be argued they are of secondary importance
(Evrard 1988).  However, we note that if such terms are included then the
expression for the gradient of the kernel becomes
\beq
  \grad W({\bf r},h) = \frac{\partial W({\bf r},h)}{\partial {\bf r}} 
                       \; + \; \frac{\partial W({\bf r},h)}
                       {\partial h} \grad h.
\eeq
This expression for the $\grad W$ is not necessarily radial for interacting
pairs of nodes, and therefore if these $\grad h$ terms are included angular
momentum is not rigorously conserved.

For the case of ASPH, using these forms of the dynamical equations again
rigorously enforces the conservation of mass and linear momentum.  However,
for an ellipsoidal kernel $\grad W$ is not necessarily radial, and
therefore angular momentum conservation is not guaranteed.  Based on our
experience with testing ASPH in rotational problems, we have found that so
long as one does not violate the basic assumptions upon which ASPH is
derived (the system studied should be smooth in nature, and the \Gt\ tensor
field should be well-behaved on scales of a few $h$), angular momentum will
be conserved acceptably (of order a few percent over several dynamical
times).  We discuss this issue in some detail in \S \ref{SHASPH.sec}, and
test the conservation of angular momentum in \S \ref{2dtests.sec}.

We parenthetically note that the obvious solution to the angular momentum
issue is to simply radialize the ASPH force in pairwise interactions between
nodes.  This would then guarantee the global conservation of angular
momentum, just as with SPH.  However, we have found that such an approach
yields unphysical results in practice.  For instance, using such radialized
forces in the Sedov problem leads to artificial, angular shapes to the
shock front.  The problem is easily understood, since forcing the ASPH
internode forces to be radial means that direction of the mutual forces
between nodes differs from the true gradient of the smoothing kernel.  The
only forces which will be correctly aligned are those along the principal
axes of the smoothing ellipses.

\subsection{(A)SPH Equations in Comoving Coordinates}
\label{codyneqs.app}
Our goal with this work is to use (A)SPH simulations to investigate the
formation and evolution of cosmological structure.  Such studies are most
naturally implemented in comoving, rather than proper, coordinates.  It is
also advantageous to convert from using time dependent equations to using a
power of the universal metric or expansion factor $a$, such that we evolve
in terms of an independent variable $p = a^\alpha$.  This is a standard
formalism used in collisionless cosmological simulations (Efstathiou \etal\
1985; Villumsen 1989), and we wish to similarly adapt (A)SPH.  In order to
accomplish this, we define several comoving quantities analogous to their
proper counterparts.  We denote comoving positions by ${\bf x}$, comoving
``velocities'' (in terms of $p$ rather than $t$) as ${\bf w} \equiv
d{\bf x}/dp$, and the comoving specific thermal energy by $\eps$.
All other comoving quantities are denoted by a superscript $c$ to avoid
confusion.  For the purposes of clarity, it is worthwhile to compile a list
of conversions for quantities between the proper time-dependent and
comoving frames.  Table \ref{prop2co.tab} provides such a summary.  Note
that in this table we relate non-relativistic quantities, appropriate for
use with our Newtonian approximation to the cosmology.  We also refer to
velocities as peculiar velocities ($\Sup{{\bf v}}{p}$ vs.\ ${\bf v}$) with
reference to the overall expansion of space.  While some of these
quantities may appear a bit strangely defined, recall that we are only
redefining the meaning of position in going from the proper to comoving
frame -- both time and mass remain unchanged.  The conversions of Table
\ref{prop2co.tab} result based on these choices.

The comoving forms of the Lagrangian conservation equations, as transformed
from their proper forms (eqs. [\ref{cont.eq}] to [\ref{therm.eq}]) are
\beq
  \frac{D\Sup{\rho}{c}}{Dp} = -\Sup{\rho}{c} \grad_x \cdot {\bf w},
\eeq
\beq
  \frac{D{\bf w}}{Dp} = -2 A(p) {\bf w} + B(p) \Sup{{\bf g}}{c}
                        - \dot{p}^{-2} \frac{\grad_x \Sup{P}{c}}{\Sup{\rho}{c}},
\eeq
\beq 
  \frac{D\eps}{Dp} = -\frac{1}{\dot{p}} \frac{\dot{a}}{a}
                     \lp 2 \eps + 3 \frac{\Sup{P}{c}}{\Sup{\rho}{c}} \rp -
                     \frac{\Sup{P}{c}}{\Sup{\rho}{c}} \grad_x \cdot {\bf w} +
                     \frac{a}{\dot{p}} \frac{\Gamma(T) - \Lambda(T)}
                                            {\Sup{\rho}{c}},
\eeq
where we use the definitions
\beq
  A(p) \equiv \frac{1 + \alpha + a \ddot{a} \dot{a}^{-2}}{2 \alpha a^\alpha},
         \quad \quad
  B(p) \equiv (\alpha^2 a^{2 \alpha + 1} \dot{a}^2)^{-1},
\eeq
and the evolution of the expansion factor is given by
\beq
  \dot{a} = \lp \frac{8 \pi G \Sup{\bar{\rho}}{c}}{3}
            \rp^{\frac{1}{2}} \left( \frac{1}{\Omega_i} 
            - 1 + \frac{1}{a} \right)^{\frac{1}{2}},
\eeq
where \Sup{\bar{\rho}}{c} is the average comoving mass density and
\Sub{\Omega}{i} the initial value of the cosmological density
parameter at $a = 1$, defined as the beginning of the simulation.

Transforming these comoving forms of the conservation equations to the
(A)SPH formalism, we find for the (A)SPH dynamical equations
\beq
  \label{cSphpos.eq}
  \frac{D{\bf x}_i}{Dp} = {\bf w}_i,
\eeq
\beq
  \label{cSphmom.eq}
  \frac{D{\bf w}_i}{Dp} = -2 A(p) {\bf w}_i + B(p) \SupSub{{\bf g}}{c}{i}
                          - \dot{p}^{-2} \sum_j m_j \left[ (\gamma - 1) \lp
                          \frac{\eps_i}{\SupSub{\rho}{c}{i}} + \frac{\eps_j}
                          {\SupSub{\rho}{c}{j}} \rp \gxWij + 
                           \SupSub{\Pi}{c}{{ij}} \gxWPij \right],
\eeq
\beqa
  \label{cSphtherm.eq}
  \frac{D\eps_i}{Dp} &=& \sum_j m_j {\bf w}_{ij} \cdot \left[ (\gamma - 1) 
                         \frac{\eps_i}{\SupSub{\rho}{c}{i}} \gxWij +
                         \frac{1}{2} \SupSub{\Pi}{c}{{ij}} \gxWPij \right] -\\
                     & & \frac{1}{\dot{p}} \frac{\dot{a}}{a} 
                         (3\gamma - 1) \eps_i +
                         \frac{a}{\dot{p}} \frac{\Gamma(T_i) - \Lambda(T_i)}
                                                {\SupSub{\rho}{c}{i}},
                         \nonumber
\eeqa
\beq
  \label{cSphcont.eq}
  \frac{D\SupSub{\rho}{c}{i}}{Dp} = -\SupSub{\rho}{c}{i} \grad_x \cdot 
                         {\bf w}_{ij}
                       = \sum_j m_j {\bf w}_{ij} \cdot \gxWij,
\eeq
\beq
  \label{cSphdens.eq}
  \SupSub{\rho}{c}{i} = \sum_j m_j \Wij,
\eeq
where we have explicitly used the ideal gas equation of state (eq.
[\ref{idealP.eq}]).

The treatment of the comoving artificial viscosity (\Sup{\Pi}{c}) deserves
some attention.  Since the artificial viscosity has units of
$\Sup{P}{{visc}}/\rho^2$, it transforms to comoving coordinates as $\Pi =
a^5 \Sup{\Pi}{c}$.  A comoving artificial viscosity term defined
analogously to equations (\ref{Sphvisc.eq}) and (\ref{Sphmu.eq}) is
therefore
\beq 
  \label{cSphvisc.eq}
  \SupSub{\Pi}{c}{i} = \left\{ \begin{array}{l@{\quad}l}
                       a^{-2} (\SupSub{\rho}{c}{i})^{-1}
                       (-\alpha_\Pi c_i \mu_i + \beta_\Pi \mu_i^2),
                     & \vij \cdot \rij = a^2 \dot{p} {\bf x}_{ij} \cdot 
                       {\bf w}_{ij} + a \dot{a} x_{ij}^2 < 0; \\
                     0, & \mbox{otherwise},
                     \end{array} \right.
\eeq
\beq
  \label{cSphmu.eq}
  \mu_i = \frac{a {\bf x}_{ij} \cdot (\dot{p} {\bf w}_{ij} + \dot{a}
                {\bf x}_{ij})}
               {\SupSub{h}{c}{i} \lp \frac{x_{ij}^2}
               {(\SupSub{h}{c}{i})^2} + \zeta^2 \rp}
        = \frac{a \Sup{\veta_i}{c} \cdot (\dot{p} {\bf w}_{ij} + 
                \dot{a} {\bf x}_{ij})}
               {\Sup{\veta_i}{c} \cdot \Sup{\veta_i}{c} + \zeta^2},
\eeq
where the sound speed is given by
\beq
  c_i = \lp \gamma \frac{P_i}{\rho_i} \rp^{1/2} = [\gamma (\gamma - 1) 
        u_i]^{1/2} = a [\gamma (\gamma -1 ) \eps_i]^{1/2}.
\eeq
Note that in equation (\ref{cSphvisc.eq}) the artificial viscosity is
switched on only for flows which are convergent in proper coordinates
rather than comoving.  This accounts for the extra term $a \dot{a}
x_{ij}^2$.

\section{Defining and Evolving the ASPH \Gt\ Tensor}
\label{Genderiv.app}
In this appendix we present the detailed mathematical derivation of the
\Gt\ tensor and its derivative $D\Gt/Dt$.  We begin by presenting these
derivations and justifications in a completely general, dimension-free
formalism.  Then in appendices \ref{2dasph.app} and \ref{3dasph.app} we
present the specific cases of 2D and 3D, respectively.

\subsection{\Gt\ as a Linear Transformation}

The \Gt\ tensor is defined as a linear transformation which maps from real
position space to normalized position space (${\bf r} \to \veta$).  In
this section we will use a superscript letter in parenthesis to denote
different coordinate frames.  A superscript $(r)$ implies the ordinary,
positional frame within which ${\bf r}$ is defined.  A superscript $(k)$
implies the primary coordinate frame of our smoothing kernel in which the
\Gt\ tensor is diagonal.  These two frames are related by a rotational
transformation.  We can find the representation of the \Gt\ tensor in the
real frame through a similarity transform
\beq
  \label{Gtensor.eq}
  \Gt^{(r)} = \Tr^{(k \to r)} \Gt^{(k)} \Tr^{(r \to k)}.
\eeq
Here $\Gt^{(r)}$ and $\Gt^{(k)}$ are the \Gt\ tensor represented in the
overall and kernel frames, and $\Tr^{(r \to k)}$ and $\Tr^{(k \to
r)}$ are the rotational transformations to and from the kernel frame,
respectively.  The advantage of this relation is that \Gt\ is trivial to
define in the kernel coordinate frame.  In this frame, $\Gt^{(k)}$ is
diagonal, with each diagonal element corresponding to the inverse smoothing
scale along that cardinal direction.

\subsection{The Evolution Equation $D\Gt/Dt$}
We will now derive the evolution equation for \Gt\ in arbitrary dimension.
The \Gt\ tensor associated with each particle can be viewed as defining an
ellipsoidal mass distribution according to $\rho({\bf r}) = m ~W(\Gt {\bf
r})$, where $m$ is the mass of the particle and $W(\Gt {\bf r})$ the
interpolation function.  For convenience, we define a new tensor \Jt\ in
terms of this density distribution as
\beq
  \label{Jdef.eq}
  J_{ij} = \frac{\int_V x_i x_j ~m ~W(\Gt {\bf r}) ~dV}
                {\int_V m ~W(\Gt {\bf r}) ~dV},
\eeq
where $x_i$ represents the $i$th coordinate of the position vector ${\bf
r}$.  Note that \Jt\ is similar to the inertia tensor, $I_{ij} = \int_V (r^2
\delta_{ij} - x_i x_j) ~\rho({\bf r}) ~dV$.  Evaluating equation
(\ref{Jdef.eq}), we find that \Jt\ is related to \Gt\ by
\beq
  \Jt = \Gt^{-2}.
\eeq

Adopting the convention that an unprimed quantity is at time $t$, and
primed at time $t + Dt$, the deformation tensor $\sigma_{ik} = \partial
v_i/\partial x_k$ maps position space as
\beq
  x_i^\prime = x_i + Dt ~\sigma_{ik} x_k.
\eeq
To first-order we then have
\beq
  x_i^\prime x_j^\prime \approx x_i x_j + x_i \sigma_{jl} x_l ~Dt + 
  x_j \sigma_{ik} x_k ~Dt,
\eeq
which implies that \Jt\ evolves as
\beq
  \frac{D J_{ij}}{Dt} = \sum_l \sigma_{jl} J_{li} + \sum_k \sigma_{ik} J_{kj}
                     = (\sigma J)_{ji} + (\sigma J)_{ij},
\eeq
so that
\beq
  \label{DJDt.eq}
  \frac{D \Jt}{Dt} = \sigt \Jt + \Jt^t \sigt^t
                   = \sigt \Jt + \Jt \sigt^t 
                   = \sigt \Gt^{-2} + \Gt^{-2} \sigt^t,
\eeq
where we denote a transposed matrix by a superscript $\null^t$.  This
relation is derived using the fact that \Gt\ and \Jt\ are symmetric, but
places no restrictions on the form of \sigt.  We can use equation
(\ref{DJDt.eq}) to find the evolution of \Gt\ as follows.
\beq
  \Delta \Jt = (\Gt + \Delta \Gt)^{-2} - \Gt^{-2}
             = (\sigt \Gt^{-2} + \Gt^{-2} \sigt^t) ~\Delta t.
\eeq
Working this out, keeping only first-order terms we get
\beq
  \Gt ~\Delta \Gt + \Delta \Gt ~\Gt = -(\Gt^2 \sigt + \sigt^t \Gt^2) ~\Delta t,
\eeq
\beq
  (\Delta \Gt + \Gt \sigt ~\Delta t) \Gt^{-1} + 
  \Gt^{-1} (\Delta \Gt + \sigt^t \Gt ~\Delta t) = 0,
\eeq
which can be rewritten
\beq
  \label{Rimply.eq}
  (\Delta \Gt + \Gt \sigt ~\Delta t) \Gt^{-1} + ((\Delta \Gt + 
  \Gt \sigt ~\Delta t) \Gt^{-1})^t = 0.
\eeq
Equation (\ref{Rimply.eq}) implies that the term $(\Delta \Gt + \Gt \sigt
~\Delta t) \Gt^{-1}$ represents an antisymmetric rotation matrix, which we
denote by $\Delta \Rt$.  The general solution for $D\Gt/Dt$ is then
\beq
  (\Delta \Gt + \Gt \sigt ~\Delta t) \Gt^{-1} = \Delta \Rt ~~~\Rightarrow~~~
  \Delta \Gt = \Delta \Rt ~\Gt - \Gt \sigt ~\Delta t,
\eeq
or
\beq
  \label{DGDt.eq}
  \frac{D\Gt}{Dt} = \frac{D\Rt}{Dt} \Gt - \Gt \sigt,
\eeq
where the infinitesimal rotational transformation $\Delta \Rt$ is uniquely
specified by
\beq
  \label{DRDt.eq}
  \Gt^{-1} \frac{D\Rt}{Dt} + \lp \Gt^{-1} \frac{D\Rt}{Dt} \rp^{t} =
  \Gt \sigt - \sigt^{t} \Gt.
\eeq
The infinitesimal rotation angles may either be solved for by expanding
equation (\ref{DRDt.eq}) or by setting the off-axis elements of $D\Gt/Dt$
from equation (\ref{DGDt.eq}) equal.  Either method yields (in $\nu$
dimensions) $\nu$ independent equations in $\nu$ unknowns.  Once these
rotational angles are found, the solution for $D\Gt/Dt$ is completely
specified.

\subsection{ASPH in 2D}
\label{2dasph.app}
We now present the full ASPH formalism for the 2D case.  The results of
this section have already been given in the body of the paper.

\subsubsection{The 2D \Gt\ tensor}
In order to define the \Gt\ tensor it is helpful to start from the
underlying geometry.  We will specify the geometry of the $\eta = 1$
isocontour (which in 2D is in general an ellipse) and define \Gt\ in terms
of this geometry.  We define the components of this isocontour as follows:
$h_1$ represents the semi-major axis, $h_2$ the semi-minor axis, and $\psi$
the position angle associated with the semi-major axis.  \Gt\ is therefore
defined in its primary (or kernel) frame by
\beq
  \Gt ^{(k)} = \lp
  \begin{array}{cc} 
     \hix & 0 \\ 
     0 & \hiy
  \end{array} 
  \rp.
\eeq
In the primary frame of \Gt, $h_1$ lies along the $x^{(k)}$ axis.  $\psi$
represents the angle of rotation for the transformation between the kernel
and real frames.  The rotational transformations relating the real and
kernel frames are then
\beq
  \Tr ^{(r \to k)} = \lp
  \begin{array}{cc}
    \cos \psi & \sin \psi \\
   -\sin \psi &  \cos \psi
  \end{array} \rp, \quad \quad
  \Tr ^{(k \to r)} = \lp
  \begin{array}{cc}
    \cos \psi & -\sin \psi \\
    \sin \psi &  \cos \psi
  \end{array} \rp.
\eeq
We can apply the similarity transform of equation (\ref{Gtensor.eq}) to
find the representation of $\Gt^{(r)}$
\beqa
  \Gt ^{(r)} &=& \Tr ^{(k \to r)} \Gt ^{(k)} \Tr ^{(r \to k)} \\
             &=& \lp
                 \begin{array}{cc}
                   \hix \cos ^{2} \psi + \hiy \sin ^{2} \psi & 
                   ( \hix - \hiy) \cos \psi \sin \psi       \\
                   ( \hix - \hiy) \cos \psi \sin \psi       & 
                   \hix \sin ^{2} \psi + \hiy \cos ^{2} \psi
                 \end{array} \rp. \nonumber
\eeqa

\subsubsection{The 2D \Gt\ Evolution Equation}
We now present the form of the 2D \Gt\ evolution equation.  First we need
the following 2D forms for \Gt, \sigt, and \Rt.
\beq 
  \Gt \equiv \lp
  \begin{array}{cc}
    G_{11} & G_{21} \\
    G_{21} & G_{22}
  \end{array} \rp,
\eeq
\beq
  \sigt \equiv \lp
  \begin{array}{cc}
    \sigma_{11} & \sigma_{12} \\
    \sigma_{21} & \sigma_{22}
  \end{array} \rp = \lp
  \begin{array}{cc}
    \vxx & \vxy \\
    \vyx & \vyy
  \end{array} \rp,
\eeq
\beq
  \frac{D\Rt}{Dt} = \lp
  \begin{array}{cc}
    0 & \dot{\theta} \\
    -\dot{\theta} & 0
  \end{array} \rp.
\eeq
We then have for $D\Gt/Dt$ via equation (\ref{DGDt.eq})
\beqa
  \frac{D\Gt}{Dt} &=& \frac{D\Rt}{Dt} \Gt - \Gt \sigt \\
  &=& \lp 
  \begin{array}{cc}
    DG_{11}/Dt & DG_{21}/Dt \\
    DG_{21}/Dt & DG_{22}/Dt
  \end{array} \rp \nonumber \\ 
  &=& \lp
  \begin{array}{cc}
    G_{21} (\dot{\theta} - \sigma_{21}) - G_{11} \sigma_{11} &
    G_{22} \dot{\theta} - G_{11} \sigma_{12} - G_{21} \sigma_{22} \\
    -G_{11} \dot{\theta} - G_{21} \sigma_{11} - G_{22} \sigma_{21} &
    -G_{21} (\dot{\theta} + \sigma_{12}) - G_{22} \sigma_{22}
  \end{array} \rp, \nonumber \\
  \dot{\theta} &=& \frac{G_{11} \sigma_{12} - G_{22} \sigma_{21} -
                       G_{21} (\sigma_{11} - \sigma_{22})}
                      {G_{11} + G_{22}},
\eeqa
where we have solved for $\dot{\theta}$ by setting $D\Gt_{12}/Dt =
D\Gt_{21}/Dt$.

\subsection{ASPH in 3D}
\label{3dasph.app}
\subsubsection{The 3D \Gt\ Tensor}
Once again we specify the form of the 3D \Gt\ tensor in terms of the
geometry of the $\eta = 1$ isocontour.  In order to specify rotations in 3D
space we adopt the so-called ``xyz'' convention as outlined in Goldstein
(1981), which has the advantage of being non-degenerate for infinitesimal
rotations.  We choose to represent the rotation angles to transform from
the real frame to the kernel's principal frame as $(\omega, \psi, \chi)$,
where $\omega$ is the yaw angle about the $z$ axis, $\psi$ is pitch angle
about the intermediate $y$ axis, and $\chi$ is the bank or roll angle about
the $x$ axis in the kernel's frame.  For the sake of notational compactness
we also adopt the convention that an angle subscripted by 1 represents the
cosine of that angle, and a subscript 2 represents the sine (\ie\ 
$\psi_1 \equiv \cos \psi$, $\psi_2 \equiv \sin \psi$).  The full rotational
transformations $\Tr^{(r \to k)}$ and $\Tr^{(k \to r)}$ can now be written
as
\beq
  \Tr^{(r \to k)} = \lp \begin{array}{ccc}
                    \psi_1 \omega_1 & \psi_1 \omega_2 & -\psi_2 \\
                    \chi_2 \psi_2 \omega_1 - \chi_1 \omega_2 &
                    \chi_2 \psi_2 \omega_2 + \chi_1 \omega_1 & 
                    \psi_1 \chi_2 \\
                    \chi_1 \psi_2 \omega_1 + \chi_2 \omega_2 &
                    \chi_1 \psi_2 \omega_2 - \chi_2 \omega_1 & 
                    \psi_1 \chi_1
                    \end{array} \rp,
\eeq
\beq
  \Tr^{(k \to r)} = \lp \begin{array}{ccc}
                    \psi_1 \omega_1 & 
                    \chi_2 \psi_2 \omega_1 - \chi_1 \omega_2 &
                    \chi_1 \psi_2 \omega_1 + \chi_2 \omega_2 \\
                    \psi_1 \omega_2 &
                    \chi_2 \psi_2 \omega_2 + \chi_1 \omega_1 &
                    \chi_1 \psi_2 \omega_2 - \chi_2 \omega_1 \\
                   -\psi_2 & \psi_1 \chi_2 & \psi_1 \chi_1
                    \end{array} \rp.
\eeq

We identify $(h_1, h_2, h_3)$ as the smoothing scales along the $(x^{(k)},
y^{(k)}, z^{(k)})$ axes in the kernel's primary frame, such that $h_1 \ge
h_2 \ge h_3$.  The \Gt\ tensor is therefore given in the kernels primary
frame by
\beq
  \Gt^{(k)} = \lp \begin{array}{ccc}
                     \hix & 0 & 0 \\
                     0 & \hiy & 0 \\
                     0 & 0 & \hiz 
                  \end{array} \rp,
\eeq
and applying the similarity transform (eq. [\ref{Gtensor.eq}]) we find
for $\Gt^{(r)}$
\beq
  \label{3dG.eq}
  \Gt^{(r)} = \Tr^{(k \to r)} \Gt^{(k)} \Tr^{(r \to k)} \equiv
              \lp \begin{array}{ccc}
                  G_{11} & G_{21} & G_{31} \\
                  G_{21} & G_{22} & G_{32} \\
                  G_{31} & G_{32} & G_{33}
                  \end{array} \rp,
\eeq
where the six unique matrix elements are given by
\beqa
  G_{11} &=& \hix \omega_1^2 \psi_1^2 + \\
         & & \hiy (\omega_1 \psi_2 \chi_2 - \omega_2 \chi_1)^2 + \nonumber \\
         & & \hiz (\omega_1 \psi_2 \chi_1 + \omega_2 \chi_2)^2, \nonumber \\
  G_{21} &=& \hix \omega_1 \omega_2 \psi_1^2 + \\
         & & \hiy (\omega_1 \psi_2 \chi_2 - \omega_2 \chi_1)
                  (\omega_2 \psi_2 \chi_2 + \omega_1 \chi_1) + \nonumber \\
         & & \hiz (\omega_2 \psi_2 \chi_1 - \omega_1 \chi_2)
                  (\omega_1 \psi_2 \chi_1 + \omega_2 \chi_2), \nonumber \\
  G_{31} &=& -\hix \omega_1 \psi_1 \psi_2 + \\
         & & \hiy \psi_1 \chi_2 
                  (\omega_1 \psi_2 \chi_2 - \omega_2 \chi_1) + \nonumber \\
         & & \hiz \psi_1 \chi_1 
                  (\omega_1 \psi_2 \chi_1 + \omega_2 \chi_2), \nonumber \\
  G_{22} &=& \hix \omega_2^2 \psi_1^2 + \\
         & & \hiy (\omega_2 \psi_2 \chi_2 + \omega_1 \chi_1)^2 + \nonumber \\
         & & \hiz (\omega_2 \psi_2 \chi_1 - \omega_1 \chi_2)^2, \nonumber \\
  G_{32} &=& -\hix \omega_2 \psi_1 \psi_2 + \\
         & & \hiy \psi_1 \chi_2 
                  (\omega_2 \psi_2 \chi_2 + \omega_1 \chi_1) + \nonumber \\
         & & \hiz \psi_1 \chi_1
                  (\omega_2 \psi_2 \chi_1 - \omega_1 \chi_2), \nonumber \\
  G_{33} &=& \hix \psi_2^2 + \hiy \psi_1^2 \chi_2^2 + \hiz \psi_1^2
             \chi_1^2.
\eeqa

Clearly these expressions for the elements of \Gt\ are quite unwieldy, and
would be most computationally expensive if we needed to evaluate these
expressions each time we wished to use the \Gt\ tensor.  Fortunately,
however, we need only use these expressions when we are initializing the
\Gt\ matrix.  Once we have the numerical values for these elements, we need
not concern ourselves with the geometry in order to use or evolve \Gt.

\subsubsection{The 3D \Gt\ Evolution Equation}
Finally, we present the form of the 3D \Gt\ evolution equation.
We define \Gt, \sigt, and \Rt as
\beq 
  \Gt \equiv \lp 
    \begin{array}{ccc}
      G_{11} & G_{21} & G_{31} \\
      G_{21} & G_{22} & G_{32} \\
      G_{31} & G_{32} & G_{33}
    \end{array} \rp,
\eeq
\beq
  \sigt \equiv \lp
  \begin{array}{ccc}
    \sigma_{11} & \sigma_{12} & \sigma_{13} \\
    \sigma_{21} & \sigma_{22} & \sigma_{23} \\
    \sigma_{31} & \sigma_{32} & \sigma_{33}
  \end{array} \rp = \lp
  \begin{array}{ccc}
    \vxx & \vxy & \vxz \\
    \vyx & \vyy & \vyz \\
    \vzx & \vzy & \vzz
  \end{array} \rp,
\eeq
\beq
  \frac{D\Rt}{Dt} = \lp
  \begin{array}{ccc}
    0 & \dgamma & -\dtheta \\
    -\dgamma & 0 & \dphi   \\
    \dtheta & -\dphi & 0
  \end{array} \rp,
\eeq
where we have used $(\dgamma, \dtheta, \dphi)$ as the infinitesimal
rotation angles for \Rt.  Then through equation (\ref{DGDt.eq}) we have for
$D\Gt/Dt$
\beq
  \frac{D\Gt}{Dt} = + \frac{D\Rt}{Dt} \Gt - \Gt \sigt = \lp
  \begin{array}{ccc}
    D\Gt_{11}/Dt & D\Gt_{12}/Dt & D\Gt_{13}/Dt \\
    D\Gt_{21}/Dt & D\Gt_{22}/Dt & D\Gt_{23}/Dt \\
    D\Gt_{31}/Dt & D\Gt_{32}/Dt & D\Gt_{33}/Dt
  \end{array} \rp,
\eeq
where the individual elements are
\beqa
  \label{3dHevolu1.eq}
  \frac{D\Gt_{11}}{Dt} &=& -G_{11} \sigma_{11} + G_{21} (\dot{\gamma} - 
    \sigma_{21}) - G_{31} (\dot{\theta} + \sigma_{31}), \\
  \frac{D\Gt_{12}}{Dt} &=& G_{22} \dot{\gamma} - G_{32} \dot{\theta} -
    G_{11} \sigma_{12} - G_{21} \sigma_{22} - G_{31} \sigma_{32}, \\
  \frac{D\Gt_{13}}{Dt} &=& G_{32} \dot{\gamma} - G_{33} \dot{\theta} - 
    G_{11} \sigma_{13} - G_{21} \sigma_{23} - G_{31} \sigma_{33}, \\
  \frac{D\Gt_{21}}{Dt} &=& -G_{11} \dot{\gamma} + G_{31} \dot{\phi} - 
    G_{21} \sigma_{11} - G_{22} \sigma_{21} - G_{32} \sigma_{31}, \\
  \frac{D\Gt_{22}}{Dt} &=& G_{32} (\dot{\phi} - \sigma_{32}) - G_{21}
    (\dot{\gamma} + \sigma_{12}) - G_{22} \sigma_{22}, \\
  \frac{D\Gt_{23}}{Dt} &=& -G_{31} \dot{\gamma} + G_{33} \dot{\phi} -
    G_{21} \sigma_{13} - G_{22} \sigma_{23} - G_{32} \sigma_{33}, \\
  \frac{D\Gt_{31}}{Dt} &=& G_{11} \dot{\theta} - G_{21} \dot{\phi} -
    G_{31} \sigma_{11} - G_{32} \sigma_{21} - G_{33} \sigma_{31}, \\
  \frac{D\Gt_{32}}{Dt} &=& G_{21} \dot{\theta} - G_{22} \dot{\phi} -
    G_{31} \sigma_{12} - G_{32} \sigma_{22} - G_{33} \sigma_{32}, \\
  \label{3dHevolu9.eq}
  \frac{D\Gt_{33}}{Dt} &=& G_{31} (\dot{\theta} - \sigma_{13}) -
    G_{32} (\dot{\phi} + \sigma_{23}) - G_{33} \sigma_{33}.
\eeqa

In order to solve for the rotation angles $(\gamma, \dtheta, \dphi)$, we
could set the symmetric off-axis elements of $D\Gt/Dt$ equal and solve the
resulting system of linear equations.
\beqa
  \dot{\gamma} &=& \frac{\gamma_c \gamma_d - \gamma_b \gamma_e}
                        {\gamma_a \gamma_c - \gamma_b^2}, \\
  \dot{\theta} &=& \frac{\gamma_b \gamma_d - \gamma_a \gamma_e}
                        {\gamma_a \gamma_c - \gamma_b^2}, \\
  \dot{\phi} &=& \frac{G_{31} \dot{\gamma} + G_{21} \dot{\theta} + C}
                      {G_{22} + G_{33}},
\eeqa
where we have defined for convenience
\beqa
  \gamma_a &\equiv& (G_{11} + G_{22}) (G_{22} + G_{33}) - G_{31}^2, \\
  \gamma_b &\equiv& (G_{22} + G_{33}) G_{32} + G_{21} G_{31}, \\
  \gamma_c &\equiv& (G_{11} + G_{33}) (G_{22} + G_{33}) - G_{21}^2, \\
  \gamma_d &\equiv& (G_{22} + G_{33}) A + G_{31} C, \\
  \gamma_e &\equiv& (G_{22} + G_{33}) B - G_{21} C,
\eeqa
\beqa
  A &\equiv& G_{11} \sigma_{12} - G_{21} (\sigma_{11} - \sigma_{22}) + 
    G_{31} \sigma_{32} - G_{22} \sigma_{21} - G_{32} \sigma_{31}, \\
  B &\equiv& G_{11} \sigma_{13} + G_{21} \sigma_{23} - G_{31} 
    (\sigma_{11} - \sigma_{33}) - G_{32} \sigma_{21} - 
    G_{33} \sigma_{31}, \\
  C &\equiv& G_{21} \sigma_{13} + G_{22} \sigma_{23} - G_{32} 
    (\sigma_{22} - \sigma_{33}) - G_{31} \sigma_{12} - 
    G_{33} \sigma_{32}.
\eeqa
We note that in the process of solving for these angles one must be
careful never to divide by off-axis elements of \Gt, as only the
diagonal elements of \Gt\ are guaranteed not to be zero.

This now completely specifies $D\Gt/Dt$ in closed (if somewhat ungainly)
form.  As before, only half (3) of the off-axis elements of $D\Gt/Dt$
need be evaluated, as $D\Gt/Dt$ is formally symmetric.  We present the
full expressions in equations (\ref{3dHevolu1.eq}) -
(\ref{3dHevolu9.eq}), since these forms are needed to find
$(\dot{\gamma}, \dot{\theta}, \dot{\phi})$.

\section{Numerical Algorithms}
\label{Spheral.app}
In this appendix we will briefly discuss some of the numerical algorithms
we have developed in the process of implementing ASPH, including time
integration (\S \ref{integrator.app}), timestep criteria (\S
\ref{timestep.app}), and neighbor selection (\S \ref{neighbor.app}).  While
these sorts of numerical details are not fundamental, we consider this
discussion worthwhile as it is imperative to code ASPH efficiently if we
wish to be able to perform large-scale, dynamic simulations.  This issue is
particularly crucial for ASPH because the increased spatial resolution (as
compared with SPH) implies correspondingly shorter timescales (and
therefore timesteps), increasing the computational demands for integrating
the system.  It is critical to an ASPH simulation that the time integration
be performed with sufficient temporal resolution, or the Courant and
related timestep criteria imply that the spatial resolution is
compromised.  The processes of integration and significant neighbor
selection typically dominate the computational time for (A)SPH simulations,
so we focus on these issues here.

\subsection{Asynchronous Time Integration}
\label{integrator.app}
In a typical cosmological structure simulation, regions of vastly different
densities and temperatures (and therefore timescales) can generically be
expected to evolve.  This follows directly from the gravitational
instability which is the driving source for this structure, as gravity
tends to form collapsed, dense, hot structures from initially cool, nearly
homogeneous gas.  In order to efficiently integrate such systems, it is
advantageous to decouple regions with such differing timescales and evolve
them independently, such that regions with small timesteps can be followed
without the penalty of having to advance the entire system at such smaller
timescales.  This is possible for (A)SPH due to the local nature of
hydrodynamic interactions, as reflected by the local sampling of the (A)SPH
interpolation kernel.  This suggests the development of an asynchronous
integrator, such that each (A)SPH node can evolve with its own timestep
and current time.  Clearly, since the evolution of each node depends upon
the state of its neighbors, neighboring nodes cannot be completely
decoupled.  However, by its nature (A)SPH is a smooth technique, and we
expect that neighboring nodes will be in physically similar states, and
thereby possess similar timescales.  This implies such an asynchronous
integration approach is consistent for (A)SPH.  We also enforce criteria in
the integration algorithm which insure that coupled nodes will remain
nearly synchronous.

Other investigators have also developed asynchronous integrators, such as
Hernquist \& Katz (HK89) who implement an asynchronous version of the
second-order accurate time-centered leapfrog algorithm in their TREESPH
code.  We choose instead to develop our integration algorithm around a
second-order Runge-Kutta scheme, which eases the conceptual complexity of
the algorithm.  This is primarily because Runge-Kutta maintains a node's
information at a single time, rather than having the variables and their
derivatives at differing times such as required by the leapfrog scheme.
The generalized second-order Runge-Kutta algorithm can be quantified as
follows.  In order to integrate a quantity $x$ at time $t$ through a
timestep $dt$ we use
\beq
  \label{RKint.eq}
  x(t + dt) = x(t) + \lp a_1 \frac{dx(t)}{dt} + a_2 \frac{dx(t + 
    \chi ~dt)}{dt}\rp ~dt,
\eeq
where
\beq
  \label{RKlimit.eq}
  a_1 = 1 - \frac{1}{2 \chi}, \quad
  a_2 = \frac{1}{2 \chi}, \quad
  \chi \in [0.5,0.9],
\eeq
and we have limits on $\chi$ in equation (\ref{RKlimit.eq}) for the sake of
stability.  Note that this scheme requires that
we be able to evaluate the derivatives for a given node $i$ at an
intermediate time $t_n = t_i + \chi_i ~dt_i$.  This requires that we have
synchronous information at time $t_n$ for all of the nodes significant to
$i$ (all nodes within a few $h$).  Fortunately, we need only integrate this
intermediate neighbor information to first-order in order to maintain the
second-order accuracy of equation (\ref{RKint.eq}), such that we can
extrapolate the set of information for all significant nodes $j$ via
\beq
  x_j(t_n) \approx x_j(t_j) + \frac{dx_j}{dt}(t_j)(t_n - t_j).
\eeq
The flexibility in the value of $\chi$ in equation (\ref{RKint.eq})
allows us to use this same set of synchronous neighbor information at $t_n$
for several nodes $i$, so long as $t_n$ falls within the desired range for
$\chi_i$.  This sort of recycling is important, as the process of
selecting neighbors and creating this synchronous information can become
computationally expensive if required too often.

We will now outline the asynchronous integration algorithm.  All nodes
start synchronously at a time $t_0$, with an allowed range of timesteps
$[\Sup{dt}{min}, \Sup{dt}{max}]$.  The goal of this algorithm is to advance
all nodes asynchronously to a ``goal time'' $\Sup{t}{goal} = t_0 +
\Sup{dt}{max}$, at which time they will all be synchronous again.  Note the
inclusion of gravity introduces a complication, in that while hydrodynamic
interactions are local in nature, gravitation is global.  Therefore, while
localized regions can be decoupled for the (A)SPH interactions, this is not
possible for the gravitational forces.  Fortunately, the gravitational
timescale (given roughly by the gravitational dynamical time
$\Sup{dt}{grav} = \Sup{\epsilon}{grav}/(G \rho)^{1/2}$, where
$\Sup{\epsilon}{grav}$ is the fraction of the gravitational timescale we
use) is typically the longest, least restrictive of the physical
timescales, and we can therefore set our overall $\Sup{dt}{max} =
\Sup{dt}{grav}$, and solve the gravitational problem synchronously.  In
order to determine the gravitational forces for the hydrodynamic
integrations in the interval $[t_0, \Sup{t}{goal}]$, we estimate the purely
gravitational forces at $t_0$, $t_0 + 0.5 \times \Sup{dt}{max}$, and
\Sup{t}{goal}, and parabolically interpolate for the forces in this
interval.  This is equivalent to integrating the gravitational problem
synchronously using the time-centered version of equation (\ref{RKint.eq}).
The asynchronous algorithm for the hydrodynamic interactions is:
\begin{enumerate}
\item All nodes start out synchronous at time $t_0$, each with a
current time $t_i$, timestep $dt_i \in [\Sup{dt}{min}, \Sup{dt}{max}]$, and
``target time'' $\SupSub{t}{targ}{i} = t_i + dt_i$.

\item A ``local goal time'' $\Sup{t}{lgoal} = \min(\SupSub{t}{targ}{i}) \;
\forall i$ is identified, which is simply the minimum available target
time.  All nodes which share this \Sup{t}{lgoal} are located and placed in
a sequential list.
\label{integ1.item}

\item The list of nodes due for integration is broken up into spatially
correlated batches, such that a single set of neighbor information can be
used for one such batch of nodes, rather than having to recalculate the
neighbor information for each node.  Such batches are identified as those
which share the same gridcell, as discussed in \S \ref{neighbor.app}.

\item We loop over each batch of nodes.  For each batch, a list of
synchronous neighbor information ($r_j, v_j, u_j,...$) is constructed at
both an intermediate time $t_n \in [t_i, \Sup{t}{lgoal}]$ and the end time
\Sup{t}{lgoal}.

\item Each node in a batch is integrated to second-order via equation
(\ref{RKint.eq}) using the intermediate neighbor information at $t_n$.
Once integrated, each nodes derivatives are updated using the neighbor
information at \Sup{t}{lgoal}.  The individual timesteps are also updated,
creating a new set of potential target times $\SupSub{t}{targ}{i}$.

\item Once all nodes with the currently targeted local goal time
$\Sup{t}{lgoal}$ have been integrated, we make another pass through all of
these nodes.  For each of the just integrated nodes the potential neighbors
are found, and the minimum target time for all these causally connected
nodes is determined.  The timestep for each of these linked nodes
(including the neighboring nodes, whether they were just integrated or not)
is then reset such that they all have this same minimum target time.  In
this way we ensure that nodes which are causally connected will be
integrated together synchronously, and that regions of small timescales
will not sweep through those with larger timescales before they can adapt.
\label{integ2.item}

\item We now loop back to step \ref{integ1.item}, and repeat this process
until all nodes have been advanced to \Sup{t}{goal}, at which time the
system is again synchronous.
\end{enumerate}

Finally, we note that for the purposes of efficiency it is useful to
try and keep as many nodes evolving at the same timestep and target time
as possible, such that the size of the batches that can be integrated
together will be increased.  In an effort to achieve this, we force all
timesteps to be integer multiples of the minimum timestep.  Of course,
forcing all nodes which are significant neighbors to one another to share
the same target time (as described in step \ref{integ2.item} above) is also
quite helpful for this purpose.

\subsection{Timestep Criteria}
\label{timestep.app}
As our integration scheme is based upon a second-order Runge-Kutta
algorithm, we have a good deal of flexibility in how to choose our
timesteps.  Typically, Runge-Kutta integrators use an accuracy limited
criterion to determine the step-size.  However, in order to determine such
an accuracy limited timestep requires trial integrations, which can become
prohibitively expensive under (A)SPH.  We therefore set the integration
timestep by using physical arguments about the timescales in the system,
leading to criteria such as the Courant time.  In general these criteria
underestimate the necessary timestep as compared with an accuracy limited
scheme, but the simplicity and speed with which these physical criteria can
be evaluated make up for the added integration cycles required.

In choosing a timestep appropriate for the gas dynamical calculations, the
basic criterion is that the timestep should be small enough such that the
fastest signal (of velocity $v_s$) can only propagate across a given
fraction $\epsilon$ of $h$, implying $\Delta t \le \epsilon h/v_s$.  Since
each ASPH node possesses an anisotropic smoothing scale embodied by \Gt, we
choose the smallest smoothing scale associated with \Gt\ to use for this
criterion.  This is given by the inverse of the maximum eigenvalue of \Gt.
There are three basic timescales which set our timestep choice.  The first
of these is the familiar sound-speed Courant condition, which can be
expressed for a given node $i$ by
\beq
  \SupSub{\Delta t}{c}{i} = \Sup{\epsilon}{c} \frac{\SupSub{h}{min}{i}}
    {\SupSub{c}{s}{i}},
\eeq
where $\Sup{\epsilon}{c}$ represents the fractional multiplier for the
Courant condition, $\SupSub{h}{min}{i}$ is the minimum smoothing scale
associated with $\Gt_i$, and $\SupSub{c}{s}{i} = (\gamma P_i/\rho_i)^{1/2} =
[(\gamma - 1) u_i]^{1/2}$ is the local sound speed.  Our second timescale
is set by the local divergence of the velocity field.  We need only concern
ourselves with the divergence of the velocity, since (A)SPH is a Lagrangian
technique and it is only the relative, rather than the bulk, velocity of
the nodes which is significant.  The eigenvalues of the symmetric part of
the local deformation tensor essentially measure this quantity on the scale
of $h$, so we can set this timescale as
\beq
  \SupSub{\Delta t}{v}{i} = \Sup{\epsilon}{v} /\min \left\{
    \mbox{Eigenvalue} \left[ \frac{1}{2}(\sigma_{\alpha \beta} + 
    \sigma_{\beta \alpha})_i \right] \right\}.
\eeq
Our final timescale limits how rapidly the \Gt\ tensor is allowed to
evolve (or equivalently the rate of density evolution), which can be set by
placing a limit on the evolution of the density.  This is accomplished
through
\beq
  \SupSub{\Delta t}{$\rho$}{i} = \Sup{\epsilon}{$\rho$} 
    \frac{\rho_i}{D\rho_i/Dt}.
\eeq
Together, these three relations determine our timestep for a given node
$i$, such that $\Delta t_i = \min(\SupSub{\Delta t}{c}{i}, \SupSub{\Delta
t}{v}{i}, \SupSub{\Delta t}{$\rho$}{i})$.  We have found experimentally
that using $\Sup{\epsilon}{c} = \Sup{\epsilon}{v} = \Sup{\epsilon}{$\rho$}
= 0.1$ is successful.  Such choices are in fact quite conservative, but our
asynchronous approach to the integration allows us to be somewhat generous
here.

There is yet one more timescale which we must consider when setting the
timestep: the gravitational timescale.  As mentioned previously in appendix
\ref{integrator.app}, we can set this timestep as a fraction of the
gravitational dynamical time, $\Sup{dt}{grav} = \Sup{\epsilon}{grav}/(G
\rho)^{1/2}$.  In practice in our current code, we choose a fixed timestep
$\Sup{dt}{max}$ such that we expect $\Sup{dt}{max} \lesssim
\Sup{\epsilon}{grav}/(G \rho)^{1/2}$ at all times, and evolve the
gravitational problem at this fixed timestep.  We only explicity evaluate
the gas dynamical times listed above, which are used to set the individual
timesteps for the ASPH particles, so that $\Delta t_i = \min(\SupSub{\Delta
t}{c}{i}, \SupSub{\Delta t}{v}{i}, \SupSub{\Delta t}{$\rho$}{i},
\Sup{dt}{max})$.  The dark matter, of course, evolves at the fixed timestep
$\Sup{dt}{max}$.

\subsection{Neighbor Selection}
\label{neighbor.app}
Hydrodynamic interactions are strongly local in nature, which is why (A)SPH
kernel estimates only sample neighboring nodes out to a few smoothing
scales.  For this reason, developing an efficient method of identifying
only those nodes which are within a few $h$ of a given nodes position can
potentially greatly increase the speed of an (A)SPH code.  For example, if
we have a simulation of $N$ nodes, each of which sample roughly $N_n \ll N$
significant neighbors, ideally the computation time should scale as ${\cal
O}(N N_n)$.  If the problem were treated as a global interaction, such that
no effort were made to identify only those neighbors which are significant
for any given interaction, the computational time would scale like ${\cal
O}(N^2)$.  This is an enormous difference, and therefore for large-scale
(A)SPH simulations it is imperative that an efficient neighbor finding
algorithm be developed.  ASPH presents two complications such an algorithm
must deal with.  First, since each node samples ellipsoidal regions, we
would like to develop an algorithm which can recognize this anisotropy when
selecting candidate neighbors.  Secondly, our symmetrization scheme
(eq. [\ref{Wij.eq}]) requires that we not only find all nodes $j$ which
fall under the influence of the node in question $i$, but also any nodes
which may happen to lie outside of the cutoff normalized radius $\eta_i >
\Sup{\eta}{cut}$ and yet still influence $i$ because $\eta_j <
\Sup{\eta}{cut}$.  The criteria for determining whether or not a given node
$j$ should be counted as significant for $i$ can be quantified as
\beq
  \SupSub{\eta}{min}{ij} \equiv \min(\eta_i, \eta_j) \le \Sup{\eta}{cut}.
\eeq

We have developed an algorithm for finding a group of such potential
nearest neighbors, which relies on a two-stage culling process.  The first
step is based upon the popular gridcell method, whereby the simulation
volume is divided up into a collection of subvolumes or gridcells.  Each
node is then associated with the gridcell within which it happens to fall,
allowing a fast but crude spatial localization of the nodes (at least to
the resolution scale of the gridcell size).  Ideally, the gridcell size
should be related to the smoothing scale, such that the length of a
gridcell is roughly the radius of influence for a given nodes influence
(\ie, a few $h$).  However, in our implementations of both SPH and ASPH the
smoothing scales vary, and for ASPH there isn't even a single, unique
smoothing scale per node.  We deal with these issues by establishing a
hierarchy of gridlevels.  On successive gridlevels the linear gridcell size
is halved, such that
\beq
  \Sup{\Delta}{g} = \frac{\Sup{\Delta}{0}}{2^g},
\eeq
where $\Sup{\Delta}{g}$ represents the length of one side of a gridcell on
gridlevel $g$, and $\Sup{\Delta}{0}$ represents the top-most gridcell size
(on gridlevel $g=0$).  Note that under this convention the gridcell sizes
decrease with increasing gridlevel, and for $N_g$ total gridlevels, $g \in
[0, N_g - 1]$.  Each node is now associated with a particular gridlevel and
gridcell on that level.  The appropriate gridlevel for a given node $i$ is
defined to be the ``deepest'' (maximum $g$) level on which that node can
only influence at most one gridcell in any direction, implying
\beq
  \SupSub{\Delta}{g}{i} \ge \Sup{\eta}{cut} \SupSub{h}{max}{i},
\eeq
giving us
\beq
  g_i \le \log_2 \lp \frac{\Sup{\Delta}{0}}{\Sup{\eta}{cut} 
    \SupSub{h}{max}{i}} \rp,
\eeq
where we have defined \SupSub{h}{max}{i} to be the maximum smoothing scale
associated with node $i$.  Once each node is associated with a gridlevel
and gridcell in this fashion, the search algorithm for finding a list of
potential neighbors for node $i$ goes as follows.  Beginning with the
topmost ($g=0$) gridlevel, we identify which gridcell contains node $i$,
and build a list of all nodes on this gridlevel which are in this or any
immediately adjacent gridcells.  We descend through gridlevels $g \le g_i$
and repeat this process.  This stage picks up any nodes with smoothing
scales greater than our node in question $\SupSub{h}{max}{j} \ge
\SupSub{h}{max}{i}$.  For $g > g_i$, we are now dealing with nodes which
possess smoothing scales $\SupSub{h}{max}{j} < \SupSub{h}{max}{i}$.  In
this case, we must check all adjacent gridcells out to a radius equivalent
to the gridcell size on level $g_i$ (a radius in gridcells on gridlevel $g$
of $2^{g - g_i}$).  Once we have descended through all gridlevels in this
fashion, we will have a list of potential neighbor nodes for node $i$
guaranteed to include all nodes which meet the criteria
$\SupSub{\eta}{min}{ij} \le \Sup{\eta}{cut}$.  Note that this list of
potential neighbors is in fact appropriate for all nodes which are members
of $i$'s gridcell.  This provides us with a logical definition for the
groups of nodes which are to be integrated together, as described in \S
\ref{integrator.app}.  Batches of nodes which are defined as those nodes
which are assigned to the same gridcell should all have the same target
time for integration, and the same set of neighbor information can be used
for them all.

While this gridcell search is quick and efficient, it is still possible to
further cull the resulting list of potential neighbors to a smaller set.
There are two reasons for this.  First, the volume per gridcell on each
gridlevel decreases as $2^\nu$ in $\nu$ dimensions, which is not a very
fine scale.  Secondly, we have not yet capitalized upon the anisotropy of
the smoothing scales, but rather have used the maximum smoothing scale
associated with each node to select neighbors.  We therefore implement a
second culling stage to the neighbor selection process.  For each $\Gt_i$,
the maximum smoothing scale in each dimension $\SupSub{{\bf h}}{max}{i}
\equiv (\SupSub{h}{max}{x}, \SupSub{h}{max}{y}, \SupSub{h}{max}{z})_i$ can
be used to more finely cull the potential neighbor list.  This process is
slightly complicated by our symmetrization scheme and the fact that we want
our list of potential neighbors to apply to all nodes in a particular
gridcell, rather than a single node.  We employ the following culling
algorithm, wherein we denote nodes which are members of our target gridcell
with the subscript $i$ and the full list of potential neighbors as $j$.
(The set $i$ is therefore a subset of $j$.)
\begin{enumerate}
\item For all nodes $i$, identify the minimum and maximum coordinates
$(\Sup{xi}{min}, \Sup{xi}{max})_\alpha \equiv (\min(x_i)_\alpha \forall i,
\max(x_i)_\alpha \forall i)$, and the minimum and maximum coordinates
influenced by these nodes $(\Sup{xih}{min}, \Sup{xih}{max})_\alpha \equiv
(\min(x_i - \Sup{\eta}{cut} \SupSub{h}{max}{i})_\alpha \forall i, \max(x_i
+ \Sup{\eta}{cut} \SupSub{h}{max}{i})_\alpha \forall i)$. 

\item Loop over all nodes $j$.  

\item For each $j$ verify whether or not the node is significant to any $i$
node (fulfilling the criterion $\eta_i < \Sup{\eta}{cut}$) by verifying
that ${\bf x}_j \in [\Sup{{\bf xih}}{min}, \Sup{{\bf xih}}{max}]$.

\item Then check whether any $i$ node can count as significant to $j$
(fulfilling the criterion $\eta_j < \Sup{\eta}{cut}$) by verifying the
volumes defined by $[{\bf x}_j - \Sup{\eta}{cut} \SupSub{{\bf h}}{max}{j}, 
{\bf x}_j + \Sup{\eta}{cut} \SupSub{{\bf h}}{max}{j}]$ and
$[\Sup{{\bf xi}}{min}, \Sup{{\bf xi}}{max}]$ overlap.

\item If a node $j$ fails both of these tests, then remove it from the list
of candidate neighbors.
\end{enumerate}

Together these two steps quickly create a relatively small list of
candidate neighbor nodes for a given gridcell.  Note that this algorithm is
guaranteed to find all significant neighbors, so long as the topmost
gridcell size meets the criterion $\Sup{\Delta}{0} \ge \Sup{\eta}{cut}
\Sup{h}{max}$.
\end{appendix} 

\clearpage
\begin{deluxetable}{rll}
\tablecaption{Simulation Parameters for Zel'dovich Pancake Simulations
\label{Zeldovich.tab}}
\tablehead{\colhead{Parameter(s)} & \colhead{2D} & \colhead{3D}}
\startdata
  $\Sup{\Omega}{bary}, \Sup{\Omega}{dm}, \Lambda$ & 0.5, 0.5, 0 \\
  $H_0$ & 50 km/sec/Mpc \\
  $a_i, a_c, a_f, a_0$ & 1, 4, 10, 1000 \\
  $z_i, z_c, z_f, z_0$ & 999, 249, 99, 0 \\
  Periodic Simulation Volume & $(x \in [0,1], y \in [0,1])$ 
                             & $(x \in [0,1], y \in [0,1], z \in [0,1])$ \\
  \SupSub{l}{phys}{i} & 1 Mpc \\
  $T_i$ & 3000 K \\
  $\gamma$ & 5/3 \\
  $\mu$ & 1 \\
  $\Sup{N}{bary}, \Sup{N}{dm}$ &  $64^2, 64^2$
                               &  $32^3, 32^3$ \\
  $N_h$ & 2 \\
  $\alpha$ & 1 \\
  $dp$ & $\in [1.0 \times 10^{-5}, 0.05]$ \\
  $h$ & $\in [1.0 \times 10^{-5}, 0.12]$ \\
  PM mesh size \Sup{N}{grid} & $128^2$ & $128^3$ \\
\enddata
%\tablecomments{\footnotesize Explanation of individual parameters:
%\Sup{\Omega}{bary} fraction of critical mass density in baryons; 
%\Sup{\Omega}{dm} fraction of critical mass density in dark matter;
%$\Lambda$ cosmological constant; $H_0$ ``current'' Hubble constant;
%$a_i, a_c, a_f$ expansion factors at time of beginning,
%caustic formation, and end of simulation, $a_0$ represents the ``current''
%expansion factor of the universe; $z_i, z_c, z_f, z_0$
%redshifts at beginning, caustic formation, end, and ``current'' time;
%\Sup{l}{phys} initial physical size of simulation (used to convert
%results to cgs units); $T_i$ initial average temperature within the
%simulation volume; $\gamma = c_p/c_v$ the ratio of the specific heats;
%$\mu$ the mean molecular weight of the gas in atomic units; $N_{bary},
%N_{dm}$ number of baryon and dark matter nodes; $N_h$ number of neighboring
%nodes per smoothing scale initially; $\alpha$ power of expansion factor
%used for integration variable $(p = a^\alpha)$; $dp$ integration step size;
%$h$ smoothing scale limits applied to simulation (internal units);
%\Sup{N}{grid} number of PM gridcells in computational volume.}
\end{deluxetable}

\clearpage
\begin{deluxetable}{rll}
\tablewidth{0pt}
\tablecaption{Simulation Parameters for Sedov Blastwave Simulations
\label{Sedov.tab}}
\tablehead{\colhead{Parameter(s)} & \colhead{2D} & \colhead{3D}}
\startdata
  Periodic simulation volume & ([-0.5,0.5], [-0.5,0.5]) 
         & ([-0.5,0.5], [-0.5,0.5], [-0.5,0.5]) \\
  $\Sub{{\bf r}}{spike}$ & (0,0,0) \\
  \Sup{E}{spike} & 1.0 \\
  \Sup{u}{bkgd} & $10^{-10}$ \\
  $\rho_0$ & 1 \\
  $\gamma$ & 1.4 \\
  $t_f$ & 0.16 & 0.1 \\
  $N$ & $128^2$ & $32^3$ \\
  $N_h$ & 2 \\
  $dt$ & $\in [1.0 \times 10^{-5}, 0.001]$ \\
  $h$ & $\in [1.0 \times 10^{-5}, 0.12]$ \\
\enddata
\end{deluxetable}

\clearpage
\begin{deluxetable}{rl}
\tablewidth{0pt}
\tablecaption{Simulation Parameters for Riemann shocktube simulations
\label{Shocktube.tab}}
\tablehead{\colhead{Parameter(s)} & \colhead{Value(s)}}
\startdata
  Periodic simulation volume & $(x \in [-1,1], y \in [-0.25,0.25])$ \\
  High Density Region $(x < 0)$: $\rho_0, P_0, {\bf v}_0$ &
                                 1.0, 1.0, 0.0 \\
  Low Density Region $(x > 0)$: $\rho_0, P_0, {\bf v}_0$ &
                                 0.25, 0.1795, 0.0 \\
  $\gamma$ & 1.4 \\
  $t_f$ & 0.15 \\
  $N$ & 6250 \\
  $N_h$ & 2 \\
  $dt$ & $\in [0.0005, 0.01]$ \\
  $h$ & $\in [10^{-5}, 0.0625]$ \\
\enddata
\end{deluxetable}

\clearpage
\begin{deluxetable}{rl}
\tablewidth{0pt}
\tablecaption{Simulation Parameters for Double Blast Wave simulations
\label{DB.tab}}
\tablehead{\colhead{Parameter(s)} & \colhead{Value(s)}}
\startdata
  Periodic simulation volume & $(x \in [-1,1], y \in [0,0.125])$ \\
  $\rho_0$ & 1 \\
  $P(|x| \leq 0.1)$ & 1000 \\
  $P(0.1 < |x| \leq 0.9)$ & 0.01 \\
  $P(|x| \leq 1)$ & 100 \\
  $\gamma$ & 1.4 \\
  $t_f$ & 0.038 \\
  $N$ & 16384 \\
  $N_h$ & 2 \\
  $dt$ & $\in [10^{-8}, 10^{-3}]$ \\
  $h$ & $\in [10^{-5}, 0.125]$ \\
\enddata
\end{deluxetable}

\clearpage
\begin{deluxetable}{rl}
\tablewidth{0pt}
\tablecaption{Simulation Parameters for Pseudo-Keplerian Disk Simulations
\label{Kep.tab}}
\tablehead{\colhead{Parameter(s)} & \colhead{Value(s)}}
\startdata  
  Non-periodic volume & $(x \in [-0.5, 0.5], y \in [-0.5, 0.5])$ \\
  Gas disk centered on $(x_d, y_d)$ & $(0,0)$ \\
  $G$ & 1 \\
  Gravitating point mass $M$ & 1 \\
  $\rho_0$ & 1 \\
  $\gamma$ & 5/3 \\
  $r_c$ & 0.05 \\
  $r_d$ & 0.25 \\
  Rotation Period at core radius $\tau(r_c)$ & 0.1 \\
  $t_f$ & 3.0 \\
  $N$ & 4096 \\
  $N_h$ & 2 \\
  $dt$ & $\in [7.8125 \times 10^{-6}, 0.001]$ \\
  $h$ & $\in [10^{-5}, 0.12]$ \\
\enddata
\end{deluxetable}

\clearpage
\begin{deluxetable}{rl}
\tablewidth{0pt}
\tablecaption{Simulation Parameters for Collapsing Disk Simulations
\label{Cdisk.tab}}
\tablehead{\colhead{Parameter(s)} & \colhead{Value(s)}}
\startdata
  Non-periodic Volume &  $(x \in [-0.5, 0.5], y \in [-0.5, 0.5])$ \\
  Gas disk centered on $(x_d, y_d)$ & $(0,0)$ \\
  $G$ & 1 \\
  \Sup{M}{disk} & 1 \\
  $r_d$ & 0.25 \\
  $\Omega/\Omega_0$ & 0.25 \\
  $f$ & 0.001 \\
  $\gamma$ & 5/3 \\
  $t_f$ & 0.5 \\
  $N$ & 4096 \\
  $N_h$ & 2 \\
  $dt$ & $\in [1.0 \times 10^{-5}, 5 \times 10^{-4}]$ \\
  $h$ &  $\in [10^{-5}, 0.12]$ \\
  PM mesh size \Sup{N}{grid} & $128^2$ \\
\enddata
\end{deluxetable}

\clearpage
\begin{deluxetable}{ccc}
\tablewidth{0pt}
\tablecaption{Proper $\to$ comoving conversions \label{prop2co.tab}}
\tablehead{\colhead{Quantity} & \colhead{Proper frame} & \colhead{Comoving
frame}}
\startdata
  Position & ${\bf r}$ & $a {\bf x}$ \\
  Peculiar Velocity & \Sup{{\bf v}}{p} & $a \Sup{{\bf v}}{pc} = 
                                          a \dot{p} {\bf w}$ \\
  Specific Thermal Energy & $u$          & $a^2 \eps$ \\
  Temperature & $T$          & $a^2 \Sup{T}{c}$ \\
  Density & $\rho$       & $a^{-3} \Sup{\rho}{c}$ \\
  Pressure & $P$          & $a^{-1} \Sup{P}{c}$ \\
  Spatial Gradient Operator &  $\grad_r$    & $a^{-1} \grad_x$ \\
\enddata
\end{deluxetable}
\end{document}